\documentclass[journal]{IEEEtran}
\usepackage[T1]{fontenc}
\usepackage{pifont}
\usepackage{array}
\usepackage{booktabs}
\usepackage{graphicx}
\usepackage{enumitem}
\usepackage{bigstrut}
\makeatletter

\newcommand{\Rmnum}[1]{\expandafter\@slowromancap\romannumeral #1@}
\makeatother
\usepackage{lineno}
\usepackage{color}
\usepackage{amssymb}
\usepackage{amsmath}
\usepackage{algorithm}
\usepackage{algorithmic}
\usepackage{subfigure}
\usepackage{multirow}
\usepackage{multicol}
\usepackage{threeparttable}
\usepackage{cite}
\usepackage[x11names]{xcolor}
\usepackage{tikz}
\usepackage[colorlinks,
            linkcolor=c1,
            anchorcolor=black,
            citecolor=cyan]{hyperref}
\usetikzlibrary{shapes,trees}
\definecolor{c1}{rgb}{1,0.5,0.75}
\usepackage{forest}
\usetikzlibrary{arrows.meta,shapes,positioning,shadows,trees}
\ifCLASSINFOpdf
\else
\fi

\begin{document}
%
\title{A Comprehensive Survey on Underwater Acoustic Target Positioning and Tracking: Progress, Challenges, and Perspectives}

\author{Zhong~Yang,~\IEEEmembership{}
        Zhengqiu~Zhu,~\IEEEmembership{}
        Yong~Zhao,~\IEEEmembership{}
        Yonglin~Tian,~\IEEEmembership{Member,~IEEE,}
        Changjun~Fan,~\IEEEmembership{} 
        Runkang~Guo,~\IEEEmembership{} \\
        Wenhao~Lu,~\IEEEmembership{}
        Jingwei~Ge,~\IEEEmembership{}
        Bin~Chen,~\IEEEmembership{}
        Yin~Zhang,~\IEEEmembership{Senior Member,~IEEE,}
        Guohua~Wu,~\IEEEmembership{Senior Member,~IEEE,}
        Rui~Wang,~\IEEEmembership{Senior Member,~IEEE,}
        Gyorgy~Eigner,~\IEEEmembership{Senior Member,~IEEE,}
        Guangquan~Cheng,~\IEEEmembership{}
        Jincai~Huang,~\IEEEmembership{} \\
        Zhong~Liu,~\IEEEmembership{Member,~IEEE,}
        Jun~Zhang,~\IEEEmembership{}
        Imre J. Rudas,~\IEEEmembership{Fellow,~IEEE,}
        and~Fei-Yue~Wang,~\IEEEmembership{Fellow,~IEEE}
\thanks{Zhong Yang, Zhengqiu Zhu, Yong Zhao, Changjun Fan, Runkang Guo, Wenhao Lu, Rui Wang, Guangquan Cheng, Jincai Huang, Zhong Liu, and Jun Zhang are 
with the College of Systems Engineering, National University of Defense Technology, Changsha 410073, Hunan Province, China. 
(e-mail:$\{$yangzhong19, zhuzhengqiu12, zhaoyong15, fanchangjun, guorunkangnudt, luwenhao20, rui$\_$wang, cgq299, huangjincai, liuzhong, zhangjun$\}$@nudt.edu.cn)
}
\thanks{Yongling Tian is with the State Key Laboratory of Multimodal Artificial Intelligence Systems, Institute of Automation, Chinese Academy of Sciences, Beijing 100190, China. (e-mail:~yonglin.tian@ia.ac.cn).}
\thanks{Jingwei Ge is with the University Research and Innovation Center, Obuda University, 1034 Budapest, Hungary and the Department of Automation, Tsinghua University, Beijing 100084, China. (e-mail:~gjw19@tsinghua.org.cn).}
\thanks{Bin Chen is with the Institute of Intelligent Computing, University of Electronic Science and Technology of China (UESTC), Chengdu 611731, Sichuan Province, China. (email:~chenbin06@uestc.edu.cn).}
\thanks{Yin Zhang is with the School of Information and Communication Engineering, University of Electronic Science and Technology of China (UESTC), Chengdu 611731, Sichuan Province, China. (email:~zhangyin123@uestc.edu.cn).}
\thanks{Guohua Wu is with the School of Automation, Central South University, Changsha 410073, China. (email:~guohuawu@csu.edu.cn).}
\thanks{Gyorgy Eigner is with the Physiological Controls Group, University Research and Innovation Center, Obuda University, 1034 Budapest, Hungary. (email:~eigner.gyorgy@uni-obuda.hu).}
\thanks{Imre J. Rudas is with the Antal Bejczy Center for Intelligent Robotics, University Research, Innovation and Service Center, Obuda University, 1034 Budapest, Hungary. (email:~rudas@uni-obuda.hu).}
\thanks{Fei-Yue Wang with the State Key Laboratory for Management and Control of Complex Systems, Chinese Academy of Sciences, Beijing 100190, and also with the Department of Engineering Science, Faculty of Innovation Engineering, Macau University of Science and Technology, Macau 999078, China. (email:~feiyue.wang@ia.ac.cn).}
}

%



\maketitle

\begin{abstract}
Underwater target tracking technology plays a pivotal role in marine resource exploration, environmental monitoring, and national defense security. Given that acoustic waves represent an effective medium for long-distance transmission in aquatic environments, underwater acoustic target tracking has become a prominent research area of underwater communications and networking. Existing literature reviews often offer a narrow perspective or inadequately address the paradigm shifts driven by emerging technologies like deep learning and reinforcement learning. To address these gaps, this work presents a systematic survey of this field and introduces an innovative multidimensional taxonomy framework based on target scale, sensor perception modes, and sensor collaboration patterns. Within this framework, we comprehensively survey the literature (more than 180 publications) over the period 2016-2025, spanning from the theoretical foundations to diverse algorithmic approaches in underwater acoustic target tracking. Particularly, we emphasize the transformative potential and recent advancements of machine learning techniques, including deep learning and reinforcement learning, in enhancing the performance and adaptability of underwater tracking systems. Finally, this survey concludes by identifying key challenges in the field and proposing future avenues based on emerging technologies such as federated learning, blockchain, embodied intelligence, and large models.
\end{abstract}


%
\IEEEpeerreviewmaketitle

\section{Introduction}
%
%
%
%
\IEEEPARstart{T}{he} ocean harbors vast resources and energy reserves, representing both a critical resource reservoir and a vital domain for sustainable human development, which has attracted growing global attention in recent decades. Simultaneously, marine ecosystems provide essential ecological services to humanity, encompassing gas regulation, nutrient cycling, and waste treatment. Moreover, maritime control constitutes a fundamental safeguard for national security and sovereignty \cite{s1.2022}. Consequently, the efficient utilization of marine resources, the enhanced conservation of marine ecosystems, and the protection of national maritime interests have emerged as an international consensus \cite{ghafoor2019}. 
\begin{figure*}[ht]
    \centering
    \includegraphics[width=0.9\linewidth]{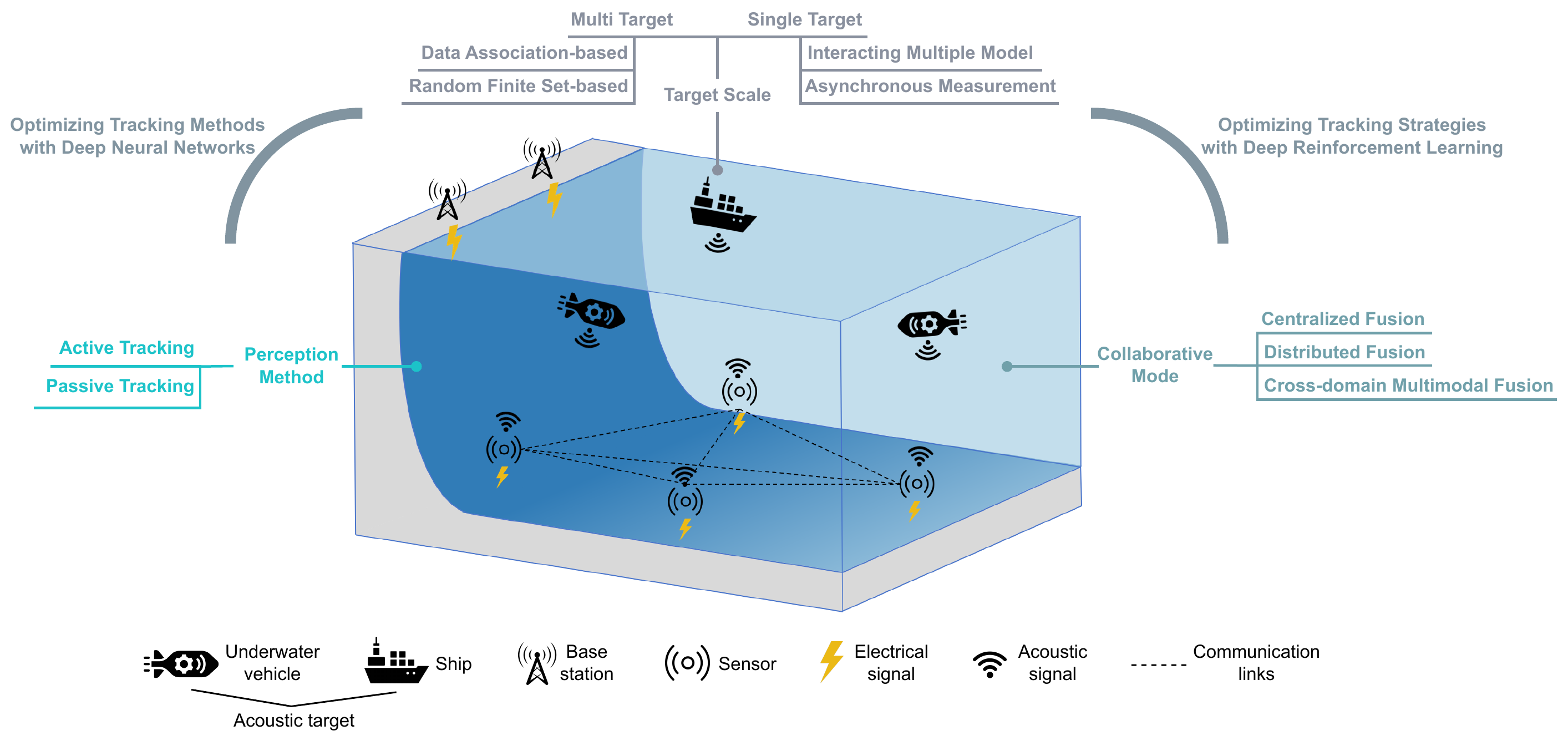}
    \caption{An illustrative scenario of underwater acoustic target tracking. The primary roles in the scenario are targets and sensors, and the taxonomy framework based on target scale, perception method, and collaboration mode is identified using them and the connections between them as the taxonomy benchmark. }
    \label{fig:changjing}
\end{figure*}

Accurate and effective tracking of underwater targets is crucial for both marine resource exploration/protection and maritime security maintenance. As a pivotal research focus in marine science and technology, underwater target tracking technology has gained prominence. Recent advancements in sensor capabilities, coupled with rapid progress in multi-source information fusion and artificial intelligence, have significantly advanced this technology, enabling its applications across diverse sectors including national defense security \cite{mansur2014}, environmental monitoring \cite{liu2021}, and resource exploration \cite{xu2019}. 

Due to the complexity of marine environments and target diversity, the development of underwater target tracking technology faces significant challenges  \cite{2023水声目标探测和识别}. The propagation characteristics of signals are substantially influenced by marine conditions \cite{2019水声目标探测技术}, while underwater targets usually demonstrate high mobility and concealment capabilities \cite{oracevic2014}. Conventional terrestrial tracking media, including electromagnetic waves and lasers, experience severe absorption and attenuation in seawater, resulting in poor performance for marine applications \cite{yan2021}. In contrast, acoustic waves remain the sole medium capable of long-range underwater propagation, making them the predominant choice for underwater target tracking.

Figure \ref{fig:changjing} depicts a typical underwater acoustic target tracking scenario, where moving targets emit acoustic signals that propagate through the aquatic medium. These signals are detected by an array of spatially distributed sensors deployed at predetermined oceanic locations. The sensors, equipped with communication modules, periodically sample sound signals and exchange both measurement data and computational results. This scenario analysis reveals three fundamental components of underwater acoustic target tracking systems: (1) the two primary entities (i.e., the underwater acoustic target and the sensor nodes); (2) the target-sensor interaction mediated by acoustic signal propagation and reception; and (3) the inter-sensor relationships established through communication links.

\subsection{Motivation for Building a Multidimensional Taxonomy Framework}
Underwater acoustic target tracking plays a pivotal role in marine exploration, attracting significant scholarly attention. Given its importance in both military and civilian applications, researchers have conducted comprehensive surveys of these technologies \cite{2021基于水下传感器,2024基于随机有限集,luo2018,kumar2021,hou2024}. In addition, the integration of artificial intelligence, big data analytics, and advanced sensor technologies has created unprecedented opportunities for technological breakthroughs in this field, positioning it as a key driver of future marine science innovation. Nevertheless, existing surveys in this domain exhibit two notable limitations \cite{jahanbakht2021}:

\begin{itemize}[leftmargin=*]
\item First, while underwater acoustic target tracking encompasses diverse research dimensions, current surveys often adopt narrow perspectives and scenario frameworks. Consequently, these surveys demonstrate limited applicability across different operational contexts, thereby hindering readers and researchers to form a systematic understanding of problem scenarios.

\item Second, the majority of existing surveys were published prior to the recent paradigm shift driven by advancements in deep learning, including the application of deep neural networks for measurement data analysis and deep reinforcement learning for sensor control coordination. Consequently, these surveys lack coverage of cutting-edge developments that are fundamentally transforming methodological approaches in this field.
\end{itemize}

This evident research gap underscores the urgent need for a comprehensive survey that systematically synthesizes both fundamental methodologies and contemporary innovations in underwater acoustic target tracking. 

\subsection{Scope of the survey}
\begin{figure*}[ht]
    \centering
    \includegraphics[width=0.65\linewidth]{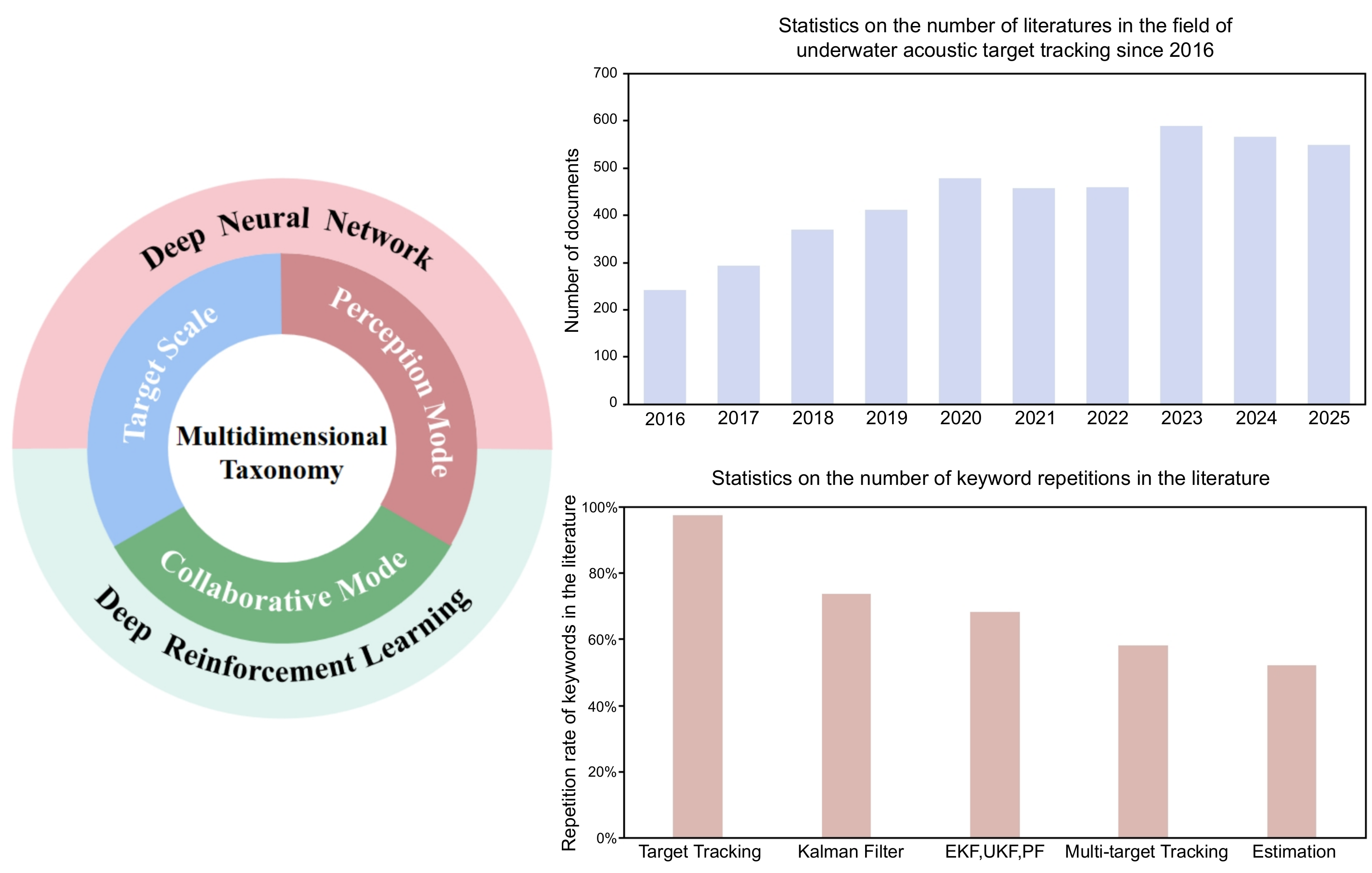}
    \caption{Results of statistical analysis of literature.  According to the number of literature published in this field in the last decade, it can be found that the research fever in the field of underwater acoustic target tracking has been increasing, while according to the results of the keyword statistics of the literature, target tracking using different filtering methods is the core research direction. }
    \label{fig:wenxain}
\end{figure*}

As illustrated in Figure \ref{fig:wenxain}, underwater acoustic target tracking has garnered increasing research attention in recent years, with a substantial number of publications emerging in this field. 
This survey systematically examines the problem scenario of underwater acoustic target tracking, analyzing its key components and proposing a comprehensive classification framework based on three dimensions: (1) underwater target scale, (2) sensor signal perception methods, and (3) inter-sensor collaboration modes. Furthermore, we synthesize recent advancements integrating deep learning, reinforcement learning, and data processing techniques with conventional tracking approaches. A systematic comparison between this survey and existing surveys or reviews is presented in Table \ref{tab:zongshu}.

\subsection{Contribution of the Survey}
To the best of our knowledge, this represents the first systematic and multidimensional survey of underwater acoustic target tracking research, comprehensively summarizing both foundational methodologies and recent technological advancements. The primary contributions of this study are fourfold:
\begin{itemize}[leftmargin=*]
    \item \textbf{We propose a novel multi-dimensional taxonomy framework.} Different from conventional single-dimensional classifications, our framework integrates three critical aspects: target scale, perception methods, and collaboration modes. This innovative approach not only elucidates the diversity and complexity of underwater acoustic tracking systems but also establishes a cross-mapping mechanism that enables researchers to examine methodologies from multiple perspectives, thereby overcoming the limitations inherent in traditional surveys. 

    \item \textbf{We provide an in-depth analysis of emerging technologies in tracking algorithms.} We examine how deep learning and reinforcement learning are revolutionizing adaptive algorithm design for dynamic underwater environments. This analysis addresses a significant gap in existing surveys by systematically documenting these technological breakthroughs, offering researchers both a reference for integrating advanced techniques and a roadmap for future innovation. 

    \item \textbf{We present a structured comparative analysis of methodological systems.} Through meticulously constructed comparison tables, technical roadmaps, and case studies, we distill the fundamental principles, distinctive features, and performance metrics of various tracking methods. This structured analytical approach not only facilitates readers’ rapid comprehension of the core methodology but also provides a practical framework for technological selection and optimization in real-world applications.

    \item \textbf{We identify current challenges and propose future research directions.} Our systematic evaluation reveals critical bottlenecks in ocean environment modeling, low signal-to-noise (SNR) signal processing, and data sharing. Building upon this analysis, we develop a multidimensional research framework encompassing algorithm refinement, technological convergence, engineering implementation, and data collaboration - providing both theoretical foundations for interdisciplinary research and actionable guidance for subsequent studies.
\end{itemize}

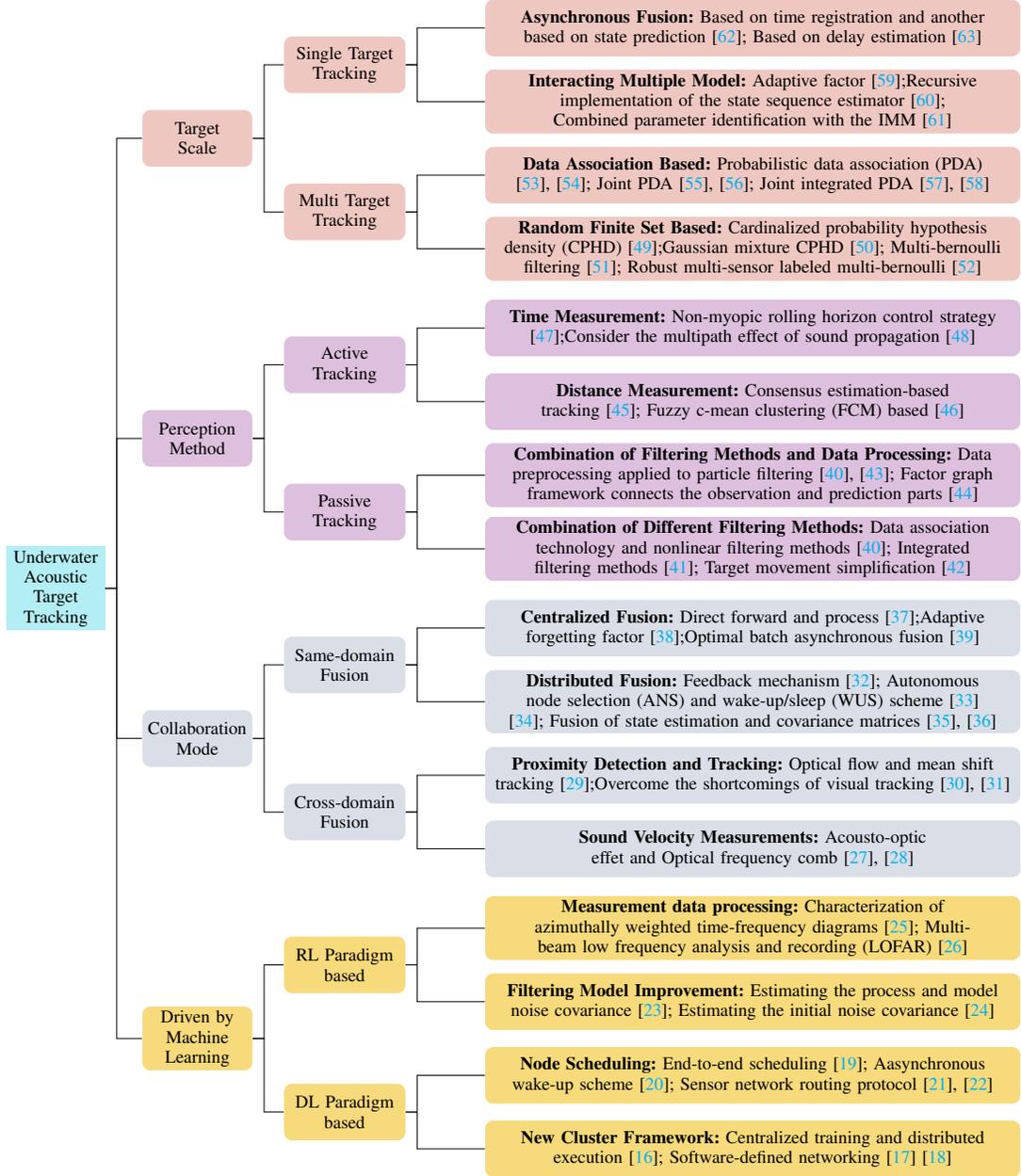
\begin{figure*}[ht]
\centering
\resizebox{0.8\textwidth}{!}{
\begin{tikzpicture}[
  draw,
  grow=right, 
  level 1/.style={level distance=25mm, sibling distance=53mm},
  level 2/.style={level distance=26mm, sibling distance=26mm},
  level 3/.style={level distance=72mm, sibling distance=13mm},
  edge from parent path={(\tikzparentnode.east) -- ++(2mm,0) |- (\tikzchildnode.west)},
  every node/.style={
    rectangle,
    minimum height=10mm,
    align=center,
    font=\small,
    rounded corners=4pt,
    text width=93mm,
    inner sep=2pt
  }
]
 \node [anchor=south,text width=16mm,rounded corners=0pt,fill={rgb,255:red,179; green,238; blue,245}]{Underwater\\Acoustic\\Target\\Tracking}
  child {node[text width=18mm,fill={rgb,255:red,247; green,218; blue,123}] {Driven by \\Machine Learning}
    child {node[text width=20mm,fill={rgb,255:red,247; green,218; blue,123}] {DL Paradigm based}
        child {node[fill={rgb,255:red,247; green,218; blue,123}] {\textbf{New Cluster Framework:} Centralized training and distributed execution\cite{yang2021a}; Software-defined networking\cite{zhu2024} \cite{wang2025}}}
        child {node[fill={rgb,255:red,247; green,218; blue,123}] {\textbf{Node Scheduling:} End-to-end scheduling\cite{zheng2023}; Aasynchronous wake-up scheme\cite{su2021}; Sensor network routing protocol\cite{jin2019}, \cite{wang2023}}}
    }
    child {node[text width=20mm,fill={rgb,255:red,247; green,218; blue,123}] {RL Paradigm based}
        child {node[fill={rgb,255:red,247; green,218; blue,123}] {\textbf{Filtering Model Improvement:} Estimating the process and model noise covariance \cite{xu2024}; Estimating the initial noise covariance\cite{hou2023}}}
        child {node[fill={rgb,255:red,247; green,218; blue,123}] {\textbf{Measurement data processing:} Characterization of azimuthally weighted time-frequency diagrams \cite{wang2021}; Multi-beam low frequency analysis and recording (LOFAR) \cite{wang2022}}}
    }
  }
  child {node[text width=18mm,fill={rgb,255:red,217; green,222; blue,231}] {Collaboration\\Mode}
    child {node[text width=20mm,fill={rgb,255:red,217; green,222; blue,231}] {Cross-domain Fusion}
          child {node[fill={rgb,255:red,217; green,222; blue,231}] {\textbf{Sound Velocity Measurements:} Acousto-optic effet and Optical frequency comb \cite{xue2018,2021基于飞秒激光}}}
          child {node[fill={rgb,255:red,217; green,222; blue,231}] {\textbf{Proximity Detection and Tracking:} Optical flow and mean shift tracking\cite{lee2012};Overcome the shortcomings of visual tracking\cite{meng-chechuang2015}, \cite{mandic2016}}}
    }
    child {node[text width=20mm,fill={rgb,255:red,217; green,222; blue,231}] {Same-domain Fusion}
        child {node[fill={rgb,255:red,217; green,222; blue,231}] {\textbf{Distributed Fusion:} Feedback mechanism \cite{xin2003}; Autonomous node selection (ANS) and wake-up/sleep (WUS) scheme\cite{yu2014} \cite{hare2017}; Fusion of state estimation and covariance matrices \cite{hare2014,braca2016}}}
        child {node[fill={rgb,255:red,217; green,222; blue,231}] {\textbf{Centralized Fusion:} Direct forward and process\cite{zhang2019a};Adaptive forgetting factor \cite{qiu2019};Optimal batch asynchronous fusion \cite{zhou2019} }}
    }
  }
  child {node[text width=18mm,fill={rgb,255:red,221; green,191; blue,222}] {Perception\\Method}
    child {node[text width=20mm,fill={rgb,255:red,221; green,191; blue,222}] {Passive Tracking}
        child {node[fill={rgb,255:red,221; green,191; blue,222}] {\textbf{Combination of Different Filtering Methods:} Data association technology and nonlinear filtering methods \cite{luo2020}; Integrated filtering methods\cite{kumar2016}; Target movement simplification\cite{qian2016} }}
        child {node[fill={rgb,255:red,221; green,191; blue,222}] {\textbf{Combination of Filtering Methods and Data Processing:} Data preprocessing applied to particle filtering \cite{ravikumar2016,luo2020}; Factor graph framework connects the observation and prediction parts \cite{cheng2021}}}
    }
    child {node[text width=20mm,fill={rgb,255:red,221; green,191; blue,222}] {Active Tracking}
        child {node[fill={rgb,255:red,221; green,191; blue,222}] {\textbf{Distance Measurement:} Consensus estimation-based tracking\cite{yan2016a}; Fuzzy c-mean clustering (FCM) based\cite{son2013}}}
        child {node[fill={rgb,255:red,221; green,191; blue,222}] {\textbf{Time Measurement:} Non-myopic rolling horizon control strategy\cite{ferri2014};Consider the multipath effect of sound propagation\cite{dehnavi2017}}}
    }
  }
  child {node[text width=18mm,fill={rgb,255:red,239; green,199; blue,193}] {Target\\Scale}
    child {node[text width=20mm,fill={rgb,255:red,239; green,199; blue,193}] {Multi Target Tracking}
        child {node[fill={rgb,255:red,239; green,199; blue,193}] {\textbf{Random Finite Set Based:} Cardinalized probability hypothesis density (CPHD) \cite{Zou2019};Gaussian mixture CPHD\cite{zhou2023}; Multi-bernoulli filtering\cite{zhang2020}; Robust multi-sensor labeled multi-bernoulli \cite{zhang2023} }}
        child {node[fill={rgb,255:red,239; green,199; blue,193}] {\textbf{Data Association Based:} Probabilistic data association (PDA) \cite{bar2009,qiu2018}; Joint PDA \cite{Puranik2007,chen2024}; Joint integrated PDA \cite{musicki2002,yao2019a}}}
    }
    child {node[text width=20mm,fill={rgb,255:red,239; green,199; blue,193}] {Single Target Tracking}
        child {node[fill={rgb,255:red,239; green,199; blue,193}] {\textbf{Interacting Multiple Model:} Adaptive factor \cite{chen2006};Recursive implementation of the state sequence estimator\cite{johnston2001}; Combined parameter identification with the IMM \cite{2001参数自适应}}}
        child {node[fill={rgb,255:red,239; green,199; blue,193}] {\textbf{Asynchronous Fusion:} Based on time registration and another based on state prediction\cite{Yang2015}; Based on delay estimation \cite{liu2017}}}
    }
  };
\end{tikzpicture}
}
\caption{Taxonomy of underwater acoustic target tracking methods.  This survey categorizes the literature in the field from four perspectives: target scale, perceptual approach, collaborative model, and machine learning driven, with the leaf nodes of the tree diagram listing the specifics and core methods of each perspective. }
\label{fig:taxonomy}
\end{figure*}

This survey is structured as follows: Chapter \ref{2} outlines fundamental theories of underwater acoustic target tracking. Chapter \ref{3} categorizes tracking methods from three perspectives: target scale, perception method, and collaboration mode. Chapter \ref{4} explores the role of deep learning and reinforcement learning technologies in the field of underwater acoustic target tracking. Finally, Chapter \ref{5} addresses current challenges and proposes future avenues and Chapter \ref{6} provides a conclusion. As illustrated in Figure \ref{fig:taxonomy}, our analysis systematically categorizes more than 180 publications over the period 2016-2025 based on these aspects and machine learning applications.

\begin{table*}[ht]
 \renewcommand{\arraystretch}{1.1}
\caption{The summary and comparison of related surveys}
\centering
\label{tab:zongshu}
\scalebox{0.85}{\begin{threeparttable}\begin{tabular}{c|c|cccc|cc|c}
\hline
\multirow{2}{*}[-14pt]{Literature} &
  \multirow{2}{*}[-14pt]{\begin{tabular}[c]{@{}c@{}}Description\end{tabular}} &
  \multicolumn{4}{c|}{\begin{tabular}[c]{@{}c@{}}Underwater\\ acoustic\\ Sensors\end{tabular}} &
  \multicolumn{2}{c|}{\begin{tabular}[c]{@{}c@{}}Underwater\\ acoustic\\ Target\end{tabular}} &
  \begin{tabular}[c]{@{}c@{}}Tracking\\ methods\end{tabular} \\ \cline{3-9} 
 &
   &
  \multicolumn{1}{c|}{\begin{tabular}[c]{@{}c@{}}Perception \\ method:\\ passive\\ tracking\end{tabular}} &
  \multicolumn{1}{c|}{\begin{tabular}[c]{@{}c@{}}Perception\\ method:\\ active\\ tracking\end{tabular}} &
  \multicolumn{1}{c|}{\begin{tabular}[c]{@{}c@{}}Cooperative\\ mode:\\ centralized\end{tabular}} &
  \begin{tabular}[c]{@{}c@{}}Cooperative\\ mode:\\ distributed\end{tabular} &
  \multicolumn{1}{c|}{\begin{tabular}[c]{@{}c@{}}Single\\ target\end{tabular}} &
  \begin{tabular}[c]{@{}c@{}}Multi\\ target\end{tabular} &
  \begin{tabular}[c]{@{}c@{}}Machine\\ learning\\ assisted\end{tabular} \\ \hline
\cite{2021基于水下传感器} &
  \begin{tabular}[c]{@{}c@{}}Perspective: system architecture,\\tracking models, energy efficiency\\optimization, single target and\\multi target tracking techniques\end{tabular} &
  \multicolumn{1}{c|}{\textcolor{red}{\ding{56}}} &
  \multicolumn{1}{c|}{\textcolor{red}{\ding{56}}} &
  \multicolumn{1}{c|}{\textcolor{green}{\ding{52}}} &
  \textcolor{green}{\ding{52}} &
  \multicolumn{1}{c|}{\textcolor{green}{\ding{52}}} &
  \textcolor{green}{\ding{52}} &
  \textcolor{red}{\ding{56}} \\ \hline
\cite{2024基于随机有限集} &
  \begin{tabular}[c]{@{}c@{}} Reviews concept of randomized\\finite set (RFS) and its\\application to multi-target tracking\end{tabular} &
  \multicolumn{1}{c|}{\textcolor{red}{\ding{56}}} &
  \multicolumn{1}{c|}{\textcolor{red}{\ding{56}}} &
  \multicolumn{1}{c|}{\textcolor{red}{\ding{56}}} &
  \textcolor{red}{\ding{56}} &
  \multicolumn{1}{c|}{\textcolor{red}{\ding{56}}} &
  \textcolor{green}{\ding{52}} &
  \textcolor{red}{\ding{56}} \\ \hline
  \cite{luo2018} &
  \begin{tabular}[c]{@{}c@{}}Perspective:instrument-assisted methods,\\model-based methods, and\\tracking optimization methods\end{tabular} &
  \multicolumn{1}{c|}{\textcolor{green}{\ding{52}}} &
  \multicolumn{1}{c|}{\textcolor{green}{\ding{52}}} &
  \multicolumn{1}{c|}{\textcolor{red}{\ding{56}}} &
  \textcolor{red}{\ding{56}} &
  \multicolumn{1}{c|}{\textcolor{green}{\ding{52}}} &
  \textcolor{red}{\ding{56}} &
  \textcolor{red}{\ding{56}} \\ \hline
  \cite{kumar2021} &
  \begin{tabular}[c]{@{}c@{}}Reviews the application of target\\tracking techniques in aeronautical\\and underwater acoustics\end{tabular} &
  \multicolumn{1}{c|}{\textcolor{green}{\ding{52}}} &
  \multicolumn{1}{c|}{\textcolor{green}{\ding{52}}} &
  \multicolumn{1}{c|}{\textcolor{red}{\ding{56}}} &
  \textcolor{red}{\ding{56}} &
  \multicolumn{1}{c|}{\textcolor{green}{\ding{52}}} &
  \textcolor{red}{\ding{56}} &
  \textcolor{red}{\ding{56}} \\ \hline
  \cite{hou2024} &
  \begin{tabular}[c]{@{}c@{}}Reviews the research progress of\\azimuth tracking technology for\\underwater acoustic targets based\\on passive sonar systems\end{tabular} &
  \multicolumn{1}{c|}{\textcolor{green}{\ding{52}}} &
  \multicolumn{1}{c|}{\textcolor{red}{\ding{56}}} &
  \multicolumn{1}{c|}{\textcolor{red}{\ding{56}}} &
  \textcolor{red}{\ding{56}} &
  \multicolumn{1}{c|}{\textcolor{green}{\ding{52}}} &
  \textcolor{green}{\ding{52}} &
  \textcolor{red}{\ding{56}} \\ \hline
\textbf{Ours} &
  \begin{tabular}[c]{@{}c@{}}\textbf{New taxonomy framework:} \\target characteristics, \\perception methods, collaboration modes, \\and machine learning-driven methods\end{tabular} &
  \multicolumn{1}{c|}{\textcolor{green}{\ding{52}}} &
  \multicolumn{1}{c|}{\textcolor{green}{\ding{52}}} &
  \multicolumn{1}{c|}{\textcolor{green}{\ding{52}}} &
  \textcolor{green}{\ding{52}} &
  \multicolumn{1}{c|}{\textcolor{green}{\ding{52}}} &
  \textcolor{green}{\ding{52}} &
  \textcolor{green}{\ding{52}} \\ \hline
\end{tabular}
\begin{tablenotes}
\footnotesize
\item[*] The table describes the differences between the surveys in the field of underwater acoustic target tracking in recent years and this survey, and compares in detail the coverage of the survey contents from the perspectives of sensors, targets, and tracking methods.
\end{tablenotes}
\end{threeparttable}
}
\end{table*}

\section{Background and Fundamental Models} \label{2}
Underwater acoustic target tracking is a technique that identifies and tracks the position and trajectory of a target using underwater acoustic signals. This process estimates the target’s current state (primarily including position, velocity, and other kinematic parameters) by analyzing data from measurement sensors. The collected measurement data typically include direct range estimation, direction-of-arrival (DOA), received signal strength (RSS), time-of-arrival (TOA), and time-difference-of-arrival (TDOA). These measurements are then associated with the target’s state through filtering methods for estimation. As an integrated process, underwater acoustic target tracking combines signal propagation and processing, target motion modeling, position estimation, and tracking \cite{luo2018}. This section systematically presents the theory of underwater acoustic target tracking from three perspectives: (1) \textbf{propagation of underwater acoustic signals}, (2) \textbf{construction of underwater acoustic target models}, and (3) \textbf{state estimation of underwater acoustic targets}. Prior to this introduction, Table \ref{tab:fuhao} lists the symbols used in this chapter.

\begin{table*}[ht]
\renewcommand{\arraystretch}{1.2} 
\setlength{\tabcolsep}{20pt}  
\caption{List of Symbols for Chapter II}
\centering
\label{tab:fuhao}
\begin{tabular}{c|c|c}
\hline
Symbol & Unit & Definition                                                  \\ \hline
  $P$     &   $N/m^2$   & Sound pressure                                              \\ \hline
  $C(\textbf{r},t)$     &   $m/s$   & Spatiotemporal variation function of sound speed            \\ \hline
  $j$     &   /   & Complex number unit          \\ \hline
  $K$     &   $rad/m^{-1}$   & Wave number                                                  \\ \hline
  $f_c$     &   $Hz$   & Acoustic frequency                                          \\ \hline
  $\nabla$     &   /   & Gradient  operator        \\ \hline
  $A$     &   $Hz^2$   & Amplitude of sound pressure          \\ \hline
  $S$     &   $rad$   & phase function          \\ \hline
  $r$     &   $m$   & Waveguide radial distance          \\ \hline
  $\phi_l(z)$     &   /   & The component of mode $l$ in the vertical direction $z$          \\ \hline
  $v$     &   $rad/m$   & Velocity factor of normal mode in mode $l$         \\ \hline
  $\phi^p(r_\perp^p,r^p)$     &   $Pa$   & Complex amplitude of the sound pressure field          \\ \hline
  $r_\perp^p$     &   /   &    Coordinates perpendicular to the direction of propagation       \\ \hline
  $r^p$     &   /   &      Coordinates in the main propagation direction     \\ \hline
  $k_0$     &   $rad/m$   & Reference wave number          \\ \hline
  $\nabla_\perp^2$     &   /   & Transversal laplace operator          \\ \hline
  $n(\cdot)$     &   /   &    Refractive index       \\ \hline
  $f(x_k,w_k)$     & /    & State transition function                                   \\ \hline
  $w_k$     &   $Hz$   & Process noise                                               \\ \hline
  $h(x_k,v_k)$     & /    & Measurement function                                        \\ \hline
  $v_k$     &   $Hz$   & Measurement noise                                           \\ \hline
  $p(x_k|z_{1:k})$     & /    & Posterior probability density function at time step $k$        \\ \hline
  $\mu$     &   $Hz$   & Mean of the noise probability density function              \\ \hline
  $\sigma$     &   /   & Variance of the noise probability density function          \\ \hline
  $\nabla^2$     &   $Hz^2$   &  Laplace operator          \\ \hline
\end{tabular}
\end{table*}

\subsection{Propagation of Underwater Acoustic Signals}
Underwater acoustic signal propagation can be characterized using either RSS \cite{chang2018} or wave equation modeling. Without accounting for the complexity of the marine environment, RSS magnitude can be expressed as a function of the straight-line distance between the transmitter and receiver \cite{sozer2000}.

Recent studies have increasingly emphasized the marine environment as a critical factor in underwater acoustic signal propagation. To accurately characterize sound propagation in water, researchers now strive to comprehensively account for the propagation medium. As mechanical waves propagating through fluids, sound waves obey fluid dynamics principles, with their propagation governed by Equation \ref{con:bodong}:
\begin{equation}
    \nabla^2P=\frac{1}{C^2(\mathbf{r},t)}\frac{\partial^2P}{\partial t^2}.\label{con:bodong}
\end{equation}

Assuming the sound source generates a harmonic signal with frequency $\omega$ i.e., $P=p\cdot\exp(-j\cdot\omega\cdot t)$ . The wave equation simplifies to the time-independent Helmholtz equation, as shown in Equation (\ref{con:Helmholtez}):
\begin{equation}
    \nabla^2p+K^2(\mathbf{r})p=0,\label{con:Helmholtez}
\end{equation}
where $K(\mathbf{r})=\frac{\omega}{C(\mathbf{r})}$ is wave number, $\omega=2\pi f_c$.

The majority of fundamental theories addressing underwater acoustic propagation problems-including \textbf{ray theory}, \textbf{normal mode theory}, and \textbf{parabolic equation theory}-are derived from wave fluctuation equations. These theoretical frameworks provide distinct approaches to modeling sound propagation in underwater environments, each with specific advantages and limitations.

\subsubsection{\textbf{Ray Theory}}
It is a high-frequency approximation method derived from geometrical optics for describing wave propagation. While originating from the wave equation, this approach does not provide complete solutions but rather serves as an effective approximation under high-frequency conditions. The theory conceptualizes acoustic energy transmission through rays, defined as wavefront normals that indicate propagation direction. The superposition of these ray clusters constitutes the observable acoustic field environment \cite{2024复杂海洋环境}. Two fundamental equations govern ray acoustics:
\begin{gather}
    (\nabla S)^2=n^2(x,y,z), \label{con:5} \\
    \nabla\cdot(A^2\nabla S)=0.  \label{con:6} 
\end{gather}

Equation (\ref{con:5}), termed the phase function, determines ray trajectories by expressing path length as a function of path endpoints. When these endpoints correspond to source and receiver positions, the resulting ray is designated as the intrinsic ray. Equation (\ref{con:6}), the intensity equation, quantifies individual ray intensity \cite{m.2013}.

Building upon these theoretical foundations, Lawrence \cite{lawrence1985} effectively employed ray theory to model the interaction of low-frequency sound waves with horizontally stratified ocean floors, demonstrating its efficacy in analyzing complex acoustic field structures. Subsequently, Etter \cite{etter2012} expanded this application to predict signal attenuation in underwater acoustic communication systems, further validating the versatility of this approach.

Recent advancements have further refined ray theory applications. Smirnov et al. \cite{smirnov2001} investigated nonlinear effects in underwater acoustic propagation, identifying scenarios requiring theoretical modifications through integration with complementary approaches. Notably, Li et al. \cite{li2023} developed a physics-informed machine learning framework that synergizes limited empirical data with ray theory principles, demonstrating robust generalization capabilities in uncharacterized regions. Most recently, Liao et al. \cite{liao2024} implemented dynamic ray analysis coupled with hybrid parallel computing strategies, effectively addressing computational bottlenecks in large-scale multipath sound field simulations.

\subsubsection{\textbf{Normal Mode Theory}}
The normal mode method provides an effective approach for representing acoustic fields through the superposition of discrete waveguide modes in sound propagation calculations. As demonstrated in \cite{pekeris1948}, this theory has been successfully applied to solve sound propagation problems in stratified ocean media. The acoustic field generated by a harmonic point source can be effectively characterized by these waveguide normal modes:
\begin{equation}
    P=\sqrt{\frac{8\pi}{r}}\cdot e^{i\frac{\pi}{4}}\sum_{l=0}^L\phi_l(z_1)\cdot \phi_l(z_2)\cdot \sqrt{\nu_l}\cdot \exp(i\cdot \nu_l \cdot r).
\end{equation}
 
While traditional normal mode theory assumes static and homogeneous media conditions, this simplification poses limitations when addressing complex marine sound propagation scenarios. The adiabatic approximation of normal mode theory addresses this by neglecting intermodal energy exchange terms, thereby substantially reducing computational requirements \cite{nagl1978}. In contrast, coupled normal mode theory employs a system of mutually interacting modes to describe acoustic system dynamics. Through the incorporation of coupling coefficients, this approach significantly improves both the accuracy and stability of sound propagation solutions in complex marine environments \cite{knobles2003}.

\subsubsection{\textbf{Parabolic Equation Theory}}
The parabolic equation method serves as an efficient approximation approach for modeling sound propagation, particularly in oceanic environments where medium parameters demonstrate both range-dependent and three-dimensional variations. The application of parabolic theory to underwater acoustics was first established in \cite{tappert1977}. The foundational parabolic equation for sound propagation is expressed as Equation (\ref{con:paowuxian}):

\begin{equation}\label{con:paowuxian}
\begin{aligned}
    \frac{\partial}{\partial r^p}\phi^p(r_\perp^p,r^p)=ik_0\{&-1+\sqrt{k_0^{-2}\nabla_\perp^2+n^2(r_\perp^p,r^p)}\} \\
    &\cdot \phi^p(r_\perp^p,r^p).
\end{aligned}
\end{equation}

In \cite{lee1983}, a method for approximating the three-dimensional Helmholtz equation was proposed, signiﬁcantly enhancing the capability to address wide-angle propagation issues. Claerbout \cite{claerbout1986} developed an acoustic wave equation model based on rational linear approximation, which characterizes underwater acoustic propagation with seaﬂoor interaction, enabling accurate handling of large-angle propagation up to $40^{\circ}$. Collins \cite{collins1989} derived and numerically solved a higher-order elastic parabolic equation to simulate sound wave propagation in ﬂuid/solid media with depth-dependent and weak lateral dependence, particularly suited for environments with elastic seabeds.

\subsection{Construction of Underwater Acoustic Target Models}  
Accurate underwater target tracking necessitates the establishment of a rigorous system model \cite{zhao2020}, as mathematically formulated in Equation (\ref{con:moxing}): 
\begin{equation}\label{con:moxing}
    \left\{ {\begin{array}{*{20}{c}}
{{x_{k + 1}} = f({x_k},{w_k})}\\
{{z_k} = h({x_k},{v_k})}
\end{array}} \right. . 
\end{equation}

In this model, $f( \cdot )$ represents the transition function, which is determined by the target motion model, and $h( \cdot )$ represents the measurement function, which is determined by the target observation model.  $x_k$ and $z_k$ represent the state vector and measurement vector at time $k$ respectively, while $w_k$ and $v_k$ represent the process noise vector and measurement noise vector at time $k$.  

A commonly adopted model assumes both process noise and measurement noise as Gaussian white noise. The amplitude of Gaussian white noise follows a Gaussian distribution, with its probability density function defined by Equation (\ref{con:gailv1}):
\begin{equation}\label{con:gailv1}
    p(x) = \frac{1}{{\sqrt {2\pi } \sigma }}\exp \left( { - \frac{{{{(x - \mu )}^2}}}{{2{\sigma ^2}}}} \right),
\end{equation}
where $\mu $ is the mean and $\sigma ^2$ is the variance.  In underwater acoustic channels, the noise mean is typically zero, i.e., $\mu=0$. 

In practical underwater environments, noise typically originates from multiple independent small-scale random factors. The Central Limit Theorem suggests that the summation of numerous independent random variables converges to a normal distribution \cite{cardone2023}. Consequently, acoustic propagation noise in underwater scenarios often demonstrates near-Gaussian characteristics. Even when individual noise sources deviate from Gaussian distribution, their combined effect through multiple independent sources yields an approximately Gaussian summation. Moreover, the favorable mathematical properties of Gaussian distribution significantly simplify the state transition and observation update processes in filtering algorithms. This simplification not only ensures computational efficiency but also accurately captures the statistical and spectral properties of real-world noise.

The research framework for underwater target tracking algorithms fundamentally relies on two model components: (1) \textbf{target motion model}, describing the target’s kinematic behavior; and (2) \textbf{target observation model}, characterizing the measurement process.

\subsubsection{\textbf{Target Motion Model}}
Motion models can be classified into two categories based on their structural complexity: single-model and multi-model approaches. The single-model approach employs a single motion model to describe the target’s kinematic state, typically using either the constant velocity (CV) or constant acceleration (CA) model \cite{sun2017}. Consider a tracking interval denoted as $t$. The CV model assumes linear motion at constant velocity, with its state transition matrix $F_k^{CV}$ expressed as:
\begin{equation}
    F_k^{CV} = \left[ {\begin{array}{*{20}{c}}
1&{\Delta t}\\
0&1
\end{array}} \right]. 
\end{equation}

The CA model assumes that the target moves in a straight line with constant acceleration.  The state transition matrix $F_k^{CA}$ can be expressed as:
\begin{equation}
    F_k^{CA} = \left[ {\begin{array}{*{20}{c}}
1&{\Delta t}&{\frac{{\Delta {t^2}}}{2}}\\
0&1&{\Delta t}\\
0&0&1
\end{array}} \right]. 
\end{equation}

Other models include time-varying models (singer model) \cite{singer1970}, semi-markov models based on Markov processes \cite{moose1975}, current statistics models based on modified rayleigh-markov processes \cite{zhou1984}, and jitter models incorporating acceleration derivatives \cite{mehrotra1997}. 

\subsubsection{\textbf{Target Observation Model}}
When developing the observation model, a thorough analysis of underwater acoustic signal propagation is essential. As previously discussed, this analysis must account for both signal attenuation and path variations. Underwater acoustic sensor measurements can be classified into three primary models: (1) TOA-based, (2) RSS-based, and (3) angle-of-arrival (AOA)-based measurement models \cite{isbitiren2011,poursheikhali2021,li2012}.

The TOA-based approach encompasses both direct TOA and TDOA methods. The TOA model calculates the distance between target and sensor by measuring acoustic signal propagation time, while TDOA localization relies on time differences observed across multiple receiving sensors \cite{yan2019}. Assuming the signal arrival times at two receivers are $t_1$ and $t_2$, the distance difference 
$\Delta d$ between the target and the two receivers is given by the formula in Equation (\ref{con:juli}):
\begin{equation}\label{con:juli}
    \Delta d = c \cdot \Delta t,
\end{equation}
where $c$ denotes the signal propagation speed.  The difference in distances from the target to the two receivers can be used to construct a hyperboloid equation, which is expressed in the form shown in Equation (\ref{con:shuangqu}):
\begin{equation}\label{con:shuangqu}
\begin{aligned}
    c \cdot &\Delta t= \sqrt {{{(x - {x_2})}^2} + {{(y - {y_2})}^2} + {{(z - {z_2})}^2}} \\
    &- \sqrt {{{(x - {x_1})}^2} + {{(y - {y_1})}^2} + {{(z - {z_1})}^2}},
\end{aligned}
\end{equation}
where $(x_1,y_1,z_1)$and $(x_2,y_2,z_2)$ represent the coordinates of the receivers.  For multiple receivers, similarly, multiple hyperboloid equations can be obtained.  By solving these equations, the position of the target can be determined. 

The RSS-based model exploits the energy attenuation characteristics of propagating acoustic signals, where signal strength exhibits a predictable relationship with propagation distance. This framework incorporates measurement equations derived from wave equation, ray theory, normal mode theory, and parabolic equation theory.

The AOA-based model employs sensor arrays for spatial sampling of acoustic signals. Through analysis of inter-sensor phase or amplitude differences, this method estimates signal incidence angles with high precision \cite{hu2017}.

\subsection{State Estimation of Underwater Acoustic Targets}
Underwater acoustic target tracking employs various state estimation methods, which can be primarily classified into two categories: batch methods and recursive methods.

\subsubsection{\textbf{Batch Processing Method}}
Batch processing methods process all observations collected during a specific time interval as a single dataset to estimate the target state. Representative batch methods include maximum likelihood estimation (MLE) \cite{jauffret2008}, least squares estimation (LS) \cite{2024单站}, gauss-newton method (GN) \cite{zhao2013}, pseudo linear estimation (PLE) \cite{2005方位／频率}, and monte carlo method (MC) \cite{hue2002}.

MLE determines the optimal target state by maximizing the likelihood function, which quantifies the probability of observed data given state parameters \cite{zhou2020}. LS estimation minimizes the sum of squared residuals between observations and model predictions through optimization \cite{wang2020}. The GN method iteratively solves nonlinear least squares problems by linearizing the model at each step  \cite{BAO2018}. PLE simplifies nonlinear problems via linear transformations or auxiliary variables \cite{gong1981}, while MC uses random sampling to approximate posterior distributions and quantify uncertainty \cite{zarai2021}.

These batch methods are essential for high-precision underwater acoustic tracking, with method selection depending on task requirements, motion models, and data characteristics. 
\subsubsection{\textbf{Recursive Method}}
However, batch methods exhibit two main limitations. First, they require complete data collection before processing, precluding real-time state updates. Second, their computational efficiency decreases significantly with large datasets. These limitations become particularly problematic given recent advancements in monitoring technologies and improvements in target maneuverability, which have increased tracking uncertainties. Such uncertainties mainly arise from three sources: (1) process noise in target motion modeling, (2) measurement noise from sensors, and (3) false alarms in multi-target and cluttered environments \cite{luo2018}. To address these challenges while maintaining real-time tracking capability, recursive methods have been widely adopted for underwater acoustic target state estimation and filtering.

The key challenge in designing target tracking algorithms is selecting an appropriate filtering algorithm or improving existing algorithms to meet the specific requirements of underwater target tracking, based on the established dynamic motion model and observation model.  Different filtering methods are all based on a fundamental recursive theory, with Bayesian filtering being the most commonly used recursive theory. 

The core design concept of the Bayesian filter is to use the observation data obtained at the current time to correct the prior probability density function (PDF) \cite{pedersen2011}, thereby obtaining the posterior probability.  Bayesian theory posits that the posterior PDF, derived from the prior PDF and current system information, better reflects the system's characteristics, and thus, system analysis should be based on the posterior PDF.  The Bayesian filter consists of two processes: prediction and update.  The prediction process aims to obtain the prior PDF based on the target motion model.  The update process introduces the measurements obtained from the observation model into the output of the prediction step to correct the probability, thereby obtaining the posterior probability of the target state.

Let ${z_{1:k}} = \{ {z_1},{z_2}, \cdots ,{z_k}\} $ represent the measurements obtained from time 1 to $ k $, and  $p({x_{0:k}}|{z_{1:k}})$ represent the posterior PDF before time $ k $, assuming that the prior probability of the target's initial state is known.  Since the state vector ${x_k}$ follows a first-order Markov process, the posterior PDF can be recursively obtained using the measurements ${z_{1:k}} = \{ {z_1},{z_2}, \cdots ,{z_k}\} $.  Given that $p({x_{k - 1}}|{z_{1:k - 1}})$ has already been obtained, the one-step prediction probability is as shown in Equation (\ref{con:yibu}):
\begin{equation}\label{con:yibu}
    p({x_k}|{z_{1:k - 1}}) = \int p ({x_k}|{x_{k - 1}})p({x_{k - 1}}|{z_{1:k - 1}}){\rm{d}}{x_{k - 1}},
\end{equation}
where $p({x_k}|{z_{1:k - 1}})$ represents the state transition probability.  The updating process uses the observations obtained at time $k$ to update the prior probability.  The posterior probability density function (PDF) can be derived from equations (\ref{con:houyan1}) and (\ref{con:houyan2}):
\begin{gather}
    p({x_k}|{y_{1:k}}) = \frac{{p({y_k}|{x_k})p({x_k}|{y_{1:k - 1}})}}{{p({y_k}|{y_{1:k - 1}})}} \label{con:houyan1},\\
    p({y_k}|{y_{1:k - 1}}) = \int p ({y_k}|{x_k})p({x_k}|{y_{1:k - 1}}){\rm{d}}{x_k} \label{con:houyan2},
\end{gather}
where $p({y_k}|{y_{1:k - 1}})$ represents the normalization constant, and $p({y_k}|{x_k})$ represents the likelihood probability.

The kalman filter (KF) provides an optimal solution to Bayesian filtering under linear gaussian conditions \cite{Kalman1960}. 
However, in nonlinear systems, obtaining an optimal solution becomes intractable due to the computational complexity of infinite-dimensional integrals. When nonlinear dynamics can be locally linearized with negligible higher-order Taylor expansion terms, the system exhibits weak nonlinearity; when the higher-order Taylor expansion terms are not negligible, the system exhibits strong nonlinearity. For such cases, the extended kalman filter (EKF) employs first-order Taylor approximation for state estimation \cite{smith2006}. Alternatively, the unscented kalman filter (UKF) preserves higher-order statistical information through sigma-point transformation \cite{xianghuiyuan2005}, whereas the particle filter (PF) utilizes Monte Carlo sampling to approximate state probability distributions, demonstrating superior performance in strongly nonlinear systems \cite{M2002}.

This chapter constructs a comprehensive theoretical framework through the synthesis of analytical models and mathematical derivations. This integrated approach not only provides a robust foundation for subsequent methodological categorization but also facilitates technological advancements in the field. The systematic integration of these theoretical components ensures logical coherence throughout the research process.

\section{Classification of underwater acoustic target tracking methods} \label{3}
Underwater acoustic target tracking constitutes a multifaceted and dynamic research domain, where classification methods are typically based on diverse criteria. A systematic taxonomy and synthesis of these approaches provide an essential foundation for in-depth investigations. This chapter categorizes current methodologies through multiple analytical lenses, including \textbf{target scale}, \textbf{perception method}, and \textbf{collaboration mode}. 
\subsection{Target Tracking Methods Based on Target Scale}
Underwater acoustic target tracking involves analyzing two key target characteristics: physical dimensions and sensor interaction capabilities. In most practical scenarios, targets exhibit non-cooperative behavior by avoiding active sensor engagement. This section classifies tracking methodologies according to target quantity, subsequently detailing both \textbf{single target} and \textbf{multi target} tracking approaches.
\subsubsection{\textbf{Single Target Tracking}}
It is pivotal in underwater acoustics, with research primarily aimed at improving tracking accuracy and response speed for dynamic targets. Recent years have witnessed significant advancements in single target tracking research. Due to the variability in target motion states, traditional single-model algorithms frequently encounter \textbf{model mismatch} issues, resulting in substantial tracking errors. To overcome this limitation, the \textbf{interacting multiple model} (IMM) algorithm has been developed. This algorithm utilizes multiple models to represent potential target motion states, thereby enabling more accurate tracking of moving targets. Additionally, measurement data from sensors often exhibit asynchrony because of \textbf{propagation delays} and \textbf{environmental interference}, making \textbf{asynchronous fusion} techniques essential for enhancing tracking performance. This section comprehensively reviews the current state and technological advancements in single target tracking, with a specific focus on the application of IMM algorithms and the significance of asynchronous fusion in improving tracking capabilities.

Recent years have witnessed extensive research on single-target tracking by scholars and institutions worldwide. For example, Liao et al.\cite{binliao2012} explored DOA estimation and tracking in uniform linear arrays with mutual coupling effects. The study proposed a subspace-based approach that treats angle-independent mutual coupling as angle-dependent complex array gain, enabling joint estimation of DOA and mutual coupling matrix through full-array data analysis. Parallel to this, Gao et al.\cite{gao2015} developed a novel sequential Bayesian algorithm named “Simultaneous Angle-Signal Update”. This innovative approach reformulates the target state update as a joint optimization problem, calculating state estimates through posterior probability maximization of measurement vectors at each timestep.

Furthermore, Kong and Chun \cite{dongkeonkong2000} introduced an EKF-based fast adaptation method for dynamic targets, overcoming the slow response limitations of conventional algorithms. In a complementary study,  Cevher et al.\cite{cevher2007} integrated target motion with Gaussian disturbances, deriving a simplified likelihood function using mutual information priors to manage signal and noise variance. Their implementation of the Independent Partitioned PF algorithm demonstrated superior flexibility and effectiveness compared to traditional linear/Gaussian-based methods. Kumar et al.\cite{kumar2021a} achieved performance enhancement under Gaussian measurements by employing Fourier-Hermite polynomials as orthogonal bases to optimize nonlinear hashing processes, outperforming conventional Taylor series linearization approaches.

\textit{1-1) \textbf{IMM algorithm} }
 
While single-model algorithms demonstrate satisfactory performance when the target’s motion state remains relatively stable, their effectiveness significantly diminishes in practical applications where target motion is unpredictable and highly variable. \textbf{The inherent limitation of employing a single, fixed model to describe dynamic target states often results in model-actual state mismatches, consequently leading to substantial tracking errors.} To overcome this challenge, IMM-based algorithms have emerged as a prominent solution in maneuvering target tracking research \cite{2003多模型}.

\begin{figure}
    \centering
    \includegraphics[width=0.8\linewidth]{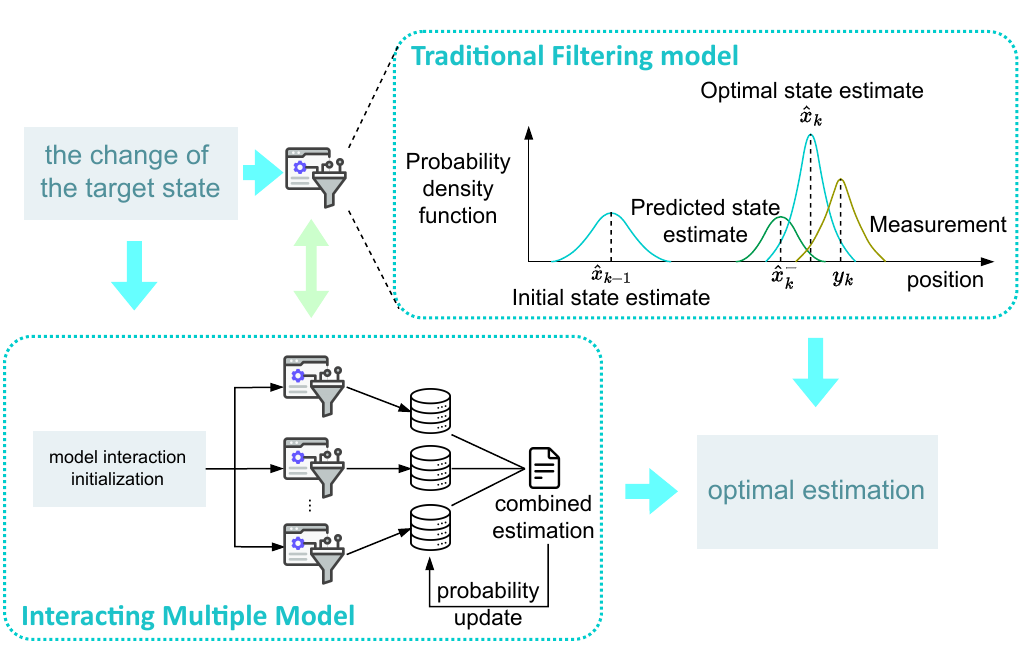}
    \caption{Flowchart of The IMM-based filtering method. It utilizes multiple base filtering models to estimate the target state, combines their estimation results with certain weights, and adjusts the weights of each filtering model according to the final estimation results.}
    \label{fig:IMM}
\end{figure}

The IMM algorithm employs multiple models to represent potential target motion states, with Markov processes governing transitions between these models. During tracking, the algorithm establishes parallel model filters corresponding to possible target states for real-time maneuvering detection. By assigning weight coefficients and model update probabilities to each filter, the system generates optimal state estimates through weighted computations, thereby achieving model-adaptive tracking \cite{Blom1984}. Figure \ref{fig:IMM} illustrates the operational flowchart of this algorithm.

The development of IMM algorithms has substantially improved maneuvering target tracking performance. Subsequent research has yielded numerous enhanced variants:

A variable-structure IMM was proposed in \cite{li1996}, establishing theoretical foundations for optimal estimators. Chen et al. \cite{chen2006} introduced adaptive factors enabling real-time model state switching based on acceleration and velocity data, simultaneously improving tracking accuracy while reducing computational load. The recursive IMM algorithm \cite{johnston2001} implemented a state sequence estimator recursively, simplifying computations by incorporating Kalman filter outputs in weight calculations. A parameter-adaptive IMM \cite{2001参数自适应} combines parameter identification to enable real-time estimation of filtering parameters including mode transition probabilities and model noise variance. Zuo et al.\cite{dongguangzuo2002}  replaced conventional probability calculations with fuzzy logic reasoning, proposing a Fuzzy-logic based maneuvering target tracking algorithm that enhances model matching accuracy through normalization processing, demonstrating significant practical utility.

\textit{1-2) \textbf{Asynchronous fusion techniques} }

\textbf{The acoustic signal’s slow propagation speed and high susceptibility to marine environmental interference, combined with communication latency between sensors, make it challenging to achieve fully synchronized target measurements}. This asynchrony manifests in two distinct fusion scenarios: in-sequence measurement (ISM) and out-of-sequence measurement (OOSM) fusion \cite{Yang2015}. ISM occurs when the fusion center receives measurements in temporal alignment with the target’s acoustic signal emission sequence, whereas OOSM arises when these sequences are misaligned.

Despite existing asynchronous fusion localization algorithms for underwater sensors \cite{yan2016,yan2018}, research on asynchronous fusion tracking remains limited. Current solutions for measurement asynchrony caused by sampling frequency disparities primarily employ time registration techniques to synchronize data. For instance, Yang et al.\cite{Yang2015} compared time-registration-based and state-prediction-based fusion algorithms, while Liu et al.\cite{liu2017} introduced a delay-estimation asynchronous particle filter that improves weight updating through backpropagation. Furthermore, Yan et al.\cite{yan2019a} presentd an innovative approach that establishes propagation delay-position relationships to compensate for clock offsets, subsequently developing a consensus Bayesian filter for continuous target tracking. This solution enhances both tracking accuracy and network longevity through combined consensus fusion and duty-cycle mechanisms.

Single target tracking technology has evolved into a well-established research framework in underwater acoustics. While traditional single-model algorithms demonstrate robust performance under stable target motion conditions, they frequently exhibit tracking inaccuracies when confronted with maneuvering targets due to model mismatch. The IMM algorithm substantially enhances tracking precision for maneuvering targets by employing parallel adaptive switching among multiple models. Additionally, techniques including time registration, delay estimation, and consensus Bayesian filtering effectively mitigate synchronization challenges in asynchronous data fusion. Collectively, these developments indicate that single-target tracking research is progressively advancing from single-model methodologies to multi-model collaborative systems, and from idealized synchronous scenarios to complex asynchronous environments. However, further optimization of model switching efficiency and asynchronous fusion accuracy remains imperative.

\subsubsection{\textbf{Multi Target Tracking}}
With the advancement of research in underwater acoustic target tracking, multi-target tracking has emerged as a critical research focus. Considering current marine environment, future targets are expected to predominantly appear in clustered formations with enhanced maneuverability and improved concealment capabilities. Furthermore, the complex marine environment and fluctuating underwater acoustic channels often generate substantial clutter, while measurements may be contaminated by uncertainties including false alarms and ambient noise \cite{yu2017}.

In these challenging underwater conditions, accurately estimating both the number of targets and their respective states based on sensor node measurements remains a fundamental challenge for underwater tracking systems. Consequently, multi-target tracking technology has become a pivotal research direction in this field. Presently, multi-target tracking methodologies can be broadly categorized into two approaches: data association-based techniques and random finite set (RFS)-based methods.

\textit{2-1) \textbf{Data association-based tracking} }

This approach employs data association algorithms to establish correspondences between targets and measurements. By integrating Bayesian filtering frameworks, the complex multi-target tracking problem is decomposed into manageable single-target tracking subproblems. The process fundamentally comprises two sequential stages: (1) data association and (2) state filtering \cite{2021基于水下传感器}.

Since state estimation validity depends on accurate data association, these techniques constitute the cornerstone of traditional multi-target tracking systems. Current methodologies are broadly categorized into:

Maximum Likelihood Methods: joint maximum likelihood (JML) and track splitting method (TSM) \cite{Mu2007}, which optimize constructed likelihood functions through batch processing.

Bayesian Methods: global nearest neighbor (GNN) \cite{kulmon2018}, multiple hypothesis tracking (MHT) \cite{coraluppi2018}, probabilistic data association (PDA) \cite{bar2009}, and their derivatives (JPDA \cite{Puranik2007}, JIPDA \cite{musicki2002}). These recursive approaches enable real-time state estimation.
Table \ref{tab:shujuguanlian2} systematically compares the trade-offs and applicability of these association techniques.

In the complex underwater acoustic environment with challenging signal processing conditions, Bayesian methods have become predominant for multi-target tracking in underwater acoustic applications. \textbf{This part specifically examines PDA, their derivatives, and MHT approach.}

\begin{table*}
\centering
\caption{Comparison of Various Data Association Based Methods}
\label{tab:shujuguanlian2}
\begin{threeparttable}
\scalebox{0.75}{\begin{tabular}{|cc|p{2.3cm}<{\centering}|p{2.3cm}<{\centering}|p{2.3cm}<{\centering}|p{2.3cm}<{\centering}|p{2.3cm}<{\centering}|p{2.3cm}<{\centering}|p{2.3cm}<{\centering}}
\hline
\multicolumn{2}{c|}{Method} &
  \begin{tabular}[c]{@{}c@{}}JML\\ \cite{Mu2007} \end{tabular} &
  \begin{tabular}[c]{@{}c@{}}TSM\\ \cite{Mu2007} \end{tabular} &
  \begin{tabular}[c]{@{}c@{}}GNN\\ \cite{kulmon2018} \end{tabular} &
  \begin{tabular}[c]{@{}c@{}}PDA\\ \cite{bar2009,qiu2018} \end{tabular} &
  \begin{tabular}[c]{@{}c@{}}MHT\\ \cite{coraluppi2018,li2019} \end{tabular}  &
  \begin{tabular}[c]{@{}c@{}}JPDA\\ \cite{Puranik2007,chen2024} \end{tabular} &
  \begin{tabular}[c]{@{}c@{}}JIPDA\\ \cite{musicki2002} \end{tabular}  \\ \hline
\multicolumn{1}{c|}{\multirow{5}{*}[-33pt]{Advantages}} &
  \begin{tabular}[c]{@{}c@{}}Theoretical\\ optimization\end{tabular} &
  \textcolor{green}{\ding{52}} &$\backslash$
   &$\backslash$
   &$\backslash$
   &$\backslash$
   &$\backslash$
   &$\backslash$
   \\ \cline{2-9} 
\multicolumn{1}{c|}{} &
  \begin{tabular}[c]{@{}c@{}}Low\\ computational\\ complexity\end{tabular} &$\backslash$
   &$\backslash$
   &
  \textcolor{green}{\ding{52}} &$\backslash$
   &$\backslash$
   &$\backslash$
   &$\backslash$
   \\ \cline{2-9} 
\multicolumn{1}{c|}{} &
  \begin{tabular}[c]{@{}c@{}}Track\\ management\end{tabular} &$\backslash$
   &
  \textcolor{green}{\ding{52}} &$\backslash$
   &$\backslash$
   &
  \textcolor{green}{\ding{52}} &$\backslash$
   &
  \textcolor{green}{\ding{52}} \\ \cline{2-9} 
\multicolumn{1}{c|}{} &
  \begin{tabular}[c]{@{}c@{}}Robustness\\ in the presence\\ of dense targets\end{tabular} &$\backslash$
   &$\backslash$
   &$\backslash$
   &
  \textcolor{green}{\ding{52}} &
  \textcolor{green}{\ding{52}} &
  \textcolor{green}{\ding{52}} &
  \textcolor{green}{\ding{52}} \\ \cline{2-9} 
\multicolumn{1}{c|}{} &
  \begin{tabular}[c]{@{}c@{}}Adaptation\\ when target\\  are added\end{tabular} &$\backslash$
   &$\backslash$
   &$\backslash$
   &$\backslash$
   &$\backslash$
   &$\backslash$
   &
  \textcolor{green}{\ding{52}} \\ \hline
\multicolumn{1}{c|}{\multirow{4}{*}[-28pt]{Disadvantages}} &
  \begin{tabular}[c]{@{}c@{}}High\\ computational\\ complexity\end{tabular} &
  \textcolor{red}{$\bigcirc$} &$\backslash$
   &$\backslash$
   &$\backslash$
   &
  \textcolor{red}{$\bigcirc$} &
  \textcolor{red}{$\bigcirc$} &
  \textcolor{red}{$\bigcirc$} \\ \cline{2-9} 
\multicolumn{1}{c|}{} &
  \begin{tabular}[c]{@{}c@{}}High\\ storage\\ requirements\end{tabular} &$\backslash$
   &
  \textcolor{red}{$\bigcirc$} &$\backslash$
   &$\backslash$
   &
  \textcolor{red}{$\bigcirc$} &$\backslash$
   &$\backslash$
   \\ \cline{2-9} 
\multicolumn{1}{c|}{} &
  \begin{tabular}[c]{@{}c@{}}Weak\\ anti-interferenc\\  capability\end{tabular} &$\backslash$
   &$\backslash$
   &
  \textcolor{red}{$\bigcirc$} &
  \textcolor{red}{$\bigcirc$} &$\backslash$
   &$\backslash$
   &$\backslash$
   \\ \cline{2-9} 
\multicolumn{1}{c|}{} &
  \begin{tabular}[c]{@{}c@{}}Poor robustness\\ when targets\\ are dense\end{tabular} &$\backslash$
   &$\backslash$
   &
  \textcolor{red}{$\bigcirc$} &
  \textcolor{red}{$\bigcirc$} &$\backslash$
   &
  \textcolor{red}{$\bigcirc$} &$\backslash$
   \\ \hline
\multicolumn{2}{c|}{Applicable Scenarios} &
  \begin{tabular}[c]{@{}c@{}}Low noise with\\ well-aligned\\ observations\\ and trajectories\end{tabular} &
  \begin{tabular}[c]{@{}c@{}}Temporarily intersect\\ trajectories or \\high uncertainty\end{tabular} &
  \begin{tabular}[c]{@{}c@{}}Few sparsely \\distributed targets\end{tabular} &
  \begin{tabular}[c]{@{}c@{}}Non-intersecting\\ trajectories with\\ low cluter ratio\end{tabular} &
  \begin{tabular}[c]{@{}c@{}}Dense targets with\\ highly intersecting\\ trajectories\end{tabular} &
  \begin{tabular}[c]{@{}c@{}}Dense targets under\\ moderate\\ noise conditions\end{tabular} &
  \begin{tabular}[c]{@{}c@{}}Well-defined \\parametric models\\ with limited targets\end{tabular} \\ \hline
\end{tabular}}
\begin{tablenotes}
 \footnotesize
  \item[*] The table summarizes the advantages, disadvantages, and applicability scenarios of the seven data association based approaches, where "\textcolor{green}{\ding{52}}" denotes\\ the method has this advantage ; "\textcolor{red}{$\bigcirc$}" denotes the method has this disadvantage; "$\backslash$" denotes does not have this advantage or disadvantage.
\end{tablenotes}
\end{threeparttable}
\end{table*}

In \cite{qiu2018}, a method combining PDA and the cubature KF (CKF) was proposed, demonstrating performance advantages in nonlinear environments.  Taking the algorithm proposed in this paper as an example, the structure of the multi-target tracking method based on data association is outlined as follows:

In the data association stage, the probability $\beta _k^{n,i}$ that each measurement $z_k^{n,i}$ from sensor $n$ originates from the target is calculated, as shown in Equation (\ref{con:beta}):
\begin{equation}\label{con:beta}
    \beta _k^{n,i} = \frac{{{\cal N}\left( {z_k^{n,i};\hat z_k^{n,i},{{\bf{S}}_k}} \right)}}{{\sum\limits_{i = 1}^{{I_k}(n)} {\cal N} \left( {z_k^{n,i};\hat z_k^{n,i},{{\bf{S}}_k}} \right)}},
\end{equation}
where $\hat z_k^{n,i} = {\bf{H}}({\widehat {\bf{x}}_{k|k - 1}})$ denotes the predicted measurement and ${{\bf{S}}_k} = {{\bf{P}}_{zz,k|k - 1}}$ represents the innovation covariance computed by the CKF.  Multiple measurements undergo probabilistically weighted fusion to yield a comprehensive innovation quantity $v_k^n$, as expressed in Equation (\ref{con:chuangxin}):
\begin{equation}\label{con:chuangxin}
    v_k^n = \sum\limits_{i = 1}^{{I_k}(n)} {\beta _k^{n,i}} \left( {z_k^{n,i} - \hat z_k^{n,i}} \right). 
\end{equation}

The filtering stage employs the combined integrated innovation metric $v_k^{}$ from all sensors to update the system state:
\begin{equation}
    {\widehat {\bf{x}}_{k|k}} = {\widehat {\bf{x}}_{k|k - 1}} + {{\bf{K}}_k}{v_k}. 
\end{equation}

Additionally, this methodology addresses multiple measurement uncertainties by substituting the conventional single measurement with a PDA-weighted measurement during the CKF update phase.  The covariance matrix is accordingly adjusted to account for association uncertainties through an additional correction term ${{\bf{K}}_k}{{\bf{W}}_k}{\bf{K}}_k^{\rm{T}}$:
\begin{equation}
    {{\bf{P}}_{k|k}} = {{\bf{P}}_{k|k - 1}} - {{\bf{K}}_k}\left( {{{\bf{W}}_k} - {{\bf{S}}_k}} \right){\bf{K}}_k^{\rm{T}},
\end{equation}
where ${{\bf{W}}_k} = {\rm{diag}}\left( {\sum\limits_{i = 1}^{{I_k}(n)} {\beta _k^{n,i}} {{\left( {v_k^{n,i}} \right)}^2} - {{\left( {v_k^n} \right)}^2}} \right)$ represents the PDA-weighted error matrix. 

Building upon the two-step paradigm of association-based filtering, Qiu et al. \cite{qiu2017} proposed a PDA filter method based on ship-radiated noise spectrum characteristics.  This approach integrates spectral features of ship-radiated noise by introducing additional characteristic information to augment conventional PDA filter, thereby enhancing tracking performance in dense environments.  Yao et al. \cite{yao2019a} incorporated target velocity components as supplementary measurements, developing a Doppler data association algorithm.  This method was further extended to linear multi-target integrated PDA techniques, with its superiority validated in marine environments. 

However, as the number of targets and valid echo counts increase, PDA algorithms exhibit exponential growth in computational complexity. To mitigate this issue, researchers have developed various suboptimal JPDA variants tailored for different application scenarios. Representative approaches include integrated JPDA \cite{musicki2002}, comprehensive JPDA \cite{Zhu1996}, and trajectory information-based JPDA \cite{chen2024}. As an exemplar optimization method, TJPDA employs multi-frame trajectory correlation analysis to screen valid measurement points \cite{chen2024}. This approach not only optimizes association probability computation but also reduces false alarm rates under low signal-to-clutter ratio conditions.

Furthermore, MHT method has been significantly enhanced for target tracking in complex environments. For instance, a novel distributed fusion method combining MHT with data fusion techniques was proposed in \cite{seget2010}. This method processes individual sensor tracking states using supplementary MHT trackers while eliminating the need for trajectory association, thereby streamlining data processing and enhancing system efficiency. Another study \cite{li2016} integrated probabilistic MHT (PMHT) with both EKF and UKF algorithms. Comparative analysis with the GNN algorithm demonstrated their superior performance in strong interference environments.

Building upon the model proposed in \cite{li2016}, Li et al. \cite{li2015} developed an enhanced PMHT algorithm capable of simultaneously resolving measurement-to-target and measurement-to-emitter association uncertainties. To overcome limitations of conventional PMHT algorithms - particularly convergence to local maxima in target posterior probabilities and initialization sensitivity - Li et al.\cite{li2019} introduced a deterministic annealing homothetic PMHT algorithm. This advanced variant demonstrates reduced dependence on initial target states and incorporates gaussian densities with identical means but distinct covariances to model measurements from single targets, thereby improving tracking accuracy.

Current data association-based tracking methodologies are fundamentally constrained by the requirement for prior knowledge of the number of targets and by computational complexity that increases exponentially with target density. These limitations present significant barriers to technological advancement. Consequently, the development of underwater multi-target tracking algorithms with superior performance and enhanced estimation precision carries both substantial practical value and strategic importance.

\textit{2-2) \textbf{RFS-based tracking}}

In multi-target tracking systems, the number of targets may vary dynamically over time, leading to mismatched dimensions between the state space and observation space. Notably, data association—a core component of multi-target tracking—is classified as an NP-hard problem \cite{gao2021}. Empirical studies indicate that over 60 \% of computational resources in tracking algorithms are consumed by this process, underscoring the necessity to optimize computational efficiency and enhance real-time performance \cite{palkki2011}. Furthermore, high clutter density poses significant challenges: traditional algorithms often misinterpret false alarms outside association thresholds as new targets, thereby exponentially increasing false tracks. Consequently, conventional data association-based methods frequently fail to satisfy practical demands for both computational efficiency and tracking accuracy \cite{mahler2007}.

To overcome the limitations of traditional multi-target tracking algorithms in computational efficiency and tracking performance, particularly in complex scenarios, the authors of \cite{mahler2007} introduced the Finite Set Statistics (FISST) theory framework. This framework, based on RFS theory, enables multisensor multi target filtering. Specifically, they proposed an efficient representation of the RFS probability density function, termed the FISST probability density function, which is determined by a symmetric joint probability density function $P({{\bf{x}}_1}, \cdots ,{{\bf{x}}_n})$ and a potential distribution $\rho (n)$, as shown in Equation (\ref{con:FISST}):
\begin{equation}\label{con:FISST}
    p\left( X \right) = p\left( {\left\{ {{{\bf{x}}_1}, \cdots ,{{\bf{x}}_n}} \right\}} \right) = n!\rho \left( n \right)P\left( {{{\bf{x}}_1}, \cdots ,{{\bf{x}}_n}} \right),
\end{equation}
where $X = \left\{ {\begin{array}{*{20}{c}}
{{{\bf{x}}_1}, \cdots ,{{\bf{x}}_n}}
\end{array}} \right\}$ represents the RFS variable. 

The RFS-based multi-target tracking algorithm models the multi-target density as an RFS. Within the Bayesian framework, the FISST-based Bayesian filtering algorithm iteratively propagates the RFS density, effectively transforming the multi-target tracking problem into a mapping problem between random sets of target states and measurements. This approach simultaneously estimates both the number of targets and their states \cite{wang}. However, the combinatorial nature of multi-target probability density and the multiple integrals in high-dimensional multi target state spaces make the implementation of an optimal RFS-based multi target Bayesian filter highly challenging. \textbf{Consequently, approximate filtering methods, such as probability hypothesis density (PHD) filter and multi-bernoulli (MB) filter, have been developed as principled approximations.}

\textbf{The PHD filter} assumes that the multi-target probability density follows a Poisson RFS distribution, defined as:
\begin{equation}
    {\pi _{k|k}}\left( X \right) = {e^{ - \lambda }}\prod\limits_{x \in X} {{\upsilon _{k|k}}} \left( {\bf{x}} \right),
\end{equation}
where $\lambda $ represents the mean number of targets in the multi-target state $X$, and the target number distribution $\rho (n)$ follows a Poisson distribution with parameter $\lambda $. 

The PHD filtering algorithm employs the PHD function - a first-order moment approximation of the multi-target probability density - to estimate the complete multi-target probability distribution. This approach achieves multi-target Bayesian filtering through recursive propagation of the PHD function \cite{mahler2003,ba-nguvo2005,vo2006}. To overcome the limitation of the PHD filter in providing higher-order potential information, the multi-target state set is modeled as a cluster point process instead of a Poisson point process. Consequently, the simultaneous propagation of both the PHD function and potential distribution yields the Cardinalized PHD (CPHD) filter \cite{Zou2019}.

Addressing the performance degradation of conventional CPHD filters under unknown clutter rates - a common scenario in practical applications - Kim \cite{kim2024} proposed an enhanced modeling approach. Their method assumes prior knowledge of the maximum target count $TM$ in the operational space. By implementing a sliding window technique to compute the discrepancy between measured values and $TM$, the algorithm dynamically estimates the clutter rate. This innovative approach leverages prior information about maximum target counts, eliminates the need for explicit data association, and maintains computational efficiency.

Further advancing the filter framework, Zhou et al. \cite{zhou2023} introduced a Gaussian Mixture CPHD (GM-CPHD) filter that integrates threshold segmentation with the GM-CPHD methodology. This improved technique extracts critical features including target regions, topological factors, and edge moments. Utilizing maximum likelihood estimation, it calculates feature contributions and selects optimal features, with automatically determined thresholds effectively distinguishing true targets from clutter.

\textbf{The MB filter} \cite{2019改进的SMC} employs a parametric approximation method. Specifically, at each time step k, the multi-target state can be mathematically represented as a union of multiple independent Bernoulli RFS, as formalized in Equation (\ref{con:RFS1}):

\begin{equation}\label{con:RFS1}
    X = \bigcup\limits_{i = 1}^{{n_k}} {{X_i}} . 
\end{equation}

The density function of ${X_i}$ is represented by the Bernoulli RFS distribution:
\begin{equation}
    {\pi _{k|k}}\left( {{X_i}} \right) = \left\{ {\begin{array}{*{20}{l}}
{1 - {r_{k|k,i}},}&{{\rm{if }}{X_i} = \emptyset }&{}\\
{{r_{k|k,i}} \cdot {p_{k|k,i}}\left( {\bf{x}} \right),}&{{\rm{if }}{X_i} = \{ {\bf{x}}\} }&{}\\
{0,}&{{\rm{else}}}&{}
\end{array}} \right. ,
\end{equation}
where ${r_{k|k,i}} \in \left( {\begin{array}{*{20}{c}}
{0,1}
\end{array}} \right)$ represents the probability of existence for the $i$-th target.  According to the definition above, the multi-target probability density can be completely represented by the MB parameter set $\left\{ {\begin{array}{*{20}{c}}
{{r_{k|k,i}},{p_{k|k,i}}({\bf{x}})}
\end{array}} \right\}_{i = 1}^{{n_k}}$.  The standard MB filtering algorithm recursively propagates the MB parameter set to replace the full multi-target probability density, thereby approximating the multi-target Bayesian filtering process. 

\begin{table*}[ht]
\centering
\caption{Comparison of Various RFS based Methods}
\label{tab:RFS2}
\begin{threeparttable}
\scalebox{0.87}{\begin{tabular}{|cc|p{2.8cm}<{\centering}|p{2.8cm}<{\centering}|p{2.8cm}<{\centering}|p{2.8cm}<{\centering}|p{2.8cm}<{\centering}}
\hline
\multicolumn{2}{c|}{Method} &
  \begin{tabular}[c]{@{}c@{}}PHD\\ \cite{mahler2003,ba-nguvo2005,vo2006}\end{tabular} &
  \begin{tabular}[c]{@{}c@{}}CPHD\\ \cite{Zou2019,kim2024,zhou2023}\end{tabular} &
  \begin{tabular}[c]{@{}c@{}}MB\\ \cite{2019改进的SMC}\end{tabular} &
  \begin{tabular}[c]{@{}c@{}}LMB\\ \cite{zhang2020,gou2024}\end{tabular} &
  \begin{tabular}[c]{@{}c@{}}PMBM\\ \cite{lin2016,legrand2018}\end{tabular} \\ \hline
\multicolumn{1}{c|}{\multirow{5}{*}[-27pt]{Advantages}}    & \begin{tabular}[c]{@{}c@{}}Low\\ computational\\ complexity\end{tabular}               & \textcolor{green}{\ding{52}} & $\backslash$  & \textcolor{green}{\ding{52}} &  $\backslash$ &  $\backslash$ \\ \cline{2-7} 
\multicolumn{1}{c|}{}                               & \begin{tabular}[c]{@{}c@{}}Quantity\\ estimates\end{tabular}                           &  $\backslash$ & \textcolor{green}{\ding{52}} & $\backslash$  &  $\backslash$ & $\backslash$  \\ \cline{2-7} 
\multicolumn{1}{c|}{}                               & \begin{tabular}[c]{@{}c@{}}track\\ management\end{tabular}                             &  $\backslash$ &  $\backslash$ & $\backslash$  & \textcolor{green}{\ding{52}} & \textcolor{green}{\ding{52}} \\ \cline{2-7} 
\multicolumn{1}{c|}{}                               & \begin{tabular}[c]{@{}c@{}}High\\ computational\\ efficiency\end{tabular}              & $\backslash$  & $\backslash$  & $\backslash$  & $\backslash$ & \textcolor{green}{\ding{52}} \\ \cline{2-7} 
\multicolumn{1}{c|}{}                               & \begin{tabular}[c]{@{}c@{}}Adaptation\\ when target\\  are added\end{tabular}          &  $\backslash$ &  $\backslash$ &  $\backslash$ & $\backslash$  & \textcolor{green}{\ding{52}} \\ \hline
\multicolumn{1}{c|}{\multirow{3}{*}[-15pt]{Disadvantages}} & \begin{tabular}[c]{@{}c@{}}High\\ computational\\ complexity\end{tabular}              & $\backslash$  & \textcolor{red}{$\bigcirc$} &  $\backslash$ & \textcolor{red}{$\bigcirc$} & \textcolor{red}{$\bigcirc$} \\ \cline{2-7} 
\multicolumn{1}{c|}{}                               & \begin{tabular}[c]{@{}c@{}}Poor robustness\\ when targets\\ are dense\end{tabular}     & \textcolor{red}{$\bigcirc$} &  $\backslash$ & \textcolor{red}{$\bigcirc$} & $\backslash$  &  $\backslash$ \\ \cline{2-7} 
\multicolumn{1}{c|}{}                               & \begin{tabular}[c]{@{}c@{}}Difficulty in\\ dealing with\\ nascent targets\end{tabular} &  $\backslash$ & \textcolor{red}{$\bigcirc$} &  $\backslash$ &  $\backslash$ &  $\backslash$ \\ \hline
\multicolumn{2}{c|}{Applicable Scenarios} &
  \begin{tabular}[c]{@{}c@{}}Scenarios with\\ dynamic target count \\and low computational \\resources\end{tabular} &
  \begin{tabular}[c]{@{}c@{}}Scenarios requiring\\ stable target\\ count and precise\\ estimation\end{tabular} &
  \begin{tabular}[c]{@{}c@{}}Scenarios with\\ sparse targets\\ requiring identity\\ distinction\end{tabular} &
  \begin{tabular}[c]{@{}c@{}}High-precision scenarios\\ with dense targets\\ and crossing trajectories\end{tabular} &
  \begin{tabular}[c]{@{}c@{}}Scenarios with clutter\\ and newly\\ emerging targets\end{tabular} \\ \hline
\end{tabular}}
\begin{tablenotes}
 \footnotesize
  \item[*] The table summarizes the advantages, disadvantages and applicability scenarios of the five RFS based approaches, where "\textcolor{green}{\ding{52}}" denotes the \\  method has this advantage ; "\textcolor{red}{$\bigcirc$}" denotes the method has this disadvantage; "$\backslash$" denotes does not have this advantage or disadvantage.
\end{tablenotes}
\end{threeparttable}
\end{table*}

To address the issue of target trajectory estimation, the labeled multi-bernoulli (LMB) filter is derived based on the theory of labeled RFS \cite{vo2013}. Notably, Zhang et al.\cite{zhang2020} enhanced the input to the LMB filter by processing raw measurement data from the sensor array directly, eliminating the need for a target detection step. This method preserves more informative data, especially under low SNR conditions or when tracking closely spaced targets, consequently improving tracking accuracy. In a similar vein, Zheng et al. \cite{zheng2024} employed raw sensor array signals to construct the sample covariance matrix as measurements, while integrating the complex Wishart distribution and its inverse counterpart for statistical modeling. In this case, the sample covariance matrix ${R_{{z_k}}} = {z_k}z_k^H$ is modeled as a complex Wishart distribution, with the specific form given by Equation (\ref{con:Wishart}):
\begin{equation}\label{con:Wishart}
    {\cal C}{{\cal W}_P}({R_{{z_k}}};M,\sigma _w^2{I_P} + \sum\limits_{x \in {X_k}} h (x)),
\end{equation}
where $P$ denotes the number of sensor elements, which corresponds to the dimensionality of the sensor array's output signal, $M$ represents the degrees of freedom, $\sigma _w^2{I_P}$ represents the noise covariance matrix, and $h(x)$ represents the nonlinear function of the target signal. The conjugacy between the complex Wishart and complex inverse Wishart distributions simplifies high-dimensional matrix integrals, reducing computational complexity.  This leads to a tractable filtering equation and improves the performance of the filter in complex scenarios such as low SNR.

In their seminal work \cite{gou2024}, the research team enhanced the data association component of the LMB filter by introducing a pre-association decision module. This innovative module adaptively selects trajectory estimation strategies based on target proximity: when targets are in close proximity, the system utilizes the trajectory prediction from the preceding time step to mitigate association errors; conversely, for distant targets, it employs the nearest neighbor method for trajectory generation. Building upon this foundation, Zhang et al. \cite{zhang2023} developed an advanced approximation of the multi-sensor LMB filter through Kullback-Leibler divergence minimization, ultimately formulating a robust multi-sensor LMB framework that demonstrates superior performance in complex scenarios.

The poisson MB mixture (PMBM) filter \cite{williams2015}, based on the poisson MB mixture density model, introduces the conjugate prior RFS into the update step, ensuring that the multi-target density form remains unchanged after prediction and update steps. This eliminates the need for the minimization approximation used in PHD-type filters, reducing computational complexity while maintaining algorithm stability.  Furthermore, the PMBM filter models detected targets using a MB mixture, while undetected targets are modeled as Poisson RFS, making it suitable for multi-target tracking scenarios where prior information about new targets is unavailable \cite{lin2016,legrand2018}.  

Table \ref{tab:RFS2} compares the advantages, disadvantages, and limitations of various multi-target tracking methods based on RFS. 

\subsection{Target Tracking Methods Based on Perception Method}
Sensor perception modalities are primarily classified into active and passive perception. Active perception involves sensors emitting detection signals and receiving the corresponding echoes, whereas passive perception entails solely receiving signals emitted by targets for non-contact sensing. Based on these two perception modalities, tracking algorithms can be categorized into \textbf{active tracking} algorithms and \textbf{passive tracking} algorithms.

\begin{figure*}
    \centering
    \includegraphics[width=0.8\linewidth]{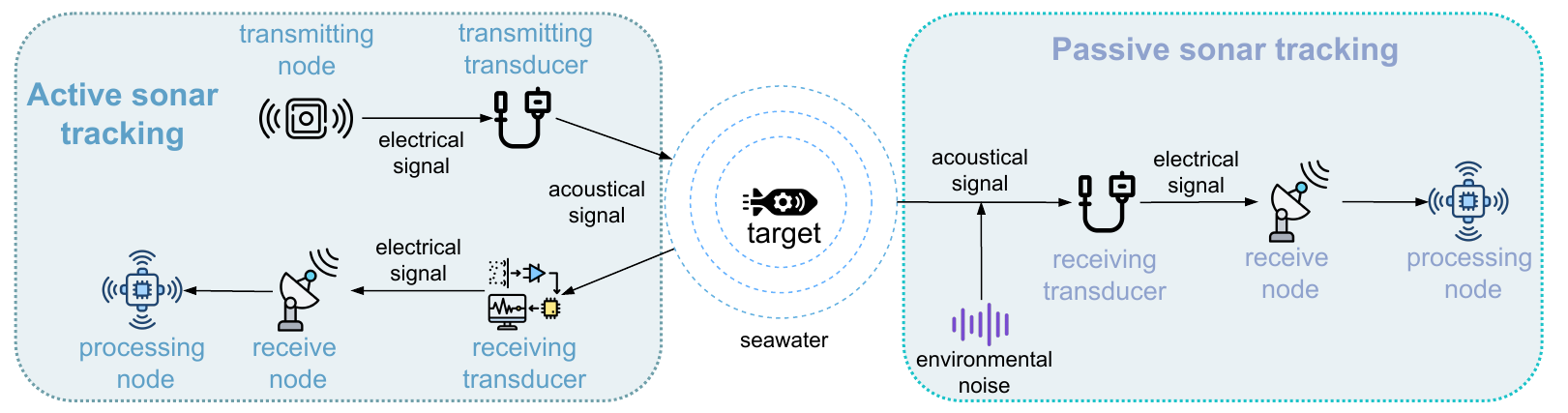}
    \caption{Schematic diagram of active and passive tracking.  The left is the process of active tracking, active tracking will actively emit acoustic signals, the tracking process starts from the sending node and ends from the receiving node; the right is the process of passive tracking, passive tracking will not actively emit acoustic signals, and only use the receiving node to receive acoustic signals from the outside world. }
    \label{fig:sonar}
\end{figure*}

Active tracking algorithms aim to estimate the target state by utilizing sensors such as active sonar, which provide information on the azimuth and range of the target based on the echoes received. These echoes originate from the system itself \cite{zhang2018}. Passive tracking utilizes sensors such as passive sonar to receive and analyze signals radiated by the target in order to track it, without transmitting detection signals. This approach avoids issues such as exposing the tracker’s information and the energy consumption associated with signal transmission \cite{guleria2021data}. The schematic of active and passive sonar perception is shown in Figure \ref{fig:sonar}.
\subsubsection{\textbf{Active Tracking}}
Active sonar systems play a pivotal role in underwater target detection and tracking, comprising three essential components: a transmitter, a receiver, and a signal processing unit. The transmitter periodically emits designed acoustic pulses, while the receiver captures the corresponding echoes. Subsequently, the processing unit analyzes these echoes to extract critical target parameters including position, velocity, and range \cite{abraham2011}. Nevertheless, active tracking presents significant operational risks as the emitted signals may reveal the sensor’s location, particularly in complex combat scenarios, potentially compromising system security \cite{han2019}. This inherent vulnerability has resulted in limited research attention in this domain.

The dual-static sonar configuration represents the most prevalent implementation for active tracking, featuring spatially separated transmitters and receivers. In this system, the transmitter emits a calibrated acoustic signal toward the target, which reflects back to the receiver. Target range estimation is then achieved through analysis of signal intensity variations, as mathematically described in Equation (\ref{con:jieshou}):
\begin{equation}\label{con:jieshou}
    SE = SL - T{L_1} - T{L_2} + TS - {L_e} - Nr{d_n},
\end{equation}
where $SE$ and $SL$ represent the signal strength received by the receiver and the signal strength transmitted by the transmitter, respectively.  $L_e$ denotes the noise level output by the beamformer, while $Nrd_n$ is related to the receiver's ability to identify noise.  $TL_1$ and $TL_2$ represent the transmission losses from the transmitter to the target and from the target to the receiver, respectively, with the distance information implicitly embedded in the transmission losses. 

In  \cite{ferri2014}, Ferri et al. developed a dual-static sonar system comprising a buoy-mounted transmitter and a mobile autonomous underwater vehicle (AUV) receiver. To enhance tracking performance, they implemented a non-myopic rolling horizon control strategy that predicts heading decisions over a five-step horizon. This approach dynamically optimizes the AUV's position by minimizing target distance while preserving wide-angle orientation, thereby reducing positioning errors. The methodology further incorporates decision tree simplification and branch-and-bound techniques to address computational constraints inherent to AUV systems.

Parallel to this work, Son et al.\cite{son2013} presented a fuzzy c-means (FCM) clustering-based algorithm for underwater target tracking in bi-static sonar configurations. The proposed method processes residuals between predicted states and actual measurements through FCM clustering, decomposing them into noise, acceleration, and deceleration components. After eliminating acceleration/deceleration perturbations, the linear dynamic components are retained as corrected measurements for subsequent KF.

Dehnavi et al. \cite{dehnavi2017} and Yan et al.\cite{yan2016a} proposed two distinct approaches to target tracking in underwater wireless sensor networks (UWSNs). The work in \cite{dehnavi2017} developed an innovative tracking method focusing on sensor network collaboration. This study systematically analyzed challenges in three-dimensional underwater acoustic wave propagation, including multipath effects, low propagation speeds, bandwidth limitations, and sensor corrosion. By employing both EKF and UKF techniques, the method effectively filters measurement noise, thereby improving dynamic target tracking accuracy and enhancing system robustness under nonlinear measurement models. To address energy consumption issues in UWSNs, the authors introduced an intelligent sensor selection scheme that activates only sensors within a specific radius of the target, significantly reducing both energy consumption and interference.

In contrast, Yan et al. \cite{yan2016a} proposed a consensus estimation-based tracking algorithm to simultaneously improve tracking accuracy and energy efficiency. This approach integrates position information from multiple sensors through an advanced data fusion mechanism, enabling more precise target localization. The algorithm not only accounts for underwater environmental factors and sensor noise but also implements a novel duty cycling strategy based on predicted target states. Sensors remain in low-power sleep mode until target detection occurs, substantially improving network energy efficiency and prolonging operational lifetime. Different from the active pulse transmission mechanism of designated projection nodes used in \cite{dehnavi2017}, this method allows each hydrophone sensor to autonomously initiate acoustic signal transmission upon target detection.

\subsubsection{\textbf{Passive Tracking}}
Passive tracking systems exhibit three notable advantages: extended detection range, enhanced operational safety, and reduced energy consumption. Nevertheless, underwater target tracking faces considerable challenges due to dynamic environmental variables, particularly water depth, salinity variations, and ocean currents \cite{tan2011}.

Recent advancements in traditional tracking methods (e.g., KF) have led to three primary research focuses: \textbf{(1) enhancing tracking accuracy in noisy marine environments through the integration of filtering methods with advanced data processing techniques; (2) optimizing multi-target tracking performance by combining multiple filtering approaches; and (3) improving tracking capability in complex dynamic environments through real-time noise covariance matrix estimation and adjustment under uncertain noise conditions.} These developments not only advance target tracking technology but also provide both theoretical foundations and practical guidance for achieving more efficient and secure underwater target monitoring.

\textit{2-1) \textbf{Filtering methods combined with data processing techniques}}

Kumar et al. \cite{ravikumar2016} implemented a preprocessing framework where raw measurements were substituted with filtered outputs before being fed into multiple UKF branches. This preprocessing stage effectively attenuates measurement noise, thereby stabilizing the input for estimation algorithms and improving tracking precision in noisy marine environments. Another significant development appears in \cite{luo2020}, where researchers introduced an improved unscented PF featuring adaptive Kalman gain adjustment based on prediction residuals. This modification addresses the inherent latency problems of fixed-gain approaches in nonlinear systems, substantially boosting tracking performance under severe nonlinear conditions. Cheng et al.\cite{cheng2021} proposed a novel approach that unifies the prediction and observation components of the EKF within a factor graph framework. This innovative integration connects the EKF’s observer and predictor through the factor graph methodology, enabling sensors to transmit only a minimal set of parameters that characterize their probability density function (PDF) attributes. Consequently, this approach significantly reduces both latency and power consumption while maintaining superior tracking performance.

\textit{2-2) \textbf{Hybrid filtering approaches}}

Another approach to algorithm integration involves combining different filtering methods. For instance, Kumar et al. \cite{kumar2016} proposed an integrated UKF that merges estimates from multiple UKFs and processes multisensor data from towed arrays in parallel. By applying the least squares method to integrate measurements from different towed array sensors, this filter retains the nonlinear processing capabilities of the UKF while significantly reducing position and velocity estimation errors.

Similarly, Luo et al. \cite{luo2020} explored the combination of filtering techniques by integrating the improved unscented PF with JPDA method. This hybrid approach leverages particle filtering to address nonlinearities and employs data association to resolve multi target conflicts, making it suitable for multi target tracking in complex scenarios.

Furthermore, although many studies employ the EKF or UKF to account for the nonlinear relationship between azimuth measurements and target states, underwater target motion often follows a simple linear model. This is because underwater targets typically exhibit limited velocities, traverse long distances, and undergo minimal angular changes during signal sampling intervals. 

Building on this observation, Qian et al. \cite{qian2016} introduced a single-dimensional and double-sided constant false alarm rate method. This technique enhances target detection at crossover points by analyzing energy differences on both sides of the target cell. By combining traditional azimuth gates with radiation intensity gates and utilizing target radiation noise intensity as a distance metric, it constructs multidimensional tracking gates to optimize measurement association. Integrating these methods with standard KF substantially reduces computational complexity.

\textit{2-3) \textbf{Filtering methods under uncertain noise conditions}}

As discussed in Chapter \ref{2}, numerous studies on underwater acoustic target tracking have traditionally assumed Gaussian-distributed environmental noise due to its favorable mathematical properties and close approximation of real-world conditions. However, actual marine environments are subject to dynamic factors such as ocean currents and temperature variations, resulting in significant temporal fluctuations that challenge this assumption.

Consequently, developing robust and accurate target tracking methods under uncertain noise conditions becomes imperative. Recent approaches to address measurement noise uncertainty have focused on integrating real-time covariance matrix estimation and updates within tracking filter algorithms.

The Sage-Husa adaptive filter is a recursive filtering method that dynamically estimates and adjusts the statistical properties of both system noise and observation noise in real time. This approach is particularly effective for underwater acoustic target tracking in scenarios with uncertain noise conditions. The algorithm’s core mechanism involves iterative updates to the observation error covariance matrix and the model error noise covariance matrix, as formalized in Equations (\ref{con:Sage1}) and (\ref{con:Sage2}):
\begin{equation}\label{con:Sage1}
\begin{aligned}
     {{\bf{R}}_k} =& (1 - d){{\bf{R}}_{k - 1}} +\\
     d( {{\bf{e}}_k}{\bf{e}}_k^T &- {{\bf{H}}_k}{{\bf{P}}_{k|k - 1}}{\bf{H}}_k^T ) ,
\end{aligned}
\end{equation}
\begin{equation}\label{con:Sage2}
\begin{aligned}
     {{\bf{Q}}_k} =& (1 - d){{\bf{Q}}_{k - 1}} + \\
     d ({{\bf{K}}_k}{{\bf{e}}_k}{\bf{e}}_k^T{\bf{K}}_k^T +&{{\bf{P}}_{k|k - 1}} - {{\bf{F}}_k}{{\bf{P}}_{k - 1|k - 1}}{\bf{F}}_k^T) ,
\end{aligned}
\end{equation}
where ${{\bf{R}}_k}$ and ${{\bf{Q}}_k}$ represent the observation error noise and model error noise at time $K$, respectively.  ${{\bf{H}}_k}$ and ${{\bf{F}}_k}$ represent the observation model and the target state model, respectively.  ${{\bf{K}}_k}$ denotes the Kalman gain, and $d$ is the forgetting factor, which is used to assign weight to historical data.  When $d$ approaches 0, the algorithm tends to fully trust historical data, at which point the Sage-Husa algorithm degenerates into a standard Kalman filter. 

To further enhance the Sage-Husa algorithm, Qu et al. \cite{qu2007} integrated a decay memory mechanism into the EKF framework. This modification improves the accuracy of both state and measurement equations by utilizing local dynamic statistical features derived from residual analysis. Their proposed adaptive algorithm combines azimuth and time-delay measurements for underwater target tracking, supplemented by a fixed sensor. This sensor configuration simplifies system design by leveraging sonar-based azimuth data and time-delay information between the sonar and sensor, thereby reducing the complexity of target tracking in challenging environments. Compared to conventional towed array systems, this method demonstrates superior implementation flexibility and practical applicability.

\begin{table*}[ht]
\renewcommand{\arraystretch}{1.2}
\centering
\caption{Comparison of the characteristics and specific methods of active and passive tracking}
\label{tab:ganzhi}
\begin{threeparttable}
\begin{tabular}{c|c|c|c|c}
\hline
Mode & Advantages & Disadvantages & Literature    & Tracking Method                                                      \\ \hline
\multirow{4}{*}{Active tracking} &
  \multirow{4}{*}{\begin{tabular}[c]{@{}c@{}}High precision,\\ controllable signal frequency,\\ long-range detection\end{tabular}} &
  \multirow{4}{*}{\begin{tabular}[c]{@{}c@{}}Prone to location exposure,\\ high energy consumption\end{tabular}} &
  {\cite{ferri2014}} &
  \begin{tabular}[c]{@{}c@{}}Control strategy\\ for AUV  a branch\\ and bound technique\end{tabular} \\ \cline{4-5} 
     &            &               & {\cite{son2013}}     & FCM,KF                                                               \\ \cline{4-5} 
     &            &               & {\cite{dehnavi2017}}     & EKF,UKF                                                              \\ \cline{4-5} 
     &            &               & {\cite{yan2016a}}     & CEUTT                                                                \\ \hline
\multirow{10}{*}{Passive tracking} &
  \multirow{10}{*}{\begin{tabular}[c]{@{}c@{}}Strong concealment,\\ low energy consumption\end{tabular}} &
  \multirow{10}{*}{\begin{tabular}[c]{@{}c@{}}Low precision,\\ susceptible to environmental\\ interference\end{tabular}} &
  {\cite{ravikumar2016}} &
  \begin{tabular}[c]{@{}c@{}}Pre-processing of noise,\\ UKF, IUKF\end{tabular} \\ \cline{4-5} 
     &            &               & {\cite{luo2020}}     & JPDA,IUPF                                                            \\ \cline{4-5} 
     &            &               & {\cite{cheng2021}}     & EKF,factor graph                                                     \\ \cline{4-5} 
     &            &               & {\cite{kumar2016}}     & \begin{tabular}[c]{@{}c@{}}consensus estimation,\\ IUKF\end{tabular} \\ \cline{4-5} 
     &            &               & {\cite{luo2020}}     & EKF,UKF                                                              \\ \cline{4-5} 
     &            &               & {\cite{qian2016}}     & SD-CFAR,KF                                                           \\ \cline{4-5} 
     &            &               & {\cite{qu2007,zhang2022}} & Sage-Husa,EKF                                                        \\ \cline{4-5} 
     &            &               & {\cite{hou2021}}     & Sage-Husa,APF                                                        \\ \cline{4-5} 
     &            &               & {\cite{sarkka2009}}     & VB,KF                                                                \\ \cline{4-5} 
     &            &               & {\cite{huang2007}}     & EF,EKF                                                               \\ \hline
\end{tabular}
\begin{tablenotes}
 \footnotesize
  \item[*] The table describes the advantages and disadvantages of active and passive tracking, along with a list of relevant literature and a summary of the tracking methods used in the literature.
\end{tablenotes}
\end{threeparttable}
\end{table*}

In \cite{zhang2022}, the Sage-Husa algorithm was further improved by dynamically adjusting the forgetting factor $d$ ,as described in Equation (\ref{con:forgetting factor}):
\begin{equation}\label{con:forgetting factor}
    {d_k} = \frac{{1 - b}}{{1 - {b^k}}},
\end{equation}
where the parameter $b$ ranges from 0 to 1.  The authors also integrated the improved Sage-Husa algorithm with a circular array EKF, thereby enhancing the stability of the linearized filtering process. 

Building on this work, Hou et al.\cite{hou2021} employed the same dynamic forgetting factor to optimize historical data weighting, thereby improving the estimation of process and measurement noise moments. The study further integrated an adaptive PF  utilizing Bayesian posterior probability and Monte Carlo techniques, demonstrating significant improvements in tracking accuracy and convergence speed under non-Gaussian conditions.

Subsequently, Sarkka \cite{sarkka2009} developed a variational bayesian approximation-based adaptive KF (VB-AKF) that estimates the joint posterior distribution of states and noise parameters via separable variational approximation. Parallel to this, Huang et al. \cite{huang2007} proposed an Expectation-Maximization-based Adaptive KF  capable of estimating non-diagonal covariance matrices, substantially expanding VB-AKF’s applicability to complex systems. An alternative approach involving auxiliary filters was presented in \cite{li2013}, where an online process noise variance estimator enabled adaptive performance adjustment based on target maneuvering intensity, enhancing filter performance across both uniform and maneuvering motion scenarios.

Table \ref{tab:ganzhi} compares the two perceptual modes and also lists the corresponding articles in order to give the reader a quick overview of the characteristics of the different perceptual modes and the general application scenarios of each paper. 


\subsection{Target Tracking Methods Based on Collaborative Mode}
When utilizing sensors for hydroacoustic target tracking, a single sensor can process the signals emitted by the target and estimate the target state, but the tracking accuracy is limited due to the very limited energy resources, bandwidth resources, and data processing capabilities of a single sensor. Therefore, it is considered to utilize the collaboration between multiple sensors to improve the tracking performance of the system and make the system more robust and reliable. The collaboration modes of sensors can be classiﬁed into \textbf{same-domain fusion} and \textbf{cross-domain multimodal fusion} based on the type of sensors. Table \ref{tab:xiezuo} provides a detailed comparison of the characteristics of these three collaborative modes. 
\begin{table*}
\centering
 \renewcommand{\arraystretch}{1.1}
\caption{Comparison of the features of three collaborative modes}
\label{tab:xiezuo}
\scalebox{0.84}{\begin{threeparttable}\begin{tabular}{cc|c|c|c}
\hline
\multicolumn{2}{c|}{Collaborative mode}  &
  Centralized Fusion &
  Distributed Fusion &
  Cross-domain Multimodal Fusion \\ \hline
\multicolumn{2}{c|}{literature} &
   {\begin{tabular}[c]{@{}c@{}}\cite{2021基于水下传感器}\cite{zhang2019a}\cite{kumar2021b} \\ \cite{qiu2019}\cite{hu2010}\cite{zhou2019}\end{tabular}}&
   {\begin{tabular}[c]{@{}c@{}}\cite{2021基于水下传感器}\cite{xin2003}\cite{yu2014}\cite{hare2017}\\\cite{hare2014}\cite{braca2016}\cite{han2019}\end{tabular}}&
   {\begin{tabular}[c]{@{}c@{}}\cite{2021基于水下传感器}\cite{jia2020}\cite{lee2012}\cite{meng-chechuang2015}\\\cite{mandic2016}\cite{snyder2012}\cite{xue2018}\cite{2021基于飞秒激光}\end{tabular}}
   
   \\ \hline
\multicolumn{1}{c}{\multirow{8}{*}[-64pt]{Feature}} &
  \begin{tabular}[c]{@{}c@{}}Data\\ Processing\end{tabular} &
  \begin{tabular}[c]{@{}c@{}}All data converges to fusion\\ node for centralized processing\end{tabular} &
  \begin{tabular}[c]{@{}c@{}}Nodes independently process \&\\ partially fuse data, share final results\end{tabular} &
  \begin{tabular}[c]{@{}c@{}}Fuses multi-type sensor data,\\ requires solving spatiotemporal\\ alignment \&feature fusion challenges\end{tabular} \\ \cline{2-5} 
\multicolumn{1}{c}{} &
  \begin{tabular}[c]{@{}c@{}}Communication\\ Needs\end{tabular} &
  \begin{tabular}[c]{@{}c@{}}\textcolor{red}{High}: Full raw data\\ transmission to center\end{tabular} &
  \begin{tabular}[c]{@{}c@{}}\textcolor{green}{Low}: Only local estimates \&\\ essential info exchange\end{tabular} &
  \begin{tabular}[c]{@{}c@{}}\textcolor{blue}{Medium}: Requires synchronized\\ multimodal streams \&\\ protocol/format conversion\end{tabular} \\ \cline{2-5} 
\multicolumn{1}{c}{} &
  Scalability &
  \begin{tabular}[c]{@{}c@{}}\textcolor{green}{Low}: Computational/communication\\ load increases with sensor expansion\end{tabular} &
  \begin{tabular}[c]{@{}c@{}}\textcolor{red}{High}: Easy node addition with\\ strong local processing\end{tabular} &
  \begin{tabular}[c]{@{}c@{}}\textcolor{blue}{Medium}: New sensor types require\\ algorithm adaptation \& hardware\\ compatibility\end{tabular} \\ \cline{2-5} 
\multicolumn{1}{c}{} &
  Fault Tolerance &
  \begin{tabular}[c]{@{}c@{}}\textcolor{green}{Low}: System failure upon\\ central node malfunction\end{tabular} &
  \begin{tabular}[c]{@{}c@{}}\textcolor{red}{High}: Single node failure has\\ minimal system impact\end{tabular} &
  \begin{tabular}[c]{@{}c@{}}\textcolor{blue}{Medium}: Relies on redundant\\ modalities with modality missing\\ robustness algorithms\end{tabular} \\ \cline{2-5} 
\multicolumn{1}{c}{} &
  \begin{tabular}[c]{@{}c@{}}Real-time\\ Performance\end{tabular} &
  \begin{tabular}[c]{@{}c@{}}\textcolor{green}{Low}: Centralized processing\\ induces latency\end{tabular} &
  \begin{tabular}[c]{@{}c@{}}\textcolor{red}{High}: Local decision-making\\ with low fusion delay\end{tabular} &
  \begin{tabular}[c]{@{}c@{}}\textcolor{blue}{Medium}: Additional computation\\ time for multimodal fusion\\ (hardware-dependent)\end{tabular} \\ \cline{2-5} 
\multicolumn{1}{c}{} &
  \begin{tabular}[c]{@{}c@{}}Computational\\ Load\end{tabular} &
  \begin{tabular}[c]{@{}c@{}}\textcolor{red}{High}: Demands high-performance\\ computing at central node\end{tabular} &
  \begin{tabular}[c]{@{}c@{}}\textcolor{green}{Low}: Distributed computation\\ across nodes\end{tabular} &
  \begin{tabular}[c]{@{}c@{}}\textcolor{red}{High}: Requires complex algorithms\\ on fusion nodes/edge devices\end{tabular} \\ \cline{2-5} 
\multicolumn{1}{c}{} &
  Management &
  \begin{tabular}[c]{@{}c@{}}Centralized control for unified\\ optimization\end{tabular} &
  \begin{tabular}[c]{@{}c@{}}Decentralized coordination with\\ synchronization complexity\end{tabular} &
  \begin{tabular}[c]{@{}c@{}}Hybrid management requiring\\ multimodal sensor parameter\\ synchronization\end{tabular} \\ \cline{2-5} 
\multicolumn{1}{c}{} &
  Data Consistency &
  \begin{tabular}[c]{@{}c@{}}Easily maintains global\\ consistency\end{tabular} &
  \begin{tabular}[c]{@{}c@{}}Requires feedback mechanisms to\\ calibrate local estimation deviations\end{tabular} &
  \begin{tabular}[c]{@{}c@{}}Demands strict timestamp\\ synchronization \& coordinate\\ alignment across modalities\end{tabular} \\ \hline
\end{tabular}
\begin{tablenotes}
\footnotesize
\item[*] The table compares the characteristics of centralized fusion, distributed fusion, and cross-domain multimodal fusion and gives the rank, and also lists the literature corresponding to the three fusion approaches.
\end{tablenotes}
\end{threeparttable}
}

\end{table*}

\subsubsection{\textbf{Same-domain Fusion}}
Same-domain fusion encompasses two primary approaches: (1) \textbf{Centralized fusion}, which involves transmitting measurement data from individual sensors to a central processing unit (fusion node) for unified processing and analysis; and (2) \textbf{Distributed fusion}, where multiple autonomous sensors or nodes share and process information while maintaining independent computing and decision-making capabilities.

\textit{1-1) \textbf{Centralized fusion} }

This collaborative approach maximizes the utilization of signal source information, thereby minimizing system information loss and achieving relatively high tracking performance. However, this architecture demands substantial communication bandwidth and exhibits reduced system reliability \cite{2021基于水下传感器}.

In underwater acoustic target tracking applications, the implementation of centralized fusion faces significant challenges due to the inherent limitations of underwater sensors, including constrained communication capabilities, limited operational endurance, and restricted computational resources. Consequently, research on centralized fusion-based target tracking in underwater environments remains relatively scarce.

Previous studies have employed various centralized fusion approaches for underwater target tracking. In \cite{zhang2019a}, sensor nodes directly transmit collected measurement data to a fusion center, where algorithms such as the EKF are utilized for high-precision state estimation through measurement model correction and node depth adjustment mechanisms. A similar centralized processing architecture is adopted in \cite{kumar2021b}, where all sensor measurements are aggregated at the fusion node and processed using Gaussian orthogonal basis linearization of nonlinear functions to achieve accurate state estimation.

Qiu et al. \cite{qiu2019} introduced two significant improvements to the conventional Centralized Fusion KF : (1) an adaptive forgetting factor was incorporated to enhance filter performance, and (2) an IMM approach was integrated with the adaptive KF. This combined approach leverages the advantages of IMM in maneuvering target tracking, the data utilization efficiency of centralized fusion, and the noise robustness provided by adaptive factors, thereby improving algorithm adaptability in complex underwater environments.

Addressing the challenge of asynchronous measurements, Hu et al. \cite{hu2010} developed a novel fusion algorithm suitable for general asynchronous multi-rate sensor systems. Their methodology involves: (1) deriving a centralized fusion algorithm based on the optimal batch asynchronous fusion approach \cite{zhou2019}, and (2) extending this to a distributed fusion framework. Notably, their proposed algorithm imposes no constraints on sensor quantity, sampling rates, or initial sampling times.

While centralized fusion architectures theoretically offer comprehensive data utilization by transmitting all measurements to a fusion node, practical implementation in underwater acoustic target tracking faces significant limitations. These include constrained communication bandwidth, limited sensor endurance, and restricted computational resources, which collectively result in reduced tracking accuracy and poor engineering feasibility.

\textit{1-2) \textbf{Distributed Fusion}}

To improve the applicability of data fusion in practical engineering while minimizing requirements for sensor communication capabilities, endurance, and computational resources, distributed fusion methods have been developed for underwater acoustic target tracking. In this framework, multiple independent sensors or nodes with autonomous computing and decision-making capabilities share and process information collectively. Each node performs local filtering based on obtained measurement data to generate a target state estimate, which is then transmitted to a fusion center \cite{wang2025a}. The fusion center subsequently applies an optimal fusion algorithm to produce a global state estimate. 

Notably, some distributed fusion systems incorporate a feedback mechanism that distributes the fused estimate back to individual sensors, thereby reducing the error covariance of local estimates. Due to its low channel capacity requirements, enhanced system survivability, and ease of engineering implementation, this approach has attracted considerable research attention in the field of information fusion.

Distributed fusion systems are distinguished by their feedback mechanisms and operational efficiency. As illustrated in \cite{xin2003}, researchers developed a feedback-based underwater distributed fusion algorithm that integrates radar system feedback mechanisms, allowing local sensors to receive information from the fusion center. Similarly, Yu et al. \cite{yu2014} proposed a distributed target tracking algorithm utilizing an IMM filter with a wake-up/sleep scheme, which selectively activates only those sensors within the target’s activity area. These activated sensors conduct local state estimation and subsequently generate globally optimal estimates through the IMM method.

The core of distributed fusion lies in integrating node data to form local estimates. For instance, Hare et atl.\cite{hare2017} describeed how sensor nodes utilize a collaborative decision processing algorithm to fuse target state estimates, covariance matrices, and classification information from neighboring nodes, thereby producing local fusion results. Subsequently, EKF is applied to predict target states for the next time step, while the autonomous node selection algorithm determines the optimal sensor set. This local information exchange enables distributed decision-making, effectively reducing energy consumption without compromising tracking accuracy.

Further developments include \cite{hare2014}, where sensor nodes receive broadcast signals via acoustic modems and employ track-to-track fusion to integrate state estimates and covariance matrices from multiple sensors. By combining the covariance matrices and cross-covariances of all neighboring nodes, a global covariance matrix is constructed. Subsequently, the fused covariance matrix $\hat \Sigma (k|k)$ and the state estimate $\hat x(k|k)$ are computed using the global covariance matrix and the state estimates from neighboring nodes:
\begin{equation}
    \hat \Sigma (k|k) = {\left( {\left[ {\begin{array}{*{20}{c}}
{{I_1}}& \cdots &{{I_n}}
\end{array}} \right]{\Sigma ^{ - 1}}\left[ {\begin{array}{*{20}{c}}
{{I_1}}\\
 \vdots \\
{{I_n}}
\end{array}} \right]} \right)^{ - 1}},
\end{equation}
\begin{equation}
    \hat x(k|k) = \hat \Sigma (k|k) \cdot \left[ {\begin{array}{*{20}{c}}
{{I_1}}& \cdots &{{I_n}}
\end{array}} \right]{\Sigma ^{ - 1}}\left[ {\begin{array}{*{20}{c}}
{{{\hat x}_1}(k|k)}\\
 \vdots \\
{{{\hat x}_n}(k|k)}
\end{array}} \right]. 
\end{equation}

Finally, based on the fused state and covariance, the target’s state and covariance at the next time step are predicted.  This fusion approach does not rely on a central node; instead, each sensor node independently performs the fusion using only local information, thereby embodying the principles of distributed data fusion. 

Additionally, Braca et al. \cite{braca2016} proposed two distinct distributed information fusion schemes: detection sharing and track sharing. The former updates the target state’s posterior probability using detection data, while the latter utilizes trajectory association test statistics to determine whether two trajectories originate from the same target. When exceeding a threshold, trajectories are considered distinct, enabling network-wide information sharing that enhances overall detection and tracking capabilities.

A notable advancement appears in \cite{han2019}, which introduces dynamic clustering into distributed fusion algorithms. Here, nodes dynamically select cluster heads and members based on real-time energy consumption and target states. This strategy significantly reduces energy consumption while maintaining tracking accuracy. The cluster head collects member node information and applies the Linear Minimum Variance  fusion criterion to perform weighted data fusion, yielding more accurate target state estimations.

\subsubsection{\textbf{Cross-domain Multimodal Fusion}}
Acoustic sensors demonstrate excellent long-range detection capabilities with strong penetration performance even in turbid water. Nevertheless, they suffer from relatively low resolution and are susceptible to interference from marine biological noise and multipath effects. In contrast, optical sensors can capture high-resolution images at close ranges, allowing for detailed target feature extraction \cite{jia2020}. However, their performance is heavily dependent on lighting conditions and deteriorates significantly in turbid water or low-light environments.

Consequently, integrating optical and acoustic sensors for underwater target tracking can effectively leverage the complementary advantages of both modalities. This fusion approach enables a balanced combination of long-range and close-range detection capabilities, thereby mitigating the inherent limitations of each individual sensing technology.

This section systematically examines \textbf{optical tracking methodologies for underwater targets}, followed by an analysis of \textbf{current applications involving acousto-optic sensing mechanisms in underwater target tracking}. Specifically, the discussion focuses on seawater sound velocity measurement techniques employing acousto-optic sensing principles.

In the field of close-range underwater target tracking, optical sensors have been extensively employed, predominantly utilizing visual analysis and underwater imaging techniques. As demonstrated in \cite{lee2012}, researchers have evaluated the applicability of two prevalent visual tracking algorithms—optical flow tracking and mean shift tracking—in underwater environments. While optical flow tracking exhibits satisfactory performance in scenarios with abundant feature points, its effectiveness significantly diminishes when feature points are scarce. Conversely, mean shift tracking, which relies on the color histogram of the target region, proves more efficient for non-rigid target tracking, though it may encounter difficulties during target overlap situations.

Subsequent studies have focused on addressing these limitations \cite{meng-chechuang2015,mandic2016}. Chuang et al. \cite{meng-chechuang2015} introduced a novel multi-fish tracking algorithm incorporating feature-based temporal matching and Viterbi data association, a technique originally developed for extended single-target tracking. This approach effectively resolves challenges posed by rapid fish movements and frequent occlusions in low-frame-rate conditions. Furthermore, the authors propose a computationally efficient block matching method that successfully accomplishes stereo matching. This innovation enables automatic compensation for the fish’s tail region, substantially reducing segmentation errors and facilitating precise length measurements. Additionally, the stereo matching results are utilized to compensate for the lower reflectivity of the fish tail, thereby minimizing associated measurement inaccuracies.

While in \cite{mandic2016}, a novel tracking filter algorithm is proposed, which integrates ultra-short baseline (USBL) and sonar image measurements by fusing multimodal data as input to a KF.  Moreover, the algorithm specifically adapts to the region of interest (ROI) in sonar images by dynamically adjusting the ROI based on covariance values ${D_x}$, ${D_y}$ and target dimensions ${S_x}$,${S_y}$ as shown in Equation (\ref{con:ROI}):
\begin{equation}\label{con:ROI}
    {\rm{RO}}{{\rm{I}}_i} = \left[ {{T_i} - \frac{{{S_i}}}{2} - 3\sqrt {{D_i}} ,{T_i} + \frac{{{S_i}}}{2} + 3\sqrt {{D_i}} } \right](i = x,y). 
\end{equation}

Acoustic sensors, when sonar is available, utilize high-precision measurements to reduce covariance, shrink the ROI, and improve positioning accuracy.  When sonar is unavailable, the system relies on low-frequency measurements from the USBL, where covariance increases due to process noise, causing the ROI to expand, while maintaining tracking continuity. 

To address nighttime tracking challenges, Snyder et al. \cite{snyder2012} proposed an innovative imaging technique as an alternative to acoustic cameras for underwater tracking. This method employs electrical impedance tomography (EIT) with self-generated electric fields. The study systematically compares two approaches: a computationally efficient cross-correlation method and a more accurate yet complex EIT-based technique. Both simulation and experimental results demonstrate the superior performance of the EIT method in estimating object velocity and position, especially when utilizing self-generated electric fields.

Acousto-optic sensing has become prevalent in precision marine measurements, particularly for sound velocity determination. As reported in \cite{xue2018}, researchers achieved seawater sound velocity measurement with  \textless 5 cm/s uncertainty by implementing an optical frequency comb sensor based on acousto-optic principles. Further advancing this technology, Liu et al.\cite{2021基于飞秒激光} developed a dual-Michelson interferometer system combined with optical frequency comb technology, attaining remarkable measurement precision of 0.023 m/s. This system represents a significant advancement in high-precision underwater sound velocity calibration devices.

\begin{figure}  
\centering
\subfigure[Orthogonal arrangement of two sonars]{
\begin{minipage}[b]{0.25\textwidth}
\includegraphics[width=1\textwidth]{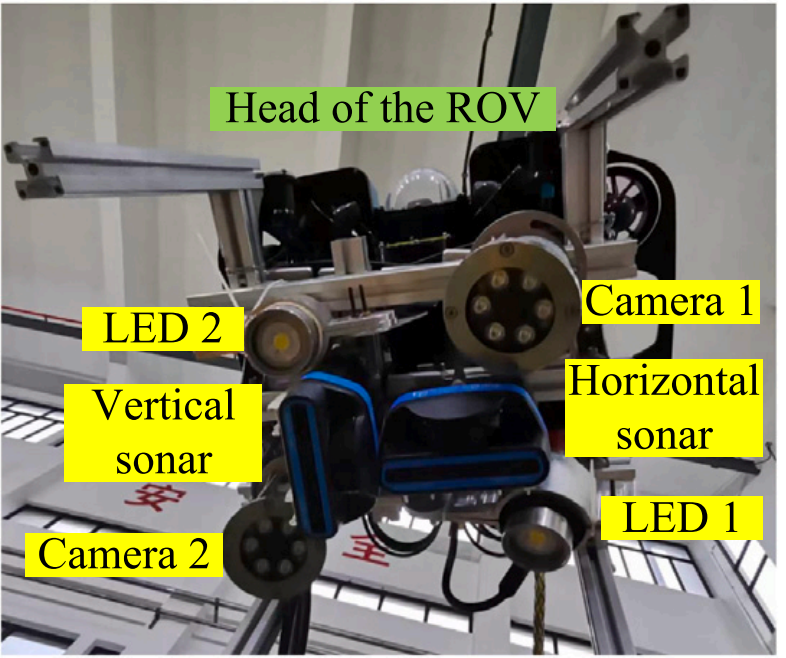} 
\end{minipage}
}
\subfigure[The installation of the sonars on the ROV]{
\begin{minipage}[b]{0.17\textwidth}
\includegraphics[width=1\textwidth]{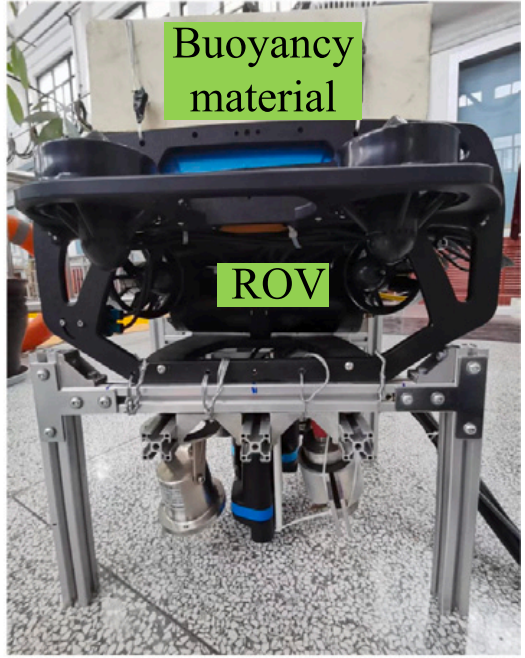} 
\end{minipage}
}
\caption{Installation of the pair of orthogonal arranged sonars on the ROV \cite{liu2024}. In (a), except for the sonars, two cameras and two LEDs with adjustable brightness are also mounted to sense the environment and calculate the position of the target based on the visual information in the short range; in (b), the sensors are installed below the ROV and a piece of buoyancy material is fixed to the top of the ROV to adjust the uneven buoyancy caused by the sensors.}
\label{fig:ronghe}
\end{figure}

The exceptional measurement accuracy and non-contact nature of acousto-optic sensing offer substantial potential for oceanographic applications. Currently, there have been audio-optical fusion remotely operated vehicles (ROV) fabricated as shown in Figure \ref{fig:ronghe}. Integrating acousto-optic methods into underwater target tracking presents not only technical challenges but also interdisciplinary system engineering opportunities, opening new possibilities for deep-sea exploration. Continued progress in data fusion technologies, sensor development, and computational power will undoubtedly enhance and promote wider adoption of acousto-optic hybrid underwater tracking systems.

\subsection{Summary}

This chapter provides a systematic classification and summary of underwater acoustic target tracking methods from three different perspectives: target scale, perception approach, and collaboration mode. 

From the perspective of target scale, tracking methods can be divided into single-target tracking and multi-target tracking.  Single-target tracking performs excellently in stable scenarios, but when dealing with maneuvering targets, its adaptability must be improved through multi-model cooperation and dynamic switching.  Multi-target tracking, on the other hand, faces challenges such as dynamic changes in the number of targets and environmental clutter interference.  It can be further categorized into deterministic methods based on association matching and stochastic set methods based on probabilistic models.  The former relies on efficient association strategies, while the latter reduces computational complexity through probability modeling.  Both methods focus on different aspects such as target density, computational efficiency, and environmental adaptability. 

From the perspective of perception approach, tracking methods can be divided into active tracking and passive tracking.  Active tracking depends on actively transmitted detection signals, offering advantages of high precision and long-range detection, but it faces risks of exposure and energy consumption.  Passive tracking, on the other hand, achieves covert detection by receiving target radiation signals, though it is susceptible to environmental noise and nonlinear interference.  In dynamic and complex environments, both methods have their unique characteristics and require the integration of adaptive filtering techniques to address the time-varying and uncertain nature of marine environmental noise.  By dynamically adjusting model parameters, the robustness of the system can be enhanced. 

From the perspective of collaboration mode, tracking methods encompass centralized, distributed, and cross-domain multimodal fusion approaches.  Centralized fusion benefits from high information utilization but faces heavy communication burdens.  Distributed fusion improves system fault tolerance and scalability through local processing and information sharing.  Cross-domain multimodal fusion combines the advantages of heterogeneous sensors such as acoustic and optical sensors, compensating for the limitations of single modalities and balancing long-range detection with close-range accuracy.  This approach represents an important direction for addressing complex underwater tracking scenarios in the future. 

In conclusion, this chapter reveals the core characteristics and applicable boundaries of different methods through multi-dimensional classification, forming a multi-layered and multi-dimensional methodological system.  This helps readers to better understand the current state of underwater acoustic target tracking technology and its future trends, providing a theoretical foundation and directional guidance for practical applications and technological innovation. 

\section{Emerging Trends in Tracking Methods Driven by Machine Learning} \label{4}
In recent years, machine learning techniques have introduced a paradigm shift in the field of underwater acoustic target tracking due to their powerful nonlinear modeling and adaptive learning capabilities. Traditional tracking methods, which rely on precise physical models and manually designed features, face challenges such as model mismatches and noise sensitivity in complex, dynamic ocean environments. 

Machine learning techniques enhance the robustness of algorithms in scenarios involving multi-target intersections and environmental disturbances by leveraging data-driven feature extraction and strategy optimization to effectively uncover hidden patterns in low signal-to-noise ratio signals. This chapter focuses on two major branches: deep neural networks (DNNs) and deep reinforcement learning (DRL)-analyzing their theoretical advancements and practical implementations in underwater acoustic target tracking. We further explore the potential for synergistic integration between machine learning approaches and conventional signal processing techniques. 

\subsection{Deep Neural Network Optimized Tracking Method}

DNNs have demonstrated remarkable effectiveness in \textbf{underwater acoustic target recognition and localization}\cite{huang2019}, indicating their substantial \textbf{potential for underwater acoustic target tracking applications}. This potential stems from DNNs' ability to process complex acoustic signals and extract discriminative features from underwater environments.

\subsubsection{\textbf{Application of DNN in Target Recognition and Localization}}
The rapid evolution of sensor technologies and intelligent information systems has revealed significant limitations in conventional approaches for intelligent underwater detection and information processing. To address these challenges, deep learning has emerged as a transformative theoretical framework that significantly enhances underwater \textbf{target recognition and localization} capabilities.

In the area of underwater target recognition, as demonstrated in \cite{yang2018}, convolutional neural networks (CNNs) were integrated with auditory perception models, where Mel-frequency cepstral coefficients \cite{abdul2022} were employed to extract features from target radiation noise. This approach simulated the full functionality of the auditory system, achieving effective recognition of ship-radiated noise. The results substantiate the applicability of deep learning in underwater acoustic research.

Further innovations are evident in \cite{yao2023}, where a data augmentation technique combined with a residual CNN architecture was proposed. The authors leveraged a Deep Convolutional Generative Adversarial Network to augment training datasets, significantly improving recognition accuracy. Complementing this, \cite{zhu2019} addressed shipwreck identification in side-scan sonar (SSS) imagery through a Gentle AdaBoost model. The study introduced a systematic workflow for the automated and precise detection of shipwrecks in SSS waterfall images.

Deep learning has demonstrated significant potential in underwater acoustic target localization, owing to its robust modeling capabilities and minimal reliance on prior environmental knowledge. For instance, Niu et al.\cite{niu2019} proposed a two-stage deep residual neural network approach: first identifying range intervals, then resolving source range and depth through interval-specific models trained on large-scale datasets, effectively mitigating uncertainties in environmental parameter acquisition.

Similarly, Huang et al.\cite{huang2018} developed a two-stage model utilizing simulated data, where feature vectors derived from covariance matrix-based modal signal space decomposition served as DNN inputs for range/depth regression. Further advancing this field, Liu et al.\cite{liu2020} implemented a Multi-Task Learning framework with adaptive weighted loss in a CNN architecture to enhance deep-sea sound source localization.

These advancements underscore deep learning’s growing adoption in underwater acoustic tasks. Notably, traditional tracking models’ heavy mathematical dependencies often introduce irreducible errors, whereas deep learning’s superior feature extraction capabilities have spurred its recent integration into underwater target tracking methodologies \cite{wu2021}.

\subsubsection{\textbf{The Promising Future of DNN-based Target Tracking}}
\begin{figure*}
    \centering
    \includegraphics[width=0.7\linewidth]{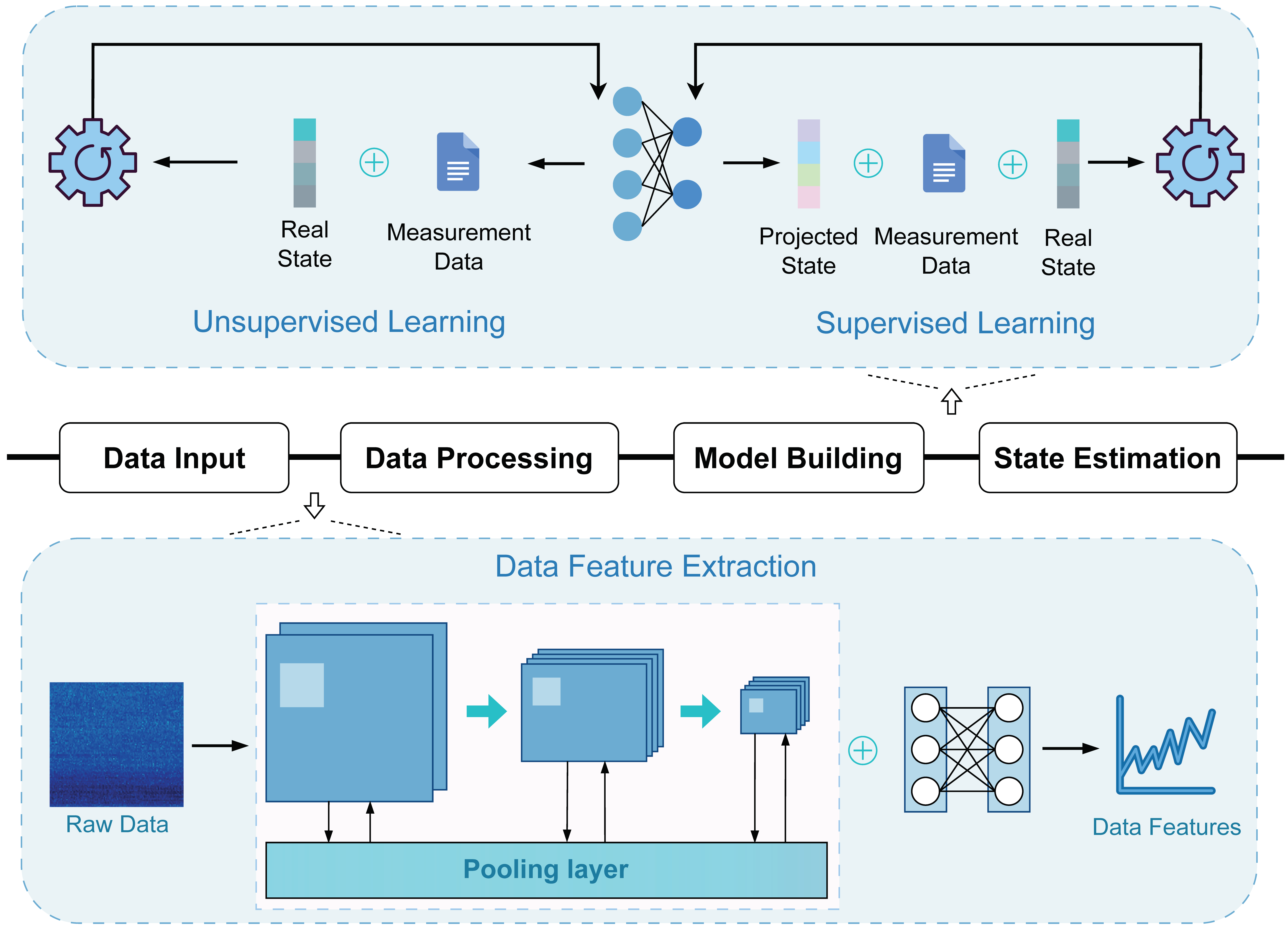}
    \caption{Schematic diagram of the application of deep learning to underwater acoustic target tracking methods.  The upper part of the figure represents deep learning applied to improve the state estimation model for target tracking, which can adaptively learn the parameters of the model using unsupervised or supervised learning, making the state estimation model more adaptable to the complex and dynamic underwater acoustic environment; the lower part of the figure represents deep learning applied to the feature extraction of measurement data for target tracking. }
    \label{fig:DL}
\end{figure*}

The underwater acoustic target tracking process comprises two primary phases: (1) \textbf{data preprocessing}, which includes data input and processing; and (2) \textbf{target state estimation}, involving model construction and state estimation (as illustrated in Figure \ref{fig:DL}). Notably, deep learning techniques demonstrate applicability in both phases.

\textit{2-1) \textbf{Application of DNNs in the data processing stage}}

During the data processing phase, the recognition and analysis of acoustic signals constitute critical factors influencing tracking performance. Deep learning demonstrates remarkable feature extraction capabilities, which can substantially improve the processing efficiency of tracking systems \cite{gao2024}. To mitigate target loss in conventional tracking systems, Wang et al.\cite{wang2021} developed a hybrid approach combining deep convolutional neural networks with Kalman filtering. This method enhances tracking robustness in challenging scenarios (e.g., weak signals or azimuthal target overlap) by integrating target azimuth information into time-frequency maps and employing azimuth-weighted feature learning alongside Kalman-based prediction-correction.

Wang et al. \cite{wang2022} proposed a passive underwater acoustic target tracking method, which integrates multi-beam Low-Frequency Analysis and Recording (LOFAR) data \cite{pala2025} with an improved LeNet-5 CNN, and further enhances tracking accuracy through an EKF. The multi-beam LOFAR spectrum is used as the CNN input, and a dropout mechanism is employed to prevent overfitting.  The model performs target recognition and azimuth estimation, followed by EKF-based correction of the prediction results.  The entire process can be represented by the Equation (\ref{con:LOFAR}):
\begin{equation}\label{con:LOFAR}
    \tilde y = {\rm{Pre}}(\omega  \otimes S),{\rm{then}}F(L,\theta ) = P(cnn(\tilde y)),
\end{equation}
where $S$ represents the raw sonar array signal, and $\tilde y$ is the resulting image data used for training the CNN model.  $Pre(\cdot)$ refers to azimuthal weighting applied to the sonar signal.  The weighted data is transformed from the frequency domain to the time domain to generate a multi-beam LOFAR spectrum.  During this process, a colormap mapping converts spectral energy values into grayscale or RGB values, which serve as inputs for CNN training and recognition.  The symbol $cnn(\cdot)$ denotes the trained CNN model, while $P(\cdot)$ indicates the model trained on azimuth-weighted data.  This model identifies target data to obtain the target state vector $F(L,\theta )$, which includes the classification label $L$ and azimuth $\theta$.

\textit{2-2) \textbf{Application of DNNs in the target state estimation stage}}

During the target state estimation phase, filtering models utilize recursive algorithms to predict the target’s state at subsequent time steps, enabling continuous tracking. Key parameters within these models—including noise characteristics and motion model parameters—significantly determine tracking accuracy. The adaptive learning capacity of deep learning can optimize these parameters effectively, consequently improving overall tracking performance \cite{dawkins2024}. Xu et al. \cite{xu2024} proposed EKFNet, an innovative learning framework for automated estimation of process and measurement noise covariance parameters in EKF. While retaining the fundamental EKF architecture (state prediction, measurement prediction, and state update), EKFNet integrates a recurrent neural network (RNN) trained with backpropagation through time to optimize noise covariance matrices.
  
EKFNet supports both supervised and unsupervised learning modes of backpropagation, each with distinct loss functions.  The state estimation error, ${\bar x_k} = x_k^G - {g_k}({\hat x_{k|k}})$ represents the difference between the true state and the estimated state from filtering.  The measurement residual ${\tilde y_k} = {z_k} - h({\hat x_{k|k - 1}})$ denotes the difference between the measured value and the result of the measurement model.  The loss function for supervised learning consists of the state estimation mean square error and the posterior log-likelihood. The loss function for unsupervised learning comprises the measurement residual error and the measurement log-likelihood.  The loss function for unsupervised learning consists of the measurement residual error and the measurement log-likelihood. 

Traditional target tracking methods exhibit high sensitivity to initial mean square error, where improper initialization may result in tracking divergence. To address this limitation, Hou et al. \cite{hou2023} proposed an attention-based deep convolutional neural network (DCNN) for robust estimation of the initial noise covariance matrix, which extends the variational Bayesian EKF framework. This approach significantly enhances the stability and precision of subsequent DOA tracking performance.

The attention-based DCNN consists of ResNet and Squeeze-and-Excitation modules.  The squeezing operation can be expressed as Equation (\ref{con:jiya}):
\begin{equation}\label{con:jiya}
    {{\bf{z}}_c} = {{\bf{F}}_{sq}}({{\bf{u}}_c}) = \frac{1}{{H \times W}}\sum\limits_H^{i = 1} {\sum\limits_W^{j = 1} {{u_c}} } (i,j),
\end{equation}
where ${u_c}$ represents the $c$-th channel of the feature map $U$, and ${F_{sq}}$ denotes the compression.  From the above equation, it can be seen that the squeezing is applied to each feature map of size $H \times W$.  After compression, each feature map $U$ is reduced to a real number, thus maximizing the compression of the information.  Subsequently, the features are excited through a fully connected layer and activation function to calculate the weight vector $s$ for each feature map.  The final output of the model is the weighted multiplication of the channel weights $s_c$ and the channels $u_c $ of the feature map. 

In conclusion, by autonomously extracting hierarchical features from the data, deep learning's powerful nonlinear modeling capability effectively distinguishes targets from background noise, enhancing adaptability to dynamic environments.  This overcomes the limitations of traditional methods that rely on handcrafted features and exhibits superior robustness in challenging conditions such as noise interference, weak signals, and overlapping targets. 

When combined with conventional filtering techniques, deep learning further improves the stability of state estimation, reduces sensitivity to initial parameters, and mitigates the risk of tracking divergence.  This presents a promising solution for underwater acoustic target tracking.  Future research should further explore the application of deep learning in more complex scenarios characterized by low signal-to-noise ratios and missing signal data. 

\subsection{Deep Reinforcement Learning Optimized Tracking Strategy}

The control strategies for underwater acoustic sensor networks are one of the critical factors determining the effectiveness of tracking strategies.  The collaborative relationships within a sensor network encompass \textbf{scheduling strategies} and \textbf{cooperative control strategies} among sensors. 

The necessity for scheduling within a sensor network arises from the challenge of recharging battery-powered sensors in the ocean's depths, which severely limits the operational duration of the sensor network.  Moreover, an excessive number of sensors or the frequent transmission of measurement data can significantly deplete energy resources \cite{islam2022}. 

Conversely, to free sensors from the constraints of fixed positions, they can be deployed on AUVs.  This approach also reduces the overall deployment and maintenance costs.  However, the substantial burden of controlling AUV swarms presents significant challenges for their coordinated control \cite{yang2021}. 

Consequently, developing efficient and rational scheduling strategies along with cooperative control mechanisms for sensor networks and AUV swarms becomes imperative to strike an optimal balance between tracking accuracy and energy efficiency \cite{zhao2021}.

As a paradigm of machine learning, reinforcement learning optimizes an agent's strategy by maximizing long-term cumulative rewards through interactions between the agent and the environment. Compared to traditional methods,DRL does not require precise prior knowledge of the environment and possesses a strong capability to adapt to dynamic environmental changes, making it particularly suitable for the complex and variable underwater environment \cite{su2021}.  Consequently, the application of reinforcement learning is gradually becoming a new trend in researching control strategies for underwater acoustic sensors.

\subsubsection{\textbf{Sensor Scheduling Optimization}}
\begin{figure*}
    \centering
    \includegraphics[width=0.85\linewidth]{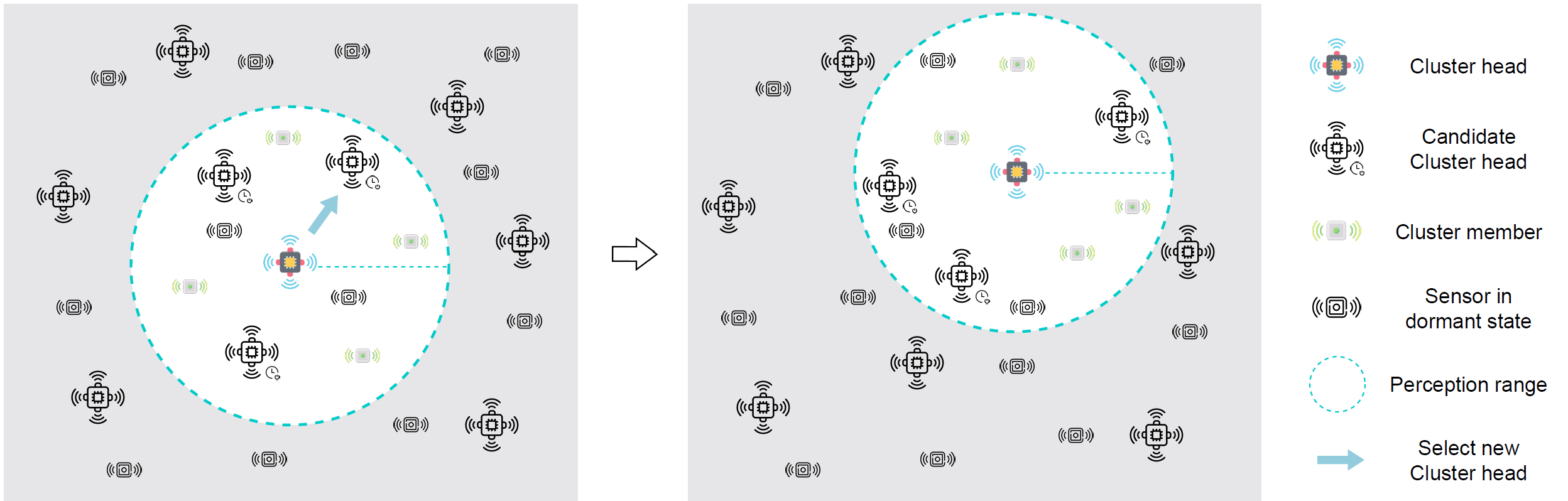}
    \caption{ Schematic diagram of the underwater acoustic sensor scheduling process.  On the left is the sensor cluster head at moment $k$, its selected cluster members and the cluster head at the next moment, and on the right is the selection and tracking process of the cluster head at moment $k+1$.  }
    \label{fig:select}
\end{figure*}

Throughout the process of underwater acoustic target tracking, the scheduling of acoustic sensors is integral and subject to continuous dynamic adjustment, as depicted in Figure \ref{fig:select}.  At time step $k$, the cluster head (CH) sensor selects appropriate sensors within its communication range as cluster members (CM) based on a specific strategy to perform the tracking task at that time.  Concurrently, the CH sensor, using the tracking results from time step $k$, selects the next CH sensor, thereby repeating the process of tracking and selection. 

\textit{1-1) \textbf{Traditional scheduling strategy}}

Recent years have witnessed significant advancements in underwater target tracking sensor scheduling methods within UWSNs. As demonstrated in \cite{zhang2014}, an adaptive sensor scheduling scheme dynamically optimizes sensor selection and sampling intervals to achieve an optimal trade-off between tracking accuracy and energy efficiency. Building upon this, Liu et al.\cite{Liu2019} proposed a distributed intelligent node scheduling approach that quantifies the influence of node deployment on tracking performance through the Posterior Cramér-Rao Lower Bound. This method employs the Group-Based Forward Orthogonal Search algorithm combined with greedy search, effectively reducing computational complexity without compromising tracking precision. 

To address the challenge of positional drift, Tian \cite{tian2022} developed a node selection algorithm that models location fluctuations as drift noise. By integrating particle filtering, the study derives the fisher information matrix (FIM) and mutual information (MI) under drift conditions as node selection criteria, with multi-objective optimization achieved through NSGA-II and TOPSIS algorithms. Further innovations include non-cooperative target tracking method in \cite{qin2023}, which combines MI with the IMM-UKF. This approach not only dynamically activates tracking nodes using MI to conserve energy but also incorporates ray tracing to mitigate ranging errors induced by ocean stratification.

\textit{1-2) \textbf{DRL-driven scheduling strategies}}

Traditional scheduling methods generally exhibit disadvantages such as limited flexibility and poor adaptability to complex environments.  In recent years, reinforcement learning techniques have gradually been applied to the research of underwater sensor scheduling strategies.  In \cite{zheng2023},  an end-to-end sensor scheduling algorithm based on dueling double deep Q network (D3QN) is proposed.  This algorithm models the scheduling strategy as a markov decision process (MDP), where the underwater passive tracking-based energy-efficient sensor scheduling process is treated as a sequential decision problem.  Since it is unrealistic to obtain ground truth for performance evaluation in non-cooperative target tracking scenarios, the authors introduce a simulation-based training approach, which does not require the true trajectory information of the target, thus addressing the issue of missing information for non-cooperative targets.

Considering the energy consumption due to the continuous operation of the sensor, Su et al. \cite{su2021} investigated an asynchronous wake-up scheme for underwater sensor networks and propose an adaptive asynchronous wake-up scheme based on DRL.  The idle listening strategy selection is modeled as an MDP, where the state space consists of four components: remaining energy, queue length, idle time, and traffic interval.  The reward function balances energy consumption and delay costs, selecting long periods to save energy under low traffic and short periods to reduce delay under high traffic.  Additionally, long short-term memory (LSTM) networks are used to predict network traffic characteristics, assisting nodes in dynamically adjusting their wake-up strategies.

The issue of high transmission latency in UWSNs poses significant challenges for sensor scheduling \cite{luo2021}. Jin et al. \cite{jin2019} introduced a reinforcement learning-based underwater acoustic sensor network routing protocol and introduced the additional cost terms: congestion-related cost $co(c)$, delay-related cost $co(t)$, and energy-related cost $co(en)$ as the reward function, which is defined in Equation (\ref{con:jiangli}):
\begin{equation}\label{con:jiangli}
    R_{{s_{i{n_j}}}}^{{a_{i \to {n_j}}}} =  - {R_0} - \varphi  \times [{\varphi _{\rm{c}}} \times co(c) + {\varphi _t} \times co(t) + co(en)]. 
\end{equation}
Furthermore, to strike a balance between convergence speed, energy balancing, and congestion avoidance, the study incorporates a dynamic virtual routing pipe with a variable radius.  The radius of the virtual routing pipe is directly proportional to the average residual energy of the sending node’s neighbors.  This design accelerates the convergence of the algorithm, mitigates congestion, and facilitates a more uniform distribution of energy across the network. 

To address the problem of void regions—areas lacking effective relay nodes due to the sparse deployment of UWSNs, Wang et al. \cite{wang2023} proposed an opportunistic routing protocol named DROR. This protocol integrates receiver-based routing decisions with Q-learning and incorporates a void recovery mechanism to bypass void nodes.  The implementation of void recovery mechanism is as follows:
\begin{itemize}[leftmargin=*]
    \item When a node detects that no candidate forwarding set exists (${\rm{CS}}(i) = \varnothing $), it sets $void_{flag} = 1$ and $recov_{flag} = 1$. 

    \item The receiving node allows downward forwarding (i.e., depth difference $\ge$ 0) and uses Q-learning to select non-void nodes as relays. 

    \item A penalty term associated with the $void_{flag} (C=2)$ is incorporated into the reward function to prevent repeated selection of void nodes. 
    
\end{itemize}

In addition, a dynamic scheduling strategy is devised based on the holding time regulated by the hyperbolic tangent (Tanh) function.  By associating the holding time with the Q-value, nodes with relatively higher Q-values are enabled to forward packets earlier.  Through dynamic adjustment of relative Q-values, DROR effectively reduces packet collisions and avoids unnecessary increases in holding time caused by decreasing Q-values.  This strategy significantly minimizes both end-to-end delay and energy consumption during data transmission. 

\subsubsection{\textbf{AUV Cooperative Control Optimization}}
Allowing AUVs to carry sensors makes the sensors independent of fixed locations, while reducing the overall deployment and maintenance costs.  In recent years, the use of reinforcement learning techniques to reduce the burden of cooperative control of AUV clusters has gradually become a research priority. \textbf{A major application of deep reinforcement learning to optimize the cooperative control of AUVs is the use of the centralized training and distributed execution (CTDE) framework for training and control of AUVs. In addition, the emerging software-defined networking (SDN) architecture provides new ideas for the application of deep reinforcement learning.}

\textit{2-1) \textbf{The CTDE framework}}

CTDE is the classical training and implementation framework for DRL, which guarantees that the training phase takes place in a secure private network and the execution phase does not require communication between agents.

In \cite{yang2021a}, a secure collaborative underwater target tracking scheme based on Multi-Agent Reinforcement Learning was proposed.  It employs the CTDE architecture. Each AUV generates control signals solely based on its own sensor data, thereby avoiding the risks of communication link attacks and ensuring the security of task execution.  

Moreover, the method also designs a reward function that includes swarm consistency, target tracking, and collision avoidance.  Group consistency treats AUV groups as graphs with algebraic connectivity, i.e., the second smallest eigenvalue of the Laplace matrix $\lambda$. When $\lambda$ is too large, the reward function is truncated to avoid collisions due to over-aggregation. 

To address the challenges of reward function design and high environmental interaction costs in DRL methods, Xu et al. \cite{xu2024a} proposed the FISHER framework, a two-stage learning paradigm that combines imitation learning (IL) and offline reinforcement learning (ORL).  This framework avoids the complexities of reward function design in traditional RL methods while enhancing sample efficiency and the generalization capability of the policy.  

In the IL stage, generative adversarial imitation learning is improved by introducing the multi-intelligent body condition and the Nash equilibrium condition to generate offline datasets using generative adversarial networks (GANs) while the Nash equilibrium constraint ensures that the strategies are locally optimal in concert by optimizing the pairwise form of the problem and the explicit condition. In the ORL stage, the multi-agent independent generalized decision transformer is proposed to extract future state features through an anti-causal attention mechanism to match potential representations of expert trajectories without relying on reward functions. 

\textit{2-2) \textbf{SDN architecture}}
\begin{figure*}[ht]
    \centering
    \includegraphics[width=0.7\linewidth]{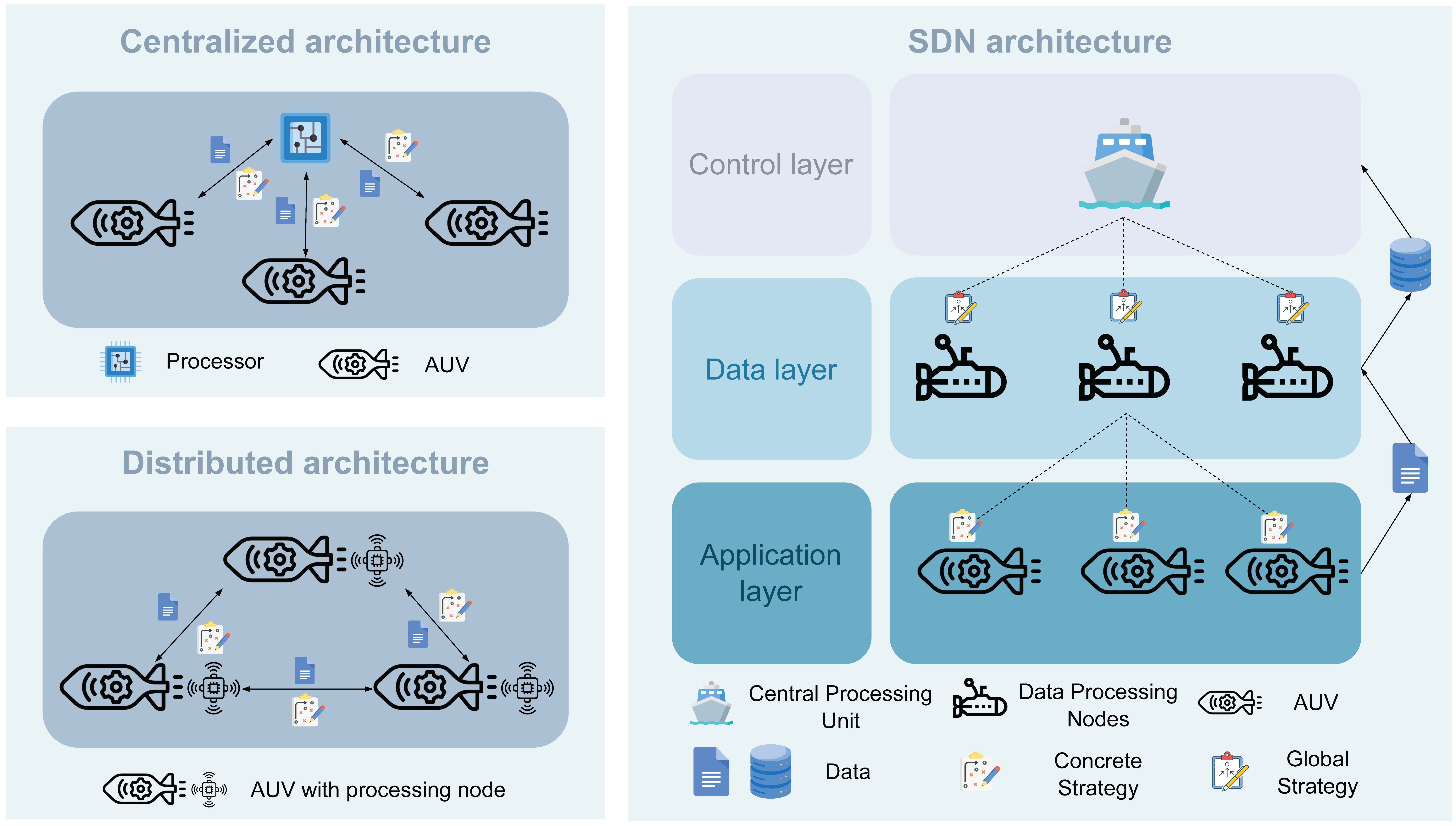}
    \caption{Schematic diagram of SDN architecture, centralized architecture, and distributed architecture.  In centralized and distributed architectures, sensors directly interact with the processor or other sensors for data and concrete Strategies; in SDN architecture, strategy generation and data processing are decoupled through the control layer and data layer, which makes the network computation burden shared and network configuration and management easier and more convenient. }
    \label{fig:SDN}
\end{figure*}

SDN represents an innovative network architecture that decouples control plane functions from data plane operations, thereby enabling programmable network management \cite{anerousis2021,espinelsarmiento2021}. This architecture has gained significant traction in underwater multi-agent network applications in recent years, offering novel solutions for underwater exploration and research operations. Within underwater acoustic sensor networks, the SDN framework distinctly segregates sensor nodes into two functional modules: a data processing unit and a decision-making unit. This architectural separation not only improves network flexibility but also simplifies configuration procedures while reducing dependence on costly specialized hardware \cite{das2020}. As demonstrated in Figure \ref{fig:SDN}, these characteristics establish clear differentiators between SDN and conventional distributed/centralized network architectures.

Lin et al. \cite{lin2020} proposed an SDN-based underwater wireless network architecture for AUVs, designed to support cooperative multi-AUV search missions.  This architecture integrates software-defined beacons, hierarchical localization, cooperative control, and a software-defined hybrid data transmission framework to enable network information synchronization, node localization, multi-AUV cooperative control, and intelligent data transmission scheduling. 

DRL-based routing method for SDN is proposed in \cite{casas-velasco2022}.  This method utilizes path state indicators and demonstrates superior efficiency and intelligence compared to traditional algorithms.  It dynamically adjusts routing strategies based on network traffic changes, showcasing both practical feasibility and outstanding performance in SDN routing.

In \cite{zhu2024} and \cite{wang2025}, the authors developed a tracking strategy framework for AUV clusters based on SDN.  In this framework, the central controller is a surface-deployed unmanned surface vehicle, which is responsible for generating globally optimal strategies and decomposing complex tracking tasks into sub-regional assignments. 

The local-training control Layer, functioning as the data layer, has as its core component the local training controller for AUVs.  This controller allocates specific actions to sub-cluster AUVs based on the global strategy and uploads training samples (such as state, action, and reward) to the Global Training Control Layer. The application execution layer is composed primarily of ordinary AUVs, which autonomously adjust their motion parameters according to the strategies assigned by the Local-Training Control Layer. 

Under this hierarchical architecture, Zhu et al.\cite{zhu2024} introduced an advantage attention mechanism and an advantage resampling method.  The advantage attention mechanism dynamically selects the most effective AUVs and expands their experiences, thereby compressing the input dimensionality in large-scale AUV clusters and enhancing the system’s scalability.  The advantage resampling method prioritizes high-reward samples to address the inefficiency of conventional experience replay buffers in utilizing advantageous samples, thus improving training efficiency. 

Similarly, Wang et al.\cite{wang2025} propose mechanisms called dynamic switching attention and dynamic switching resampling.  These methods focus on adaptively adjusting the sample selection strategy to increase the proportion of high-reward samples, thereby accelerating convergence while avoiding local optima.  This approach is particularly suited for complex and interference-prone environments.  In contrast, \cite{zhu2024} emphasized architectural scalability, making it more applicable to large-scale AUV clusters. 

In summary, the introduction of reinforcement learning technology provides an intelligent solution for the efficient control of underwater sensor networks. By dynamically optimizing sensor scheduling strategies and AUV collaborative control, it significantly enhances the energy efficiency and robustness of target tracking.  These collaborative strategies complement the sensor cooperation models presented in Chapter \ref{3}, together forming a multi-layered sensor cooperation tracking system. 

In the future, with the advancement of multi-modal sensor technologies and the improvement of reinforcement learning algorithms, underwater acoustic target tracking systems will exhibit greater adaptability in communication-limited and environmentally dynamic underwater scenarios, offering more reliable technological support for ocean monitoring and resource development. 

\subsection{Summary}
This chapter systematically explores the cutting-edge applications of machine learning technologies in underwater acoustic target tracking.  Deep learning leverages the powerful feature extraction capabilities of neural networks to achieve adaptive extraction of complex signal features and dynamic calibration of state estimation parameters, significantly enhancing tracking accuracy in low SNR environments.  Reinforcement learning, with intelligent decision-making at its core, optimizes sensor scheduling strategies and multi-agent cooperative control to achieve globally optimal tracking performance under energy constraints and dynamic environments. 

The introduction of these two technologies not only breaks through the theoretical limitations of traditional methods but also drives the transition of tracking systems from passive adaptation to active optimization.  In the future, with the advancement of multimodal data fusion and lightweight model design, machine learning technologies will further empower underwater acoustic target tracking systems, providing more efficient solutions for their practical deployment in complex marine environments. 
\section{Challenges and future Avenues} \label{5}
\begin{figure*}
    \centering
    \includegraphics[width=0.9\linewidth]{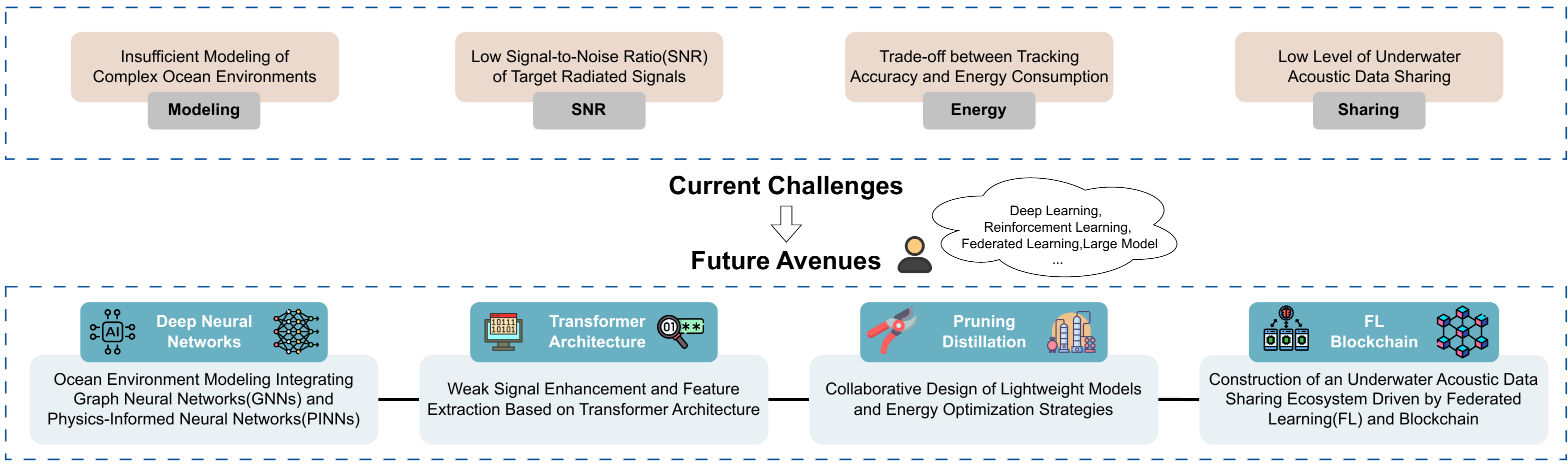}
    \caption{Current challenges and future avenues. The first half of the picture describes the current challenges, including four aspects: modeling, SNR, energy, and sharing, and the second half provides corresponding future avenue in conjunction with deep learning, reinforcement learning, federated learning, and large model techniques.}
    \label{fig:challenge and direction}
\end{figure*}

This chapter systematically examines the key challenges in underwater acoustic target tracking research, focusing on four critical aspects: environmental modeling, SNR, energy efficiency, and data sharing. Furthermore, potential solutions leveraging emerging technologies (e.g., physics-informed machine learning and federated learning) are proposed. The interrelationship between current challenges and future research directions is illustrated in Figure \ref{fig:challenge and direction}.
\subsection{Current Challenges}
\subsubsection{\textbf{Insufficient Modeling of Complex Ocean Environments}}
Contemporary underwater acoustic target tracking algorithms predominantly rely on oversimplified environmental assumptions, significantly limiting their effectiveness in dynamic marine environments. The underwater acoustic channel exhibits distinct time-varying characteristics influenced by multiple factors, including temperature variations, salinity gradients, depth changes, and pressure fluctuations. These environmental variables collectively affect sound propagation through four primary mechanisms: (1) restricted communication bandwidth, (2) prolonged propagation delays, (3) increased interference and clutter levels, and (4) elevated bit error rates. However, conventional modeling approaches inadequately address the complex interactions among these variables, resulting in suboptimal performance of tracking algorithms in practical scenarios.

\subsubsection{\textbf{Low Signal-to-Noise Ratio of Target Radiated Signals}}

The continuous advancement of noise reduction technologies has paradoxically exacerbated detection challenges. While vessel radiated noise and echo intensity have progressively decreased, anthropogenic activities have concurrently amplified ambient ocean noise. This inverse relationship has precipitated a substantial decline in target SNR. Under such conditions, conventional signal processing methods prove inadequate for reliable feature extraction, as target signals become indistinguishable from background noise, thereby compromising detection and tracking efficacy.

\subsubsection{\textbf{Trade-off between Tracking Accuracy and Energy Consumption}}

Underwater target tracking inherently involves a trade-off between tracking accuracy and energy efficiency. Achieving higher tracking accuracy generally necessitates data fusion from multiple sensors, which substantially elevates energy consumption. However, both stationary sensors and mobile AUVs are subject to energy constraints, making prolonged high-energy tracking operations unsustainable.

\subsubsection{\textbf{Limited Underwater Acoustic Data Sharing}}

The field suffers from severe data fragmentation due to three primary barriers: (1) exorbitant data acquisition costs, (2) domain sensitivity, and (3) lack of standardized protocols. These constraints hinder cross-institutional and international data integration, resulting in duplicated research efforts and inadequate algorithm training datasets. Consequently, the generalization capacity of tracking algorithms remains severely limited, stifling technological progress in the field.

\subsection{Future Avenues}
\subsubsection{\textbf{Ocean Environment Modeling Integrating Graph Neural Networks (GNNs) and Physics-Informed Neural Networks (PINNs)}}

To address the challenges of modeling complex ocean environments, the integrated approach combining GNNs and PINNs is feasible. GNNs, with their powerful capability to process graph-structured data, effectively characterize intricate spatiotemporal relationships among nodes (e.g., sensors measuring sound speed and temperature at different locations). By constructing a graph representation of the ocean environment, GNNs can identify latent associations between environmental variables and capture dynamic patterns of environmental changes.   

Meanwhile, PINNs incorporate the physical laws governing acoustic propagation as regularization terms within the neural network architecture. This integration ensures that the model simultaneously learns data features while adhering to fundamental physical constraints.  

The synergy of these two methodologies establishes a physics-informed, data-driven framework that accurately simulates acoustic propagation characteristics in complex ocean environments. Consequently, it provides reliable environmental priors for underwater target tracking algorithms.

\subsubsection{\textbf{Weak Signal Enhancement and Feature Extraction Based on Transformer Architecture}}

To address the limitations of target detection under low SNR conditions, the Transformer architecture demonstrates unique advantages in underwater acoustic signal processing. Its self-attention mechanism and capability to model long-range dependencies enable effective feature extraction from noisy signals.  

Specifically, variants like the Swin Transformer utilize a hierarchical sliding window mechanism, which facilitates efficient extraction of subtle features from underwater acoustic signals while capturing long-range dependencies across both temporal and frequency domains. Furthermore, by integrating Conditional Generative Adversarial Networks (CGANs), this framework can augment low-SNR signals through the generation of virtual samples containing target features, thereby expanding the training dataset.  

The proposed pipeline consists of two key stages: (1) CGAN-based generation of diverse enhanced signals, followed by (2) Transformer-based feature extraction. This combined approach significantly improves weak target signal enhancement and boosts the accuracy of target detection and tracking systems.

\subsubsection{\textbf{Collaborative Design of Lightweight Models and Energy Optimization Strategies}}

In the context of the prevailing trend toward large-scale models, lightweight techniques—including model pruning and knowledge distillation—have emerged as pivotal solutions to address the energy constraints of underwater nodes. Model pruning eliminates redundant connections and parameters within tracking algorithm models, whereas knowledge distillation facilitates the transfer of knowledge from complex large-scale models to compact lightweight models, thereby preserving performance while significantly reducing model size.

Moreover, reinforcement learning is employed to formulate dynamic sensor scheduling strategies, empowering underwater nodes to adaptively adjust sensor operating modes and data transmission frequencies in response to target movement and environmental variations. Additionally, the sparse attention mechanism inherent in Transformer architectures can substantially mitigate computational demands, effectively minimizing energy consumption without compromising tracking performance. This integrated approach ultimately achieves an optimal balance between tracking accuracy and energy efficiency.

\subsubsection{\textbf{Construction of an Underwater Acoustic Data Sharing Ecosystem Driven by Federated Learning and Blockchain}}

To address the challenges in underwater acoustic data sharing, federated learning presents a novel approach. Participating institutions can locally train target tracking models using their proprietary datasets while uploading only encrypted model parameters or data summaries to the platform. This methodology effectively safeguards the confidentiality of sensitive raw data. By employing aggregation strategies such as federated averaging, these distributed models can be collaboratively optimized across multiple institutions, ultimately yielding a high-performance global tracking model.

The integration of blockchain technology with federated learning further enhances this ecosystem. All data usage records and model training processes are immutably recorded on the blockchain. Smart contracts autonomously execute data-sharing protocols and benefit-distribution mechanisms, thereby ensuring data credibility, traceability, and regulatory compliance. This synergistic technological framework not only facilitates international collaboration in underwater acoustic data utilization but also provides extensive datasets for algorithm training. Consequently, it improves model generalizability and robustness, significantly advancing the field of underwater acoustic target tracking.

\section{Conclusion} \label{6}
This comprehensive survey proposed a novel multi-dimensional taxonomy for underwater acoustic target tracking, categorizing recent advances by target scale, sensing modality, and collaboration strategy, while systematically reviewing both theoretical foundations and algorithmic developments. Emphasizing the integration of machine learning, especially deep learning and reinforcement learning, we highlighted how these methods are driving significant progress in tracking accuracy, adaptability, and intelligence. However, challenges persist, including constrained communication bandwidth, prolonged propagation delays, intensiﬁed interference and clutter, and elevated bit error rates. To address these issues, further application of artificial intelligence approaches—including GNNs, PINNs, transformer architectures, and embodied intelligence—is crucial for enhancing modeling accuracy and algorithm performance. Additionally, it is essential to fully leverage federated learning and blockchain technology to strengthen the security and comprehensiveness of data sharing. We hope this survey will deepen understanding within the field and steer future research towards developing more intelligent, reliable, and versatile underwater target tracking systems.

\bibliographystyle{./IEEEtran}
\bibliography{reference}

\begin{thebibliography}{100}
\providecommand{\url}[1]{#1}
\csname url@samestyle\endcsname
\providecommand{\newblock}{\relax}
\providecommand{\bibinfo}[2]{#2}
\providecommand{\BIBentrySTDinterwordspacing}{\spaceskip=0pt\relax}
\providecommand{\BIBentryALTinterwordstretchfactor}{4}
\providecommand{\BIBentryALTinterwordspacing}{\spaceskip=\fontdimen2\font plus
\BIBentryALTinterwordstretchfactor\fontdimen3\font minus \fontdimen4\font\relax}
\providecommand{\BIBforeignlanguage}[2]{{%
\expandafter\ifx\csname l@#1\endcsname\relax
\typeout{** WARNING: IEEEtran.bst: No hyphenation pattern has been}%
\typeout{** loaded for the language `#1'. Using the pattern for}%
\typeout{** the default language instead.}%
\else
\language=\csname l@#1\endcsname
\fi
#2}}
\providecommand{\BIBdecl}{\relax}
\BIBdecl

\bibitem{s1.2022}
A.~S. S. and S.~C. Dhongdi, ``Review of underwater mobile sensor network for ocean phenomena monitoring,'' \emph{J. Network Comput. Appl.}, vol. 205, p. 103418, Sep. 2022.

\bibitem{ghafoor2019}
H.~Ghafoor and Y.~Noh, ``An {{Overview}} of {{Next-Generation Underwater Target Detection}} and {{Tracking}}: {{An Integrated Underwater Architecture}},'' \emph{IEEE Access}, vol.~7, pp. 98\,841--98\,853, 2019.

\bibitem{mansur2014}
P.~Mansur and S.~Sreedharan, ``Survey of prediction algorithms for object tracking in wireless sensor networks,'' in \emph{2014 {{IEEE International Conference}} on {{Computational Intelligence}} and {{Computing Research}}}.\hskip 1em plus 0.5em minus 0.4em\relax Coimbatore, India: IEEE, Dec. 2014, pp. 1--4.

\bibitem{liu2021}
Y.~Liu, H.~Wang, L.~Cai, X.~Shen, and R.~Zhao, ``Fundamentals and {{Advancements}} of {{Topology Discovery}} in {{Underwater Acoustic Sensor Networks}}: {{A Review}},'' \emph{IEEE Sensors J.}, vol.~21, no.~19, pp. 21\,159--21\,174, Oct. 2021.

\bibitem{xu2019}
G.~Xu, Y.~Shi, X.~Sun, and W.~Shen, ``Internet of {{Things}} in {{Marine Environment Monitoring}}: {{A Review}},'' \emph{Sensors}, vol.~19, no.~7, p. 1711, Apr. 2019.

\bibitem{2023水声目标探测和识别}
Y.~Zhang, C.~Wang, Q.~Zhang, Q.~Li, and L.~Da, ``The development review of fusion technology for underwater acoustic target detection and recognition{(in Chinese)},'' \emph{Signal Processing}, pp. 1711--1727, 2023.

\bibitem{2019水声目标探测技术}
h.~Huang and Y.~Li, ``Research status and prospects of underwater acoustic target detection technology{(in Chinese)},'' \emph{The journal of the Chinese Academy of Sciences}, pp. 264--271, 2019.

\bibitem{oracevic2014}
A.~Oracevic and S.~Ozdemir, ``A survey of secure target tracking algorithms for wireless sensor networks,'' in \emph{2014 {{World Congress}} on {{Computer Applications}} and {{Information Systems}} ({{WCCAIS}})}.\hskip 1em plus 0.5em minus 0.4em\relax Hammamet, Tunisia: IEEE, Jan. 2014, pp. 1--6.

\bibitem{yan2021}
J.~Yan, Y.~Meng, X.~Luo, and X.~Guan, ``To {{Hide Private Position Information}} in {{Localization}} for {{Internet}} of {{Underwater Things}},'' \emph{IEEE Internet Things J.}, vol.~8, no.~18, pp. 14\,338--14\,354, Sep. 2021.

\bibitem{2021基于水下传感器}
M.~Liu, X.~Han, S.~Zhang, R.~Zheng, and J.~Lan, ``Research status and prospect of target tracking technologies via underwater sensor networks{(in Chinese)},'' \emph{ACTA AUTOMATICA SINICA}, pp. 235--251, 2021.

\bibitem{2024基于随机有限集}
L.~Yan, J.~Gu, Y.~Jiang, M.~Xu, Z.~Gao, B.~Tian, and T.~Zhang, ``Overview of multi-target tracking technology based on random finite set{(in Chinese)},'' \emph{Electronic Information Warfare Technology}, pp. 81--88, 2024.

\bibitem{luo2018}
J.~Luo, Y.~Han, and L.~Fan, ``Underwater {{Acoustic Target Tracking}}: {{A Review}},'' \emph{Sensors}, vol.~18, no.~1, p. 112, Jan. 2018.

\bibitem{kumar2021}
M.~Kumar and S.~Mondal, ``Recent developments on target tracking problems: {{A}} review,'' \emph{Ocean Engineering}, vol. 236, p. 109558, Sep. 2021.

\bibitem{hou2024}
X.~Hou, Y.~Chen, B.~Zhang, and Y.~Yang, ``Review of {{Research Progress}} on {{Passive Direction-of-Arrival Tracking Technology}} for {{Underwater Targets}},'' \emph{Remote Sensing}, vol.~16, no.~23, p. 4511, Dec. 2024.

\bibitem{jahanbakht2021}
M.~Jahanbakht, W.~Xiang, L.~Hanzo, and M.~Rahimi~Azghadi, ``Internet of {{Underwater Things}} and {{Big Marine Data Analytics}}---{{A Comprehensive Survey}},'' \emph{IEEE Commun. Surv. Tutorials}, vol.~23, no.~2, pp. 904--956, 2021.

\bibitem{yang2021a}
Z.~Yang, J.~Du, Z.~Xia, C.~Jiang, A.~Benslimane, and Y.~Ren, ``Secure and {{Cooperative Target Tracking}} via {{AUV Swarm}}: {{A Reinforcement Learning Approach}},'' in \emph{2021 {{IEEE Global Communications Conference}} ({{GLOBECOM}})}.\hskip 1em plus 0.5em minus 0.4em\relax Madrid, Spain: IEEE, Dec. 2021, pp. 1--6.

\bibitem{zhu2024}
S.~Zhu, G.~Han, C.~Lin, and Q.~Tao, ``Underwater {{Target Tracking Based}} on {{Hierarchical Software-Defined Multi-AUV Reinforcement Learning}}: {{A Multi-AUV Advantage-Attention Actor-Critic Approach}},'' \emph{IEEE Trans. on Mobile Comput.}, vol.~23, no.~12, pp. 13\,639--13\,653, Dec. 2024.

\bibitem{wang2025}
S.~Wang, C.~Lin, G.~Han, S.~Zhu, Z.~Li, Z.~Wang, and Y.~Ma, ``Multi-{{AUV Cooperative Underwater Multi-Target Tracking Based}} on {{Dynamic-Switching-Enabled Multi-Agent Reinforcement Learning}},'' \emph{IEEE Trans. on Mobile Comput.}, vol.~24, no.~5, pp. 4296--4311, May 2025.

\bibitem{zheng2023}
L.~Zheng, M.~Liu, and S.~Zhang, ``An end-to-end sensor scheduling method based on {{D3QN}} for underwater passive tracking in {{UWSNs}},'' \emph{Journal of Network and Computer Applications}, vol. 219, p. 103730, Oct. 2023.

\bibitem{su2021}
R.~Su, Z.~Gong, D.~Zhang, C.~Li, Y.~Chen, and R.~Venkatesan, ``An {{Adaptive Asynchronous Wake-Up Scheme}} for {{Underwater Acoustic Sensor Networks Using Deep Reinforcement Learning}},'' \emph{IEEE Trans. Veh. Technol.}, vol.~70, no.~2, pp. 1851--1865, Feb. 2021.

\bibitem{jin2019}
Z.~Jin, Q.~Zhao, and Y.~Su, ``{{RCAR}}: {{A Reinforcement-Learning-Based Routing Protocol}} for {{Congestion-Avoided Underwater Acoustic Sensor Networks}},'' \emph{IEEE Sensors J.}, vol.~19, no.~22, pp. 10\,881--10\,891, Nov. 2019.

\bibitem{wang2023}
C.~Wang, X.~Shen, H.~Wang, H.~Zhang, and H.~Mei, ``Reinforcement {{Learning-Based Opportunistic Routing Protocol Using Depth Information}} for {{Energy-Efficient Underwater Wireless Sensor Networks}},'' \emph{IEEE Sensors J.}, vol.~23, no.~15, pp. 17\,771--17\,783, Aug. 2023.

\bibitem{xu2024}
L.~Xu and R.~Niu, ``{{EKFNet}}: {{Learning System Noise Covariance Parameters}} for {{Nonlinear Tracking}},'' \emph{IEEE Trans. Signal Process.}, vol.~72, pp. 3139--3152, 2024.

\bibitem{hou2023}
X.~Hou, Y.~Qiao, B.~Zhang, and Y.~Yang, ``Robust {{Underwater Direction-of-Arrival Tracking Based}} on {{AI-Aided Variational Bayesian Extended Kalman Filter}},'' \emph{Remote Sensing}, vol.~15, no.~2, p. 420, Jan. 2023.

\bibitem{wang2021}
M.~Wang, B.~Qiu, Z.~Zhu, H.~Xue, and C.~Zhou, ``Study on {{Active Tracking}} of {{Underwater Acoustic Target Based}} on {{Deep Convolution Neural Network}},'' \emph{Applied Sciences}, vol.~11, no.~16, p. 7530, Aug. 2021.

\bibitem{wang2022}
M.~Wang, B.~Qiu, Z.~Zhu, L.~Ma, and C.~Zhou, ``Passive tracking of underwater acoustic targets based on multi-beam {{LOFAR}} and deep learning,'' \emph{PLoS ONE}, vol.~17, no.~12, p. e0273898, Dec. 2022.

\bibitem{xue2018}
B.~Xue, Z.~Wang, K.~Zhang, H.~Zhang, Y.~Chen, L.~Jia, H.~Wu, and J.~Zhai, ``Direct measurement of the sound velocity in seawater based on the pulsed acousto-optic effect between the frequency comb and the ultrasonic pulse,'' \emph{Opt. Express}, vol.~26, no.~17, p. 21849, Aug. 2018.

\bibitem{2021基于飞秒激光}
C.~Liu, B.~Xue, X.~Xu, Z.~Qian, and Y.~Wei, ``Traceable accuracy measurement of underwater sound velocit based on femtosecond laser {(in Chinese)},'' \emph{Laser \& Optoelectronics Progress}, vol.~58, no.~11, p. 1112006, 2021.

\bibitem{lee2012}
D.~Lee, G.~Kim, D.~Kim, H.~Myung, and H.-T. Choi, ``Vision-based object detection and tracking for autonomous navigation of underwater robots,'' \emph{Ocean Engineering}, vol.~48, pp. 59--68, Jul. 2012.

\bibitem{meng-chechuang2015}
M.~Chuang, J.~Hwang, K.~Williams, and R.~Towler, ``Tracking {{Live Fish From Low-Contrast}} and {{Low-Frame-Rate Stereo Videos}},'' \emph{IEEE Trans. Circuits Syst. Video Technol.}, vol.~25, no.~1, pp. 167--179, Jan. 2015.

\bibitem{mandic2016}
F.~Mandi{\'c}, I.~Renduli{\'c}, N.~Mi{\v s}kovi{\'c}, and {\DJ}.~Na{\dj}, ``Underwater {{Object Tracking Using Sonar}} and {{USBL Measurements}},'' \emph{Journal of Sensors}, vol. 2016, pp. 1--10, 2016.

\bibitem{xin2003}
G.~Xin, H.~You, and Y.~Xiao, ``Bearings-only underwater distributed fusion algorithm with feedback information,'' \emph{Journal of System Simulation}, vol.~15, no.~7, pp. 947--949, 2003.

\bibitem{yu2014}
C.~H. Yu and J.~W. Choi, ``Interacting multiple model filter-based distributed target tracking algorithm in underwater wireless sensor networks,'' \emph{Int. J. Control Autom. Syst.}, vol.~12, no.~3, pp. 618--627, Jun. 2014.

\bibitem{hare2017}
J.~Z. Hare, S.~Gupta, J.~Song, and T.~A. Wettergren, ``Classification induced distributed sensor scheduling for energy-efficiency in underwater target tracking sensor networks,'' in \emph{OCEANS 2017-Anchorage}.\hskip 1em plus 0.5em minus 0.4em\relax IEEE, 2017, pp. 1--7.

\bibitem{hare2014}
J.~Hare, S.~Gupta, and J.~Song, ``Distributed smart sensor scheduling for underwater target tracking,'' in \emph{2014 {{Oceans}} - {{St}}. {{John}}'s}.\hskip 1em plus 0.5em minus 0.4em\relax St. John's, NL: IEEE, Sep. 2014, pp. 1--6.

\bibitem{braca2016}
P.~Braca, R.~Goldhahn, G.~Ferri, and K.~D. LePage, ``Distributed {{Information Fusion}} in {{Multistatic Sensor Networks}} for {{Underwater Surveillance}},'' \emph{IEEE Sensors J.}, vol.~16, no.~11, pp. 4003--4014, Jun. 2016.

\bibitem{zhang2019a}
D.~Zhang, M.~Liu, S.~Zhang, and Q.~Zhang, ``Non-{{Myopic Energy Allocation}} for {{Target Tracking}} in {{Energy Harvesting UWSNs}},'' \emph{IEEE Sensors J.}, vol.~19, no.~10, pp. 3772--3783, May 2019.

\bibitem{qiu2019}
J.~Qiu, Z.~Xing, C.~Zhu, K.~Lu, J.~He, Y.~Sun, and L.~Yin, ``Centralized {{Fusion Based}} on {{Interacting Multiple Model}} and {{Adaptive Kalman Filter}} for {{Target Tracking}} in {{Underwater Acoustic Sensor Networks}},'' \emph{IEEE Access}, vol.~7, pp. 25\,948--25\,958, 2019.

\bibitem{zhou2019}
Z.~Zhou, ``Optimal {{Batch Distributed Asynchronous Multisensor Fusion With Feedback}},'' \emph{IEEE Trans. Aerosp. Electron. Syst.}, vol.~55, no.~1, pp. 46--56, Feb. 2019.

\bibitem{luo2020}
J.~Luo, Z.~Wang, Y.~Chen, M.~Wu, and Y.~Yang, ``An {{Improved Unscented Particle Filter Approach}} for {{Multi-Sensor Fusion Target Tracking}},'' \emph{Sensors}, vol.~20, no.~23, p. 6842, Nov. 2020.

\bibitem{kumar2016}
D.~Kumar, S.~Rao, and K.~Raju, ``Integrated {{Unscented Kalman}} filter for underwater passive target tracking with towed array measurements,'' \emph{Optik}, vol. 127, no.~5, pp. 2840--2847, Mar. 2016.

\bibitem{qian2016}
Y.~Qian, Y.~Chen, X.~Cao, J.~Wu, and J.~Sun, ``An underwater bearing-only multi-target tracking approach based on enhanced {{Kalman}} filter,'' in \emph{2016 {{IEEE International Conference}} on {{Electronic Information}} and {{Communication Technology}} ({{ICEICT}})}.\hskip 1em plus 0.5em minus 0.4em\relax Harbin, China: IEEE, Aug. 2016, pp. 203--207.

\bibitem{ravikumar2016}
D.~Ravi~Kumar, S.~Koteswara~Rao, and K.~Padma~Raju, ``A novel stochastic estimator using pre-processing technique for long range target tracking in heavy noise environment,'' \emph{Optik}, vol. 127, no.~10, pp. 4520--4530, May 2016.

\bibitem{cheng2021}
M.~Cheng and M.~R.~K. Aziz, ``{{DOA-Based 3D Tracking With Factor Graph Technique}} for a {{Multi-Sensor System}},'' \emph{IEEE Sensors J.}, vol.~21, no.~22, pp. 25\,853--25\,861, Nov. 2021.

\bibitem{yan2016a}
J.~Yan, Z.~Xu, X.~Luo, and Y.~Sun, ``Consensus estimation based underwater target tracking with acoustic sensor networks,'' in \emph{2016 35th {{Chinese Control Conference}} ({{CCC}})}.\hskip 1em plus 0.5em minus 0.4em\relax Chengdu, China: IEEE, Jul. 2016, pp. 7837--7842.

\bibitem{son2013}
H.~S. Son, J.~B. Park, and Y.~H. Joo, ``The study on tracking algorithm for the underwater target: {{Applying}} to noise limited bi-static sonar model,'' in \emph{2013 9th {{Asian Control Conference}} ({{ASCC}})}.\hskip 1em plus 0.5em minus 0.4em\relax Istanbul, Turkey: IEEE, Jun. 2013, pp. 1--6.

\bibitem{ferri2014}
G.~Ferri, A.~Munafo, R.~Goldhahn, and K.~LePage, ``A non-myopic, receding horizon control strategy for an {{AUV}} to track an underwater target in a bistatic sonar scenario,'' in \emph{53rd {{IEEE Conference}} on {{Decision}} and {{Control}}}.\hskip 1em plus 0.5em minus 0.4em\relax Los Angeles, CA, USA: IEEE, Dec. 2014, pp. 5352--5358.

\bibitem{dehnavi2017}
S.~M. Dehnavi, M.~Ayati, and M.~R. Zakerzadeh, ``Three dimensional target tracking via {{Underwater Acoustic Wireless Sensor Network}},'' in \emph{2017 {{Artificial Intelligence}} and {{Robotics}} ({{IRANOPEN}})}.\hskip 1em plus 0.5em minus 0.4em\relax Qazvin, Iran: IEEE, Apr. 2017, pp. 153--157.

\bibitem{Zou2019}
Z.~Z.~Z. Zou, L.~S.~L. Song, and X.~C.~X. Cheng, ``Labeled box-particle cphd filter for multiple extended targets tracking,'' \emph{Journal of Systems Engineering and Electronics}, vol.~30, no.~1, p.~57, 2019.

\bibitem{zhou2023}
T.~Zhou, Y.~Wang, L.~Zhang, B.~Chen, and X.~Yu, ``Underwater {{Multitarget Tracking Method Based}} on {{Threshold Segmentation}},'' \emph{IEEE J. Oceanic Eng.}, vol.~48, no.~4, pp. 1255--1269, Oct. 2023.

\bibitem{zhang2020}
G.-p. Zhang, C.~Zheng, L.-h. Qiu, and S.-b. Sun, ``Multi-{{Bernoulli Filter}} for {{Tracking Multiple Targets Using Sensor Array}},'' \emph{China Ocean Eng}, vol.~34, no.~2, pp. 245--256, Apr. 2020.

\bibitem{zhang2023}
Y.~Zhang, Y.~Li, S.~Li, J.~Zeng, Y.~Wang, and S.~Yan, ``Multi-{{Target Tracking}} in {{Multi-Static Networks}} with {{Autonomous Underwater Vehicles Using}} a {{Robust Multi-Sensor Labeled Multi-Bernoulli Filter}},'' \emph{JMSE}, vol.~11, no.~4, p. 875, Apr. 2023.

\bibitem{bar2009}
Y.~Bar-Shalom, F.~Daum, and J.~Huang, ``The probabilistic data association filter,'' \emph{IEEE Control Systems Magazine}, vol.~29, no.~6, pp. 82--100, 2009.

\bibitem{qiu2018}
W.~Qiu, L.~Li, P.~Lei, and Z.~Wang, ``Multiple {{Targets Tracking}} by {{Using Probability Data Association}} and {{Cubature Kalman Filter}},'' in \emph{2018 10th {{International Conference}} on {{Wireless Communications}} and {{Signal Processing}} ({{WCSP}})}.\hskip 1em plus 0.5em minus 0.4em\relax Hangzhou: IEEE, Oct. 2018, pp. 1--5.

\bibitem{Puranik2007}
S.~Puranik and J.~Tugnait, ``Tracking of multiple maneuvering targets using multiscan jpda and 1mm filtering,'' \emph{IEEE Transactions on Aerospace and Electronic Systems}, pp. 23--35, 2007.

\bibitem{chen2024}
Q.~Chen, P.~Wang, and H.~Wei, ``An algorithm for multi-target tracking in low-signal-to-clutter-ratio underwater acoustic scenes,'' \emph{AIP Advances}, vol.~14, no.~10, p. 105121, Oct. 2024.

\bibitem{musicki2002}
D.~Musicki and R.~Evans, ``Joint {{Integrated Probabilistic Data Association}} - {{JIPDA}},'' in \emph{Proceedings of the {{Fifth International Conference}} on {{Information Fusion}}. {{FUSION}} 2002. ({{IEEE Cat}}.{{No}}.{{02EX5997}})}, vol.~2.\hskip 1em plus 0.5em minus 0.4em\relax Annapolis, MD, USA: Int. Soc. Inf. Fusion, 2002, pp. 1120--1125.

\bibitem{yao2019a}
Y.~Yao, J.~Zhao, and L.~Wu, ``Doppler data association scheme for multi-target tracking in an active sonar system,'' \emph{Sensors}, vol.~19, no.~9, p. 2003, Apr. 2019.

\bibitem{chen2006}
J.~Chen, J.~Han, and X.~Zhang, ``An {{Adaptive Single Model}} of {{Maneuvering Target Tracking}},'' in \emph{2006 {{International Conference}} on {{Computational Intelligence}} and {{Security}}}.\hskip 1em plus 0.5em minus 0.4em\relax Guangzhou, China: IEEE, Nov. 2006, pp. 714--718.

\bibitem{johnston2001}
L.~Johnston and V.~Krishnamurthy, ``An improvement to the interacting multiple model ({{IMM}}) algorithm,'' \emph{IEEE Trans. Signal Process.}, vol.~49, no.~12, pp. 2909--2923, Dec. 2001.

\bibitem{2001参数自适应}
Y.~Liang, Y.~Jia, Q.~Pan, and H.~Zhang, ``Parameter identification in switching multiple model estimation and adaptive interacting multiple model estimator{in Chinese},'' \emph{CONTROL THEORY AND APPLICATIONS}, pp. 653--656, 2001.

\bibitem{Yang2015}
R.~Yang, G.~W. Ng, and Y.~Bar-Shalom, ``Bearings-only tracking with fusion from heterogenous passive sensors: Esm/eo and acoustic,'' in \emph{2015 18th International Conference on Information Fusion (Fusion)}, 2015.

\bibitem{liu2017}
M.~Liu, L.~Zhao, and S.~Zhang, ``Delay-estimation-based asynchronous particle filtering for passive target tracking in underwater wireless sensor networks,'' in \emph{2017 36th {{Chinese Control Conference}} ({{CCC}})}.\hskip 1em plus 0.5em minus 0.4em\relax Dalian, China: IEEE, Jul. 2017, pp. 8929--8934.

\bibitem{chang2018}
S.~Chang, Y.~Li, Y.~He, and H.~Wang, ``Target {{Localization}} in {{Underwater Acoustic Sensor Networks Using RSS Measurements}},'' \emph{Applied Sciences}, vol.~8, no.~2, p. 225, Feb. 2018.

\bibitem{sozer2000}
E.~Sozer, M.~Stojanovic, and J.~Proakis, ``Underwater acoustic networks,'' \emph{IEEE J. Oceanic Eng.}, vol.~25, no.~1, pp. 72--83, Jan. 2000.

\bibitem{2024复杂海洋环境}
X.~Ji, L.~Liu, F.~Ji, G.~Li, and s.~Lu, ``Review of calculation methods for low-frequency acoustic transmission loss of underwater vehicle in complex marine environments{(in Chinese)},'' \emph{SHIP SCIENCE AND TECHNOLOGY}, pp. 13--18, 2024.

\bibitem{m.2013}
J.~M., ``Ray {{Trace Modeling}} of {{Underwater Sound Propagation}},'' in \emph{Modeling and {{Measurement Methods}} for {{Acoustic Waves}} and for {{Acoustic Microdevices}}}, M.~G. Beghi, Ed.\hskip 1em plus 0.5em minus 0.4em\relax InTech, Aug. 2013.

\bibitem{lawrence1985}
M.~W. Lawrence, ``Ray theory modeling applied to low-frequency acoustic interaction with horizontally stratified ocean bottoms,'' \emph{The Journal of the Acoustical Society of America}, vol.~78, no.~2, pp. 649--658, Aug. 1985.

\bibitem{etter2012}
P.~C. Etter, ``Advanced {{Applications}} for {{Underwater Acoustic Modeling}},'' \emph{Advances in Acoustics and Vibration}, vol. 2012, pp. 1--28, May 2012.

\bibitem{smirnov2001}
I.~P. Smirnov, A.~L. Virovlyansky, and G.~M. Zaslavsky, ``Theory and applications of ray chaos to underwater acoustics,'' \emph{Phys. Rev. E}, vol.~64, no.~3, p. 036221, Aug. 2001.

\bibitem{li2023}
K.~Li and M.~Chitre, ``Data-{{Aided Underwater Acoustic Ray Propagation Modeling}},'' \emph{IEEE J. Oceanic Eng.}, vol.~48, no.~4, pp. 1127--1148, Oct. 2023.

\bibitem{liao2024}
S.~Liao, W.~Xiao, and Y.~Wang, ``Dynamic hybrid parallel computing of the {{Ray Model}} for solving underwater acoustic fields in vast sea,'' \emph{Sci Rep}, vol.~14, no.~1, p. 25385, Oct. 2024.

\bibitem{pekeris1948}
C.~L. Pekeris, ``{{THEORY OF PROPAGATION OF EXPLOSIVE SOUND IN SHALLOW WATER}},'' in \emph{Geological {{Society}} of {{America Memoirs}}}.\hskip 1em plus 0.5em minus 0.4em\relax Geological Society of America, 1948, vol.~27, pp. 1--116.

\bibitem{nagl1978}
A.~Nagl, H.~{\"U}berall, A.~J. Haug, and G.~L. Zarur, ``Adiabatic mode theory of underwater sound propagation in a range-dependent environment,'' \emph{The Journal of the Acoustical Society of America}, vol.~63, no.~3, pp. 739--749, Mar. 1978.

\bibitem{knobles2003}
D.~P. Knobles, S.~A. Stotts, and R.~A. Koch, ``Low frequency coupled mode sound propagation over a continental shelf,'' \emph{The Journal of the Acoustical Society of America}, vol. 113, no.~2, pp. 781--787, Feb. 2003.

\bibitem{tappert1977}
F.~D. Tappert, ``The parabolic approximation method,'' in \emph{Wave {{Propagation}} and {{Underwater Acoustics}}}, J.~B. Keller and J.~S. Papadakis, Eds.\hskip 1em plus 0.5em minus 0.4em\relax Berlin, Heidelberg: Springer Berlin Heidelberg, 1977, vol.~70, pp. 224--287.

\bibitem{lee1983}
D.~Lee, M.~H. Schultz, and K.~R. Jackson, ``A wide-angle three-dimensional parabolic wave equation,'' \emph{The Journal of the Acoustical Society of America}, vol.~74, no.~S1, pp. S96--S96, Nov. 1983.

\bibitem{claerbout1986}
J.~F. Claerbout, ``Fundamentals of {{Geophysical Data Processing}}, 2nd edn,'' \emph{Geophys J Int}, vol.~86, no.~1, pp. 217--219, Jul. 1986.

\bibitem{collins1989}
M.~D. Collins, ``A higher-order parabolic equation for wave propagation in an ocean overlying an elastic bottom,'' \emph{The Journal of the Acoustical Society of America}, vol.~86, no.~4, pp. 1459--1464, Oct. 1989.

\bibitem{zhao2020}
H.~Zhao, J.~Yan, X.~Luo, and X.~Gua, ``Privacy preserving solution for the asynchronous localization of underwater sensor networks,'' \emph{IEEE/CAA J. Autom. Sinica}, vol.~7, no.~6, pp. 1511--1527, Nov. 2020.

\bibitem{cardone2023}
M.~Cardone, A.~Dytso, and C.~Rush, ``Entropic {{Central Limit Theorem}} for {{Order Statistics}},'' \emph{IEEE Trans. Inform. Theory}, vol.~69, no.~4, pp. 2193--2205, Apr. 2023.

\bibitem{sun2017}
W.~Sun and Y.~Yang, ``Adaptive maneuvering frequency method of current statistical model,'' \emph{IEEE/CAA J. Autom. Sinica}, vol.~4, no.~1, pp. 154--160, Jan. 2017.

\bibitem{singer1970}
R.~Singer, ``Estimating {{Optimal Tracking Filter Performance}} for {{Manned Maneuvering Targets}},'' \emph{IEEE Trans. Aerosp. Electron. Syst.}, vol. AES-6, no.~4, pp. 473--483, Jul. 1970.

\bibitem{moose1975}
R.~Moose, ``An adaptive state estimation solution to the maneuvering target problem,'' \emph{IEEE Trans. Automat. Contr.}, vol.~20, no.~3, pp. 359--362, Jun. 1975.

\bibitem{zhou1984}
H.~Zhou and K.~Kumar, ``A 'current' statistical model and adaptive algorithm for estimating maneuvering targets,'' \emph{Journal of Guidance, Control, and Dynamics}, vol.~7, no.~5, pp. 596--602, Sep. 1984.

\bibitem{mehrotra1997}
K.~Mehrotra and P.~Mahapatra, ``A jerk model for tracking highly maneuvering targets,'' \emph{IEEE Trans. Aerosp. Electron. Syst.}, vol.~33, no.~4, pp. 1094--1105, Oct. 1997.

\bibitem{isbitiren2011}
G.~Isbitiren and O.~B. Akan, ``Three-{{Dimensional Underwater Target Tracking With Acoustic Sensor Networks}},'' \emph{IEEE Trans. Veh. Technol.}, vol.~60, no.~8, pp. 3897--3906, Oct. 2011.

\bibitem{poursheikhali2021}
S.~Poursheikhali and H.~{Zamiri-Jafarian}, ``Source localization in inhomogeneous underwater medium using sensor arrays: {{Received}} signal strength approach,'' \emph{Signal Processing}, vol. 183, p. 108047, Jun. 2021.

\bibitem{li2012}
T.~Li and A.~Nehorai, ``Direction-of-{{Arrival Estimation}} of {{Hydroacoustic Signals From Marine Vessels Containing Random}} and {{Sinusoidal Components}},'' \emph{IEEE Signal Process. Lett.}, vol.~19, no.~8, pp. 503--506, Aug. 2012.

\bibitem{yan2019}
J.~Yan, H.~Zhao, B.~Pu, X.~Luo, C.~Chen, and X.~Guan, ``Energy-{{Efficient Target Tracking With UASNs}}: {{A Consensus-Based Bayesian Approach}},'' \emph{IEEE Trans. Automat. Sci. Eng.}, pp. 1--15, 2019.

\bibitem{hu2017}
R.~Hu, Y.~Fu, Z.~Chen, J.~Xu, and J.~Tang, ``Robust {{DOA Estimation}} via {{Sparse Signal Reconstruction With Impulsive Noise}},'' \emph{IEEE Commun. Lett.}, vol.~21, no.~6, pp. 1333--1336, Jun. 2017.

\bibitem{jauffret2008}
C.~Jauffret, P.~Blanc-Benon, and D.~Pillon, ``Multi frequencies and bearing target motion analysis : Properties and sonar applications,'' in \emph{2008 11th International Conference on Information Fusion}, 2008.

\bibitem{2024单站}
D.~Sun, Y.~Zhang, T.~Teng, and Z.~Hu, ``A single-platform underwater maneuvering target motion analysis method based on bearing and frequency measurements{(in Chinese)},'' \emph{ACTA ACUSTICA}, pp. 683--695, 2024.

\bibitem{zhao2013}
Z.~Zhao, B.~Huang, and F.~Liu, ``Bayesian method for state estimation of batch process with missing data,'' \emph{Computers \& Chemical Engineering}, vol.~53, pp. 14--24, Jun. 2013.

\bibitem{2005方位／频率}
Q.~Hu and X.~Gong, ``Experimental research of bearing/frequency target motion analysis{(in Chinese)},'' \emph{ACTA ACUSTICA}, pp. 120--124, 2005.

\bibitem{hue2002}
C.~Hue, J.-P. Le~Cadre, and P.~Perez, ``Sequential {{Monte Carlo}} methods for multiple target tracking and data fusion,'' \emph{IEEE Trans. Signal Process.}, vol.~50, no.~2, pp. 309--325, Feb. 2002.

\bibitem{zhou2020}
C.~Zhou, R.~Wang, and C.~Hu, ``Equivalent point estimation for small target groups tracking based on maximum group likelihood estimation,'' \emph{Sci. China Inf. Sci.}, vol.~63, no.~8, p. 189302, Aug. 2020.

\bibitem{wang2020}
T.~Wang, X.~Wang, W.~Shi, Z.~Zhao, Z.~He, and T.~Xia, ``Target localization and tracking based on improved {{Bayesian}} enhanced least-squares algorithm in wireless sensor networks,'' \emph{Computer Networks}, vol. 167, p. 106968, Feb. 2020.

\bibitem{BAO2018}
B.~Zhichao and J.~Qiuxi, ``Properties of gauss-newton filter in linear cases,'' \emph{Journal of Systems Engineering and Electronics}, pp. 899--907, 2018.

\bibitem{gong1981}
G.~Gong and F.~J. Samaniego, ``Pseudo {{Maximum Likelihood Estimation}}: {{Theory}} and {{Applications}},'' \emph{The Annals of Statistics}, vol.~9, no.~4, pp. 861--869, 1981.

\bibitem{zarai2021}
K.~Zarai and A.~Cherif, ``Adaptive filter based on {{Monte Carlo}} method to improve the non-linear target tracking in the radar system,'' \emph{AS}, vol.~4, no.~1, pp. 67--74, Mar. 2021.

\bibitem{pedersen2011}
M.~Pedersen, U.~Thygesen, and H.~Madsen, ``Nonlinear tracking in a diffusion process with a {{Bayesian}} filter and the finite element method,'' \emph{Computational Statistics \& Data Analysis}, vol.~55, no.~1, pp. 280--290, Jan. 2011.

\bibitem{Kalman1960}
R.~E. Kalman, ``A new approach to linear filtering and prediction problems,'' \emph{Journal of Fluids Engineering}, pp. 35--45, 1960.

\bibitem{smith2006}
D.~Smith and S.~Singh, ``Approaches to {{Multisensor Data Fusion}} in {{Target Tracking}}: {{A Survey}},'' \emph{IEEE Trans. Knowl. Data Eng.}, vol.~18, no.~12, pp. 1696--1710, Dec. 2006.

\bibitem{xianghuiyuan2005}
{Xianghui Yuan}, {Chongzhao Han}, {Zhansheng Duan}, and {Ming Lei}, ``Comparison and choice of models in tracking target with coordinated turn motion,'' in \emph{2005 7th {{International Conference}} on {{Information Fusion}}}.\hskip 1em plus 0.5em minus 0.4em\relax Philadelphia, PA, USA: IEEE, 2005, p. 6 pp.

\bibitem{M2002}
M.~S. Arulampalam, S.~Maskell, N.~J. Gordon, and T.~Clapp, ``A tutorial on particle filters for online nonlinear/non-gaussian bayesian tracking{(Article)},'' \emph{IEEE Transactions on Signal Processing}, pp. 174--188, 2002.

\bibitem{binliao2012}
B.~Liao, Z.~Zhang, and S.~Chan, ``{{DOA Estimation}} and {{Tracking}} of {{ULAs}} with {{Mutual Coupling}},'' \emph{IEEE Trans. Aerosp. Electron. Syst.}, vol.~48, no.~1, pp. 891--905, Jan. 2012.

\bibitem{gao2015}
X.~Gao, X.~Li, F.~Jason, and W.~Dai, ``A {{Sequential Bayesian Algorithm}} for {{DOA Tracking}} in {{Time}}-{{Varying Environments}},'' \emph{Chin. j. electron.}, vol.~24, no.~1, pp. 140--145, Jan. 2015.

\bibitem{dongkeonkong2000}
D.~Kong and J.~Chun, ``A fast {{DOA}} tracking algorithm based on the extended {{Kalman}} filter,'' in \emph{Proceedings of the {{IEEE}} 2000 {{National Aerospace}} and {{Electronics Conference}}. {{NAECON}} 2000. {{Engineering Tomorrow}} ({{Cat}}. {{No}}.{{00CH37093}})}.\hskip 1em plus 0.5em minus 0.4em\relax Dayton, OH, USA: IEEE, 2000, pp. 235--238.

\bibitem{cevher2007}
V.~Cevher, R.~Velmurugan, and J.~H. McClellan, ``Acoustic {{Multitarget Tracking Using Direction-of-Arrival Batches}},'' \emph{IEEE Trans. Signal Process.}, vol.~55, no.~6, pp. 2810--2825, Jun. 2007.

\bibitem{kumar2021a}
K.~Kumar, S.~Bhaumik, and P.~Date, ``Extended {{Kalman Filter Using Orthogonal Polynomials}},'' \emph{IEEE Access}, vol.~9, pp. 59\,675--59\,691, 2021.

\bibitem{2003多模型}
H.~Li and X.~Feng, ``A survey of the development of multiple - model algorithm for target tracking{in Chinese},'' \emph{Telecommunication technology}, pp. 12--15, 2003.

\bibitem{Blom1984}
H.~A.~P. Blom, ``An efficient filter for abruptly changing systems,'' in \emph{The 23rd IEEE Conference on Decision and Control}, 1984.

\bibitem{li1996}
X.-R. Li and Y.~Bar-Shalom, ``Multiple-model estimation with variable structure,'' \emph{IEEE Transactions on Automatic Control}, vol.~41, no.~4, pp. 478--493, 1996.

\bibitem{dongguangzuo2002}
D.~Zuo, C.~Han, Z.~Lin, H.~Zhu, and H.~Hong, ``Fuzzy multiple model tracking algorithm for manoeuvring target,'' in \emph{Proceedings of the {{Fifth International Conference}} on {{Information Fusion}}. {{FUSION}} 2002. ({{IEEE Cat}}.{{No}}.{{02EX5997}})}, vol.~2.\hskip 1em plus 0.5em minus 0.4em\relax Annapolis, MD, USA: Int. Soc. Inf. Fusion, 2002, pp. 818--823.

\bibitem{yan2016}
J.~Yan, X.~Zhang, X.~Luo, Y.~Sun, and X.~Guan, ``{{AUV}} assisted asynchronous localization for underwater sensor networks,'' in \emph{2016 35th {{Chinese Control Conference}} ({{CCC}})}.\hskip 1em plus 0.5em minus 0.4em\relax Chengdu, China: IEEE, Jul. 2016, pp. 7291--7296.

\bibitem{yan2018}
J.~Yan, X.~Zhang, X.~Luo, Y.~Wang, C.~Chen, and X.~Guan, ``Asynchronous {{Localization With Mobility Prediction}} for {{Underwater Acoustic Sensor Networks}},'' \emph{IEEE Trans. Veh. Technol.}, vol.~67, no.~3, pp. 2543--2556, Mar. 2018.

\bibitem{yan2019a}
J.~Yan, H.~Zhao, B.~Pu, X.~Luo, C.~Chen, and X.~Guan, ``Energy-{{Efficient Target Tracking With UASNs}}: {{A Consensus-Based Bayesian Approach}},'' \emph{IEEE Trans. Automat. Sci. Eng.}, pp. 1--15, 2019.

\bibitem{yu2017}
Y.~Yu, ``Distributed {{Target Tracking}} in {{Wireless Sensor Networks With Data Association Uncertainty}},'' \emph{IEEE Commun. Lett.}, vol.~21, no.~6, pp. 1281--1284, Jun. 2017.

\bibitem{Mu2007}
D.~d. Mušicki, B.~F. La~Scala, and R.~J. Evans, ``Integrated track splitting filter--efficient multi-scan single target tracking in clutter,'' \emph{IEEE Transactions on Aerospace and Electronic Systems}, pp. 1409--1425, 2007.

\bibitem{kulmon2018}
P.~Kulmon and P.~Stukovska, ``Assessing {{Multiple-Target Tracking Performance Of GNN Association Algorithm}},'' in \emph{2018 19th {{International Radar Symposium}} ({{IRS}})}.\hskip 1em plus 0.5em minus 0.4em\relax Bonn, Germany: IEEE, Jun. 2018, pp. 1--10.

\bibitem{coraluppi2018}
S.~Coraluppi, C.~Carthel, and A.~Coon, ``An {{MHT Approach}} to {{Multi-Sensor Passive Sonar Tracking}},'' in \emph{2018 21st {{International Conference}} on {{Information Fusion}} ({{FUSION}})}.\hskip 1em plus 0.5em minus 0.4em\relax Cambridge: IEEE, Jul. 2018, pp. 480--487.

\bibitem{li2019}
X.~Li, C.~Zhao, X.~Lu, and W.~Wei, ``Underwater {{Bearings-Only Multitarget Tracking Based}} on {{Modified PMHT}} in {{Dense-Cluttered Environment}},'' \emph{IEEE Access}, vol.~7, pp. 93\,678--93\,689, 2019.

\bibitem{qiu2017}
W.~Qiu, W.~Wang, Z.~Zhuang, P.~Lei, and L.~Li, ``Using ship radiated noise spectrum feature for data association in underwater target tracking,'' in \emph{2017 20th {{International Conference}} on {{Information Fusion}} ({{Fusion}})}.\hskip 1em plus 0.5em minus 0.4em\relax Xi'an, China: IEEE, Jul. 2017, pp. 1--5.

\bibitem{Zhu1996}
Z.~Ziqian, ``Iterated joint probabilistic data association,'' in \emph{Proceedings of International Radar Conference}.\hskip 1em plus 0.5em minus 0.4em\relax IEEE, 1996, pp. 434--438.

\bibitem{seget2010}
K.~Seget, A.~Schulz, and U.~Heute, ``Multi-{{Hypothesis Tracking}} and fusion techniques for multistatic active sonar systems,'' in \emph{2010 13th {{International Conference}} on {{Information Fusion}}}.\hskip 1em plus 0.5em minus 0.4em\relax Edinburgh: IEEE, Jul. 2010, pp. 1--8.

\bibitem{li2016}
X.~Li, P.~Willett, M.~Baum, and Y.~Li, ``{{PMHT Approach}} for {{Underwater Bearing-Only Multisensor}}--{{Multitarget Tracking}} in {{Clutter}},'' \emph{IEEE J. Oceanic Eng.}, vol.~41, no.~4, pp. 831--839, Oct. 2016.

\bibitem{li2015}
X.~Li, Y.~Li, J.~Yu, X.~Chen, and M.~Dai, ``{{PMHT Approach}} for {{Multi-Target Multi-Sensor Sonar Tracking}} in {{Clutter}},'' \emph{Sensors}, vol.~15, no.~11, pp. 28\,177--28\,192, Nov. 2015.

\bibitem{gao2021}
L.~Gao, G.~Battistelli, L.~Chisci, and A.~Farina, ``Fusion-{{Based Multidetection Multitarget Tracking With Random Finite Sets}},'' \emph{IEEE Trans. Aerosp. Electron. Syst.}, vol.~57, no.~4, pp. 2438--2458, Aug. 2021.

\bibitem{palkki2011}
R.~D. Palkki, A.~D. Lanterman, and W.~D. Blair, ``Addressing {{Track Hypothesis Coalescence}} in {{Sequential}} \${{K}}\$-{{Best Multiple Hypothesis Tracking}},'' \emph{IEEE Trans. Aerosp. Electron. Syst.}, vol.~47, no.~3, pp. 1551--1563, Jul. 2011.

\bibitem{mahler2007}
R.~Mahler, \emph{Statistical multisource-multitarget information fusion}.\hskip 1em plus 0.5em minus 0.4em\relax Artech, 2007.

\bibitem{wang}
Y.~Wang, H.~Li, and J.~Lu, ``Passive multi-station joint positioning and tracking algorithm based on stochastic finite set theory{(in Chinese)},'' \emph{Electronic information countermeasures technology}, pp. 1--15, 2015.

\bibitem{mahler2003}
R.~Mahler, ``Multitarget bayes filtering via first-order multitarget moments,'' \emph{IEEE Trans. Aerosp. Electron. Syst.}, vol.~39, no.~4, pp. 1152--1178, Oct. 2003.

\bibitem{ba-nguvo2005}
{Ba-Ngu Vo}, S.~Singh, and A.~Boucet, ``Sequential monte carlo methods for multi-target filtering with random finite sets,'' \emph{IEEE Trans. Aerosp. Electron. Syst.}, vol.~41, no.~4, pp. 1224--1245, Oct. 2005.

\bibitem{vo2006}
B.-N. Vo and W.-K. Ma, ``The {{Gaussian Mixture Probability Hypothesis Density Filter}},'' \emph{IEEE Trans. Signal Process.}, vol.~54, no.~11, pp. 4091--4104, Nov. 2006.

\bibitem{kim2024}
J.~Kim, ``Tracking {{Multiple Underwater Targets Using Adaptive Gaussian Mixture Probability Hypothesis Density Filter With Unknown Clutter Rate}},'' \emph{IEEE Trans. Aerosp. Electron. Syst.}, vol.~60, no.~6, pp. 9154--9162, Dec. 2024.

\bibitem{2019改进的SMC}
J.~Pei, Y.~Huang, Y.~Dong, and X.~Chen, ``Improved smc cardinality-balanced multi-bernoulli forwardbackward smoothing track-before-detect algorithm,'' \emph{Journal on Communications}, pp. 102--113, 2019.

\bibitem{gou2024}
J.~Gou, X.~Liu, Y.~Yang, C.~Sun, R.~Zhang, and Q.~Bao, ``Underwater {{Multiple Targets Tracking Using Multi-Bernoulli Filter}} and {{Improved Data Association}},'' in \emph{2024 7th {{International Conference}} on {{Information Communication}} and {{Signal Processing}} ({{ICICSP}})}.\hskip 1em plus 0.5em minus 0.4em\relax Zhoushan, China: IEEE, Sep. 2024, pp. 853--857.

\bibitem{lin2016}
S.~Lin, B.~T. Vo, and S.~E. Nordholm, ``Measurement driven birth model for the generalized labeled multi-{{Bernoulli}} filter,'' in \emph{2016 {{International Conference}} on {{Control}}, {{Automation}} and {{Information Sciences}} ({{ICCAIS}})}.\hskip 1em plus 0.5em minus 0.4em\relax Ansan, South Korea: IEEE, Oct. 2016, pp. 94--99.

\bibitem{legrand2018}
K.~LeGrand and K.~J. DeMars, ``The data-driven delta-generalized labeled multi-{{Bernoulli}} tracker for automatic birth initialization,'' in \emph{Signal {{Processing}}, {{Sensor}}/{{Information Fusion}}, and {{Target Recognition XXVII}}}, I.~Kadar, Ed.\hskip 1em plus 0.5em minus 0.4em\relax Orlando, United States: SPIE, Apr. 2018, p.~4.

\bibitem{vo2013}
B.-T. Vo and B.-N. Vo, ``Labeled {{Random Finite Sets}} and {{Multi-Object Conjugate Priors}},'' \emph{IEEE Trans. Signal Process.}, vol.~61, no.~13, pp. 3460--3475, Jul. 2013.

\bibitem{zheng2024}
C.~Zheng, Y.~Chen, Q.~Wang, X.~Li, S.~Liu, and C.~Dong, ``Track - {{Before- Detect Labeled Multi}} - {{Bernoulli Filter}} for {{Multi- Target Bearing-Only Tracking}} using an {{Autonomous Underwater Vehicle}},'' in \emph{2024 {{IEEE}} 13rd {{Sensor Array}} and {{Multichannel Signal Processing Workshop}} ({{SAM}})}.\hskip 1em plus 0.5em minus 0.4em\relax Corvallis, OR, USA: IEEE, Jul. 2024, pp. 1--5.

\bibitem{williams2015}
J.~L. Williams, ``Marginal multi-bernoulli filters: {{RFS}} derivation of {{MHT}}, {{JIPDA}}, and association-based member,'' \emph{IEEE Trans. Aerosp. Electron. Syst.}, vol.~51, no.~3, pp. 1664--1687, Jul. 2015.

\bibitem{zhang2018}
D.~Zhang, M.-q. Liu, S.-l. Zhang, Z.~Fan, and Q.-f. Zhang, ``Mutual-information based weighted fusion for target tracking in underwater wireless sensor networks,'' \emph{Frontiers Inf Technol Electronic Eng}, vol.~19, no.~4, pp. 544--556, Apr. 2018.

\bibitem{guleria2021data}
K.~Guleria, S.~B. Atham, and A.~Kumar, ``Data fusion in underwater wireless sensor networks and open research challenges,'' in \emph{Energy-Efficient Underwater Wireless Communications and Networking}.\hskip 1em plus 0.5em minus 0.4em\relax IGI Global, 2021, pp. 67--84.

\bibitem{abraham2011}
D.~A. Abraham, J.~M. Gelb, and A.~W. Oldag, ``Background and {{Clutter Mixture Distributions}} for {{Active Sonar Statistics}},'' \emph{IEEE J. Oceanic Eng.}, vol.~36, no.~2, pp. 231--247, Apr. 2011.

\bibitem{han2019}
X.~Han, M.~Liu, S.~Zhang, and Q.~Zhang, ``A {{Multi-Node Cooperative Bearing-Only Target Passive Tracking Algorithm}} via {{UWSNs}},'' \emph{IEEE Sensors J.}, vol.~19, no.~22, pp. 10\,609--10\,623, Nov. 2019.

\bibitem{tan2011}
H.-P. Tan, R.~Diamant, W.~K. Seah, and M.~Waldmeyer, ``A survey of techniques and challenges in underwater localization,'' \emph{Ocean Engineering}, vol.~38, no. 14-15, pp. 1663--1676, Oct. 2011.

\bibitem{qu2007}
Y.~Qu, Z.~Liu, and S.~Sun, ``The {{Research}} of {{Underwater Target Tracking Adaptive Algorithm Based}} on {{Bearings}} and {{Time-Delay}},'' in \emph{2007 {{IEEE International Conference}} on {{Integration Technology}}}.\hskip 1em plus 0.5em minus 0.4em\relax Shenzhen, China: IEEE, Mar. 2007, pp. 530--533.

\bibitem{zhang2022}
B.~Zhang, X.~Hou, and Y.~Yang, ``Robust underwater direction-of-arrival tracking with uncertain environmental disturbances using a uniform circular hydrophone array,'' \emph{The Journal of the Acoustical Society of America}, vol. 151, no.~6, pp. 4101--4113, Jun. 2022.

\bibitem{hou2021}
X.~Hou, J.~Zhou, Y.~Yang, L.~Yang, and G.~Qiao, ``{{3D Underwater Uncooperative Target Tracking}} for a {{Time-Varying Non-Gaussian Environment}} by {{Distributed Passive Underwater Buoys}},'' \emph{Entropy}, vol.~23, no.~7, p. 902, Jul. 2021.

\bibitem{sarkka2009}
S.~Sarkka and A.~Nummenmaa, ``Recursive {{Noise Adaptive Kalman Filtering}} by {{Variational Bayesian Approximations}},'' \emph{IEEE Trans. Automat. Contr.}, vol.~54, no.~3, pp. 596--600, Mar. 2009.

\bibitem{huang2007}
D.~Huang, H.~Leung, and E.-S. Naser, ``Expectation maximization based gps/ins integration for land-vehicle navigation,'' \emph{IEEE Transactions on Aerospace and Electronic Systems}, vol.~43, no.~3, pp. 1168--1177, 2007.

\bibitem{li2013}
W.~Li, Y.~Li, S.~Ren, and X.~Feng, ``Tracking an underwater maneuvering target using an adaptive {{Kalman}} filter,'' in \emph{2013 {{IEEE International Conference}} of {{IEEE Region}} 10 ({{TENCON}} 2013)}.\hskip 1em plus 0.5em minus 0.4em\relax Xi'an, China: IEEE, Oct. 2013, pp. 1--4.

\bibitem{kumar2021b}
K.~Kumar, S.~Bhaumik, and P.~Date, ``Extended {{Kalman Filter Using Orthogonal Polynomials}},'' \emph{IEEE Access}, vol.~9, pp. 59\,675--59\,691, 2021.

\bibitem{hu2010}
Y.~Hu, Z.~Duan, and D.~Zhou, ``Estimation {{Fusion}} with {{General Asynchronous Multi-Rate Sensors}},'' \emph{IEEE Trans. Aerosp. Electron. Syst.}, vol.~46, no.~4, pp. 2090--2102, Oct. 2010.

\bibitem{jia2020}
J.~Jia, Z.~Lai, Y.~Qian, and Z.~Yao, ``Aerial {{Video Trackers Review}},'' \emph{Entropy}, vol.~22, no.~12, p. 1358, Nov. 2020.

\bibitem{snyder2012}
J.~Snyder, Y.~Silverman, Y.~Bai, and M.~A. MacIver, ``Underwater object tracking using electrical impedance tomography,'' in \emph{2012 {{IEEE}}/{{RSJ International Conference}} on {{Intelligent Robots}} and {{Systems}}}.\hskip 1em plus 0.5em minus 0.4em\relax Vilamoura-Algarve, Portugal: IEEE, Oct. 2012, pp. 520--525.

\bibitem{wang2025a}
K.~Wang, Y.~Tang, T.~Q. Duong, S.~R. Khosravirad, O.~A. Dobre, and G.~K. Karagiannidis, ``Multi-{{Tier Distributed Computing Systems}} by {{Leveraging Digital Twin}}: {{Challenges}}, {{Techniques}}, and {{Research Directions}},'' \emph{IEEE Wireless Commun.}, pp. 1--9, 2025.

\bibitem{liu2024}
W.~Liu, Y.~Li, L.~Li, W.~Zhang, and W.~Huang, ``Target positioning of dual forward looking sonars based on orthogonal detection,'' \emph{Mechatronics}, vol.~98, p. 103135, Apr. 2024.

\bibitem{huang2019}
Z.~Huang, J.~Xu, Z.~Gong, H.~Wang, and Y.~Yan, ``Multiple {{Source Localization}} in a {{Shallow Water Waveguide Exploiting Subarray Beamforming}} and {{Deep Neural Networks}},'' \emph{Sensors}, vol.~19, no.~21, p. 4768, Nov. 2019.

\bibitem{yang2018}
H.~Yang, S.~Shen, X.~Yao, M.~Sheng, and C.~Wang, ``Competitive {{Deep-Belief Networks}} for {{Underwater Acoustic Target Recognition}},'' \emph{Sensors}, vol.~18, no.~4, p. 952, Mar. 2018.

\bibitem{abdul2022}
Z.~K. Abdul and A.~K. {Al-Talabani}, ``Mel {{Frequency Cepstral Coefficient}} and its {{Applications}}: {{A Review}},'' \emph{IEEE Access}, vol.~10, pp. 122\,136--122\,158, 2022.

\bibitem{yao2023}
Q.~Yao, Y.~Wang, and Y.~Yang, ``Underwater {{Acoustic Target Recognition Based}} on {{Data Augmentation}} and {{Residual CNN}},'' \emph{Electronics}, vol.~12, no.~5, p. 1206, Mar. 2023.

\bibitem{zhu2019}
B.~Zhu, X.~Wang, Z.~Chu, Y.~Yang, and J.~Shi, ``Active {{Learning}} for {{Recognition}} of {{Shipwreck Target}} in {{Side-Scan Sonar Image}},'' \emph{Remote Sensing}, vol.~11, no.~3, p. 243, Jan. 2019.

\bibitem{niu2019}
H.~Niu, Z.~Gong, E.~Ozanich, P.~Gerstoft, H.~Wang, and Z.~Li, ``Deep-learning source localization using multi-frequency magnitude-only data,'' \emph{The Journal of the Acoustical Society of America}, vol. 146, no.~1, pp. 211--222, Jul. 2019.

\bibitem{huang2018}
Z.~Huang, J.~Xu, Z.~Gong, H.~Wang, and Y.~Yan, ``A {{Deep Neural Network Based Method}} of {{Source Localization}} in a {{Shallow Water Environment}},'' in \emph{2018 {{IEEE International Conference}} on {{Acoustics}}, {{Speech}} and {{Signal Processing}} ({{ICASSP}})}.\hskip 1em plus 0.5em minus 0.4em\relax Calgary, AB: IEEE, Apr. 2018, pp. 3499--3503.

\bibitem{liu2020}
Y.~Liu, H.~Niu, and Z.~Li, ``A multi-task learning convolutional neural network for source localization in deep ocean,'' \emph{The Journal of the Acoustical Society of America}, vol. 148, no.~2, pp. 873--883, Aug. 2020.

\bibitem{wu2021}
F.~Wu, H.~Luo, H.~Jia, F.~Zhao, Y.~Xiao, and X.~Gao, ``Predicting the {{Noise Covariance With}} a {{Multitask Learning Model}} for {{Kalman Filter-Based GNSS}}/{{INS Integrated Navigation}},'' \emph{IEEE Trans. Instrum. Meas.}, vol.~70, pp. 1--13, 2021.

\bibitem{gao2024}
W.~Gao, Y.~Liu, and D.~Chen, ``A {{Dual-Stream Deep Learning-Based Acoustic Denoising Model}} to {{Enhance Underwater Information Perception}},'' \emph{Remote Sensing}, vol.~16, no.~17, p. 3325, Sep. 2024.

\bibitem{pala2025}
S.~Pala, K.~Singh, C.-P. Li, O.~A. Dobre, and T.~Q. Duong, ``Joint {{Beamforming Design}} and {{Sensing}} in {{Satellite}} and {{RIS-Enhanced Terrestrial Networks}}: {{A Federated Learning Approach}},'' \emph{IEEE Trans. Cogn. Commun. Netw.}, pp. 1--1, 2025.

\bibitem{dawkins2024}
M.~Dawkins and et~al., ``{{FishTrack23}}: {{An Ensemble Underwater Dataset}} for {{Multi-Object Tracking}},'' in \emph{2024 {{IEEE}}/{{CVF Winter Conference}} on {{Applications}} of {{Computer Vision}} ({{WACV}})}.\hskip 1em plus 0.5em minus 0.4em\relax Waikoloa, HI, USA: IEEE, Jan. 2024, pp. 7152--7161.

\bibitem{islam2022}
K.~Y. Islam, I.~Ahmad, D.~Habibi, and A.~Waqar, ``A survey on energy efficiency in underwater wireless communications,'' \emph{Journal of Network and Computer Applications}, vol. 198, p. 103295, Feb. 2022.

\bibitem{yang2021}
Y.~Yang, Y.~Xiao, and T.~Li, ``A {{Survey}} of {{Autonomous Underwater Vehicle Formation}}: {{Performance}}, {{Formation Control}}, and {{Communication Capability}},'' \emph{IEEE Commun. Surv. Tutorials}, vol.~23, no.~2, pp. 815--841, 2021.

\bibitem{zhao2021}
H.~Zhao, J.~Yan, X.~Luo, and X.~Guan, ``Ubiquitous {{Tracking}} for {{Autonomous Underwater Vehicle With IoUT}}: {{A Rigid-Graph-Based Solution}},'' \emph{IEEE Internet Things J.}, vol.~8, no.~18, pp. 14\,094--14\,109, Sep. 2021.

\bibitem{zhang2014}
S.~Zhang, H.~Chen, and M.~Liu, ``Adaptive sensor scheduling for target tracking in underwater wireless sensor networks,'' in \emph{2014 {{International Conference}} on {{Mechatronics}} and {{Control}} ({{ICMC}})}.\hskip 1em plus 0.5em minus 0.4em\relax Jinzhou, China: IEEE, Jul. 2014, pp. 55--60.

\bibitem{Liu2019}
M.~Liu, Q.~Zhang, and S.~Zhang, ``Computationally efficient target-node geometry selection for target tracking in {{UWSNs}},'' in \emph{19 th Int Conf on Information Fusion}, 2016.

\bibitem{tian2022}
S.~Tian and Z.~Zhang, ``A {{Node Selection Algorithm Based}} on {{Multi-Objective Optimization Under Position Floating}},'' \emph{IEEE Access}, vol.~10, pp. 41\,863--41\,873, 2022.

\bibitem{qin2023}
Y.~Qin, H.~Liu, R.~Yin, S.~Zhao, and M.~Dong, ``Non-cooperative target tracking method based on underwater acoustic sensor networks,'' \emph{J Supercomput}, vol.~79, no.~17, pp. 19\,227--19\,253, Nov. 2023.

\bibitem{luo2021}
J.~Luo, Y.~Chen, M.~Wu, and Y.~Yang, ``A {{Survey}} of {{Routing Protocols}} for {{Underwater Wireless Sensor Networks}},'' \emph{IEEE Commun. Surv. Tutorials}, vol.~23, no.~1, pp. 137--160, 2021.

\bibitem{xu2024a}
J.~Xu, G.~Xie, Z.~Zhang, X.~Hou, D.~Ma, S.~Zhang, Y.~Ren, and D.~Niyato, ``Is {{FISHER All You Need}} in {{The Multi-AUV Underwater Target Tracking Task}}?'' Dec. 2024.

\bibitem{anerousis2021}
N.~Anerousis, P.~Chemouil, A.~A. Lazar, N.~Mihai, and S.~B. Weinstein, ``The {{Origin}} and {{Evolution}} of {{Open Programmable Networks}} and {{SDN}},'' \emph{IEEE Commun. Surv. Tutorials}, vol.~23, no.~3, pp. 1956--1971, 2021.

\bibitem{espinelsarmiento2021}
D.~Espinel~Sarmiento, A.~Lebre, L.~Nussbaum, and A.~Chari, ``Decentralized {{SDN Control Plane}} for a {{Distributed Cloud-Edge Infrastructure}}: {{A Survey}},'' \emph{IEEE Commun. Surv. Tutorials}, vol.~23, no.~1, pp. 256--281, 2021.

\bibitem{das2020}
R.~K. Das, N.~Ahmed, F.~H. Pohrmen, A.~K. Maji, and G.~Saha, ``{{6LE-SDN}}: {{An Edge-Based Software-Defined Network}} for {{Internet}} of {{Things}},'' \emph{IEEE Internet Things J.}, vol.~7, no.~8, pp. 7725--7733, Aug. 2020.

\bibitem{lin2020}
C.~Lin, G.~Han, M.~Guizani, Y.~Bi, J.~Du, and L.~Shu, ``An {{SDN Architecture}} for {{AUV-Based Underwater Wireless Networks}} to {{Enable Cooperative Underwater Search}},'' \emph{IEEE Wireless Commun.}, vol.~27, no.~3, pp. 132--139, Jun. 2020.

\bibitem{casas-velasco2022}
D.~M. {Casas-Velasco}, O.~M.~C. Rendon, and N.~L.~S. Da~Fonseca, ``{{DRSIR}}: {{A Deep Reinforcement Learning Approach}} for {{Routing}} in {{Software-Defined Networking}},'' \emph{IEEE Trans. Netw. Serv. Manage.}, vol.~19, no.~4, pp. 4807--4820, Dec. 2022.

\end{thebibliography}

\end{document}